\documentclass[fleqn,usenatbib]{mnras}
\usepackage{subcaption}

\usepackage{newtxtext,newtxmath}

\usepackage[T1]{fontenc}

\usepackage{graphicx}   
\usepackage{amsmath}    
\usepackage{comment}
\usepackage{epsfig}
\usepackage{xcolor}
\usepackage{xargs}
\usepackage{tabularx}
\usepackage{longtable}
\usepackage{threeparttable}
\usepackage{hyperref}
\hypersetup{colorlinks=true, citecolor=blue, linkcolor=blue}
\usepackage{xspace}
\let\oldAA\AA
\renewcommand{\AA}{\text{\oldAA}\xspace}

\newcommand{\redtxt}[1]{\textcolor{black}{#1}}
\newcommand{\xj}[1]{\textcolor{orange}{#1}}
\newcommand{\fde}[1]{\textcolor{blue}{#1}}

\newcommand{\hii}{H\,{\sc ii}}
\newcommand{\hei}{He\,{\sc i}\xspace}
\newcommand{\heii}{He\,{\sc ii}}
\newcommand{\neiii}{[Ne\,{\sc iii}]}
\newcommand{\oiii}{[O\,{\sc iii}]}

\newcommand{\oii}{[O\,{\sc ii}]}
\newcommand{\oi}{[O\,{\sc i}]}

\newcommand{\civ}{C\,{\sc iv}}
\newcommand{\nai}{Na\,{\sc i}\xspace}
\newcommand{\feii}{Fe\,{\sc ii}}
\newcommand{\fev}{[Fe\,{\sc v}]}
\newcommand{\caii}{Ca\,{\sc ii}}
\newcommand{\ariii}{[Ar\,\textsc{iii}]}

\newcommand{\nii}{[N\,{\sc ii}]}
\newcommand{\sii}{[S\,{\sc ii}]}

\newcommand{\ha}{H\ensuremath{\alpha}\xspace}
\newcommand{\hb}{H$\beta$\xspace}
\newcommand{\hg}{H$\gamma$\xspace}
\newcommand{\hd}{H$\delta$\xspace}
\newcommand{\jwst}{\textit{JWST}\xspace}
\newcommand{\chandra}{\textit{Chandra}\xspace}
\newcommand{\ergs}{$\rm erg~s^{-1}$}
\newcommand{\ergscm}{$\rm erg~s^{-1}~cm^{-2}$}
\newcommand{\kms}{\ensuremath{\rm km~s^{-1}}\xspace}
\newcommandx{\fluxdcgs}[1][1=-20]{$\times 10^{[#1]}$~erg~s$^{-1}$~cm$^{-2}$~\AA$^{-1}$\xspace}
\newcommandx{\fluxcgs}[2][1=-20,2=\ensuremath{\times}]{${#2}10^{#1}$~erg~s$^{-1}$~cm$^{-2}$\xspace}
\newcommand\sbullet[1][.5]{\mathbin{\vcenter{\hbox{\scalebox{#1}{$\bullet$}}}}}

\newcommand{\mbh}{\ensuremath{M_{\sbullet[0.85]}}\xspace}
\newcommand{\Msun}{\ensuremath{{\rm M}_\odot}\xspace}
\newcommand{\ledd}{\ensuremath{\lambda_\mathrm{Edd}}\xspace}

\newcommand{\target}{J1025+1402\xspace}

\newcommand{\cloudy}{\textsc{Cloudy}\xspace}
\newcommand{\pyneb}{\textsc{PyNeb}\xspace}

\newcommand{\galex}{\textit{GALEX}\xspace}
\newcommand{\nustar}{\textit{NuSTAR}\xspace}

\title[Little Red Dot at $z=0.1$]{\centering Lord of LRDs: Insights into a ``Little Red Dot'' with a low-ionization spectrum at $z=0.1$}

\author[Ji et al.]{Xihan Ji,$^{1,2}$\thanks{E-mail: \href{mailto:xj274@cam.ac.uk}{xj274@cam.ac.uk}}
Francesco D'Eugenio,$^{1,2}$
Ignas Juodžbalis,$^{1,2}$
Dominic J. Walton,$^{3}$
Andrew C. Fabian,$^{4}$
\newauthor
Roberto Maiolino,$^{1,2,5}$
Cristina Ramos Almeida,$^{6,7}$
Jose A. Acosta Pulido,$^{6,7}$
Vasily A. Belokurov,$^{4}$
\newauthor
Yuki Isobe,$^{1,2,8}$
Gareth Jones,$^{1,2}$
Claudia Maraston,$^{9}$
Jan Scholtz,$^{1,2}$
Charlotte Simmonds,$^{1,2}$
Sandro Tacchella,$^{1,2}$
\newauthor
Elena Terlevich,$^{10,11,12}$
Roberto Terlevich$^{10,4,11}$
\\
$^{1}$Kavli Institute for Cosmology, University of Cambridge, Madingley Road, Cambridge, CB3 0HA, UK\\
$^{2}$Cavendish Laboratory, University of Cambridge, 19 JJ Thomson Avenue, Cambridge, CB3 0HE, UK\\
$^{3}$Centre for Astrophysics Research, University of Hertfordshire, College Lane, Hatfield AL10 9AB, UK\\
$^{4}$Institute of Astronomy, University of Cambridge, Madingley Road, Cambridge CB3 0HA, UK\\
$^{5}$Department of Physics and Astronomy, University College London, Gower Street, London WC1E 6BT, UK\\
$^{6}$ Instituto de Astrofísica de Canarias, Calle Vía Láctea, s/n, E-38205, La Laguna, Tenerife, Spain\\
$^{7}$ Departamento de Astrofísica, Universidad de La Laguna, E-38206, La Laguna, Tenerife, Spain\\
$^{8}$ Waseda Research Institute for Science and Engineering, Faculty of Science and Engineering, Waseda University, 3-4-1, Okubo, Shinjuku, Tokyo 169-8555, Japan\\
$^{9}$ Institute of Cosmology, University of Portsmouth, Burnaby Road, Portsmouth PO1 3FX, UK\\
$^{10}$ Instituto Nacional de Astrof\'{ı}sica, Optica y Electr\'onica, Tonantzintla, Puebla, M\'exico\\
$^{11}$ Facultad de Astronom\'{ı}a y Geof\'{ı}sica, Universidad de La Plata, La Plata, Argentina\\
$^{12}$ Visiting Astronomer, Institute of Astronomy, University of Cambridge, Madingley Road, Cambridge CB3 0HA, UK
}

\begin{document} 
\label{firstpage}
\pagerange{\pageref{firstpage}--\pageref{lastpage}}
\maketitle
 
\begin{abstract}
  Recent observations \redtxt{by the \textit{James Webb Space Telescope} (\jwst)} have revealed a puzzling population of optically red and compact galaxies with peculiar ``V’’-shaped spectra at high redshift, known as ``Little Red Dots'' (LRDs).
  Until now, most spectroscopically confirmed LRDs are found at $z>4$ and it has been speculated that LRDs are tracing the early stages of black hole evolution.
  We report an independent rediscovery of a broad-line active galactic nucleus (AGN), SDSS J102530.29+140207.3, at $z=0.1$, which shows spectral features matching those of LRDs seen in the early Universe, including the V-shaped spectrum, broad Balmer lines (with widths of 1000\,-\,2000 \kms), and deep Balmer absorption.
  We present a new GTC observation of this LRD, which reveals an optical continuum similar to those of G-to-K giant stars including an unambiguous G-band absorption originating from the CH molecule.
  In addition, this local LRD shows a series of absorption lines potentially related to low-ionization ions or atoms but are deeper than what is observed in empirical stellar templates.
  We further identify a series of [\feii] emission lines indicative of low-ionization gas, which we find also present in a \jwst-selected LRD at $z=2.26$.
  We find small but statistically significant variability in \redtxt{the} \ha\ \redtxt{of SDSS J102530.29+140207.3} consistent with previous findings.
  Finally, \redtxt{we report} new \redtxt{observations with} \textit{NuSTAR}. We confirm the extreme X-ray weakness of this LRD, which might imply Compton-thick gas obscuration with $N_{\rm H}>10^{24}~{\rm cm^{-2}}$.
  All evidence suggests SDSS J102530.29+140207.3 has a complex gaseous environment and the strong ionic, atomic, and molecular absorptions are hard to explain with typical stellar and AGN models.
\end{abstract}

%

\section{Introduction}

One of the most exciting and puzzling discoveries of \textit{James Webb Space Telescope} \citep[\jwst,][]{jwst0,jwst1} is the abundant population of compact and optically red galaxies known as ``Little Red Dots'' (LRDs) typically found at \redtxt{$z>4$} \citep[e.g.,][]{kocevski2023,Kocevski_lrd_2024,harikane2023,matthee2024,greene2024,wang_break_2024,Kokorev_lrd_2024,labbe_monster,Labbe_lrd_2025,taylor_agn_2024}.
These galaxies ubiquitously show red optical colors and blue \redtxt{ultraviolet (UV)} slopes, leading to a peculiar `V'-shaped spectral turnover near the Balmer limit \citep[i.e., $\sim 3646$ \AA, see, e.g.,][]{greene2024,setton_lrd_2024}.
{In the local Universe, UV upturn is observed in early-type galaxies, but in this case the turnover point typically occurs at $\sim 2500$ \AA, which is likely produced by old and evolved low-mass stars \citep[e.g.,][]{maraston_2000,LeCras_2016,Martocchia_2025}, different from that observed in LRDs.
Also, the V-shaped turnover of LRDs is not observed in the typical continua of accreting supermassive black holes in the local Universe \citep{vandenberk2001}.
}

Another key characteristic of LRDs is their compact morphology, and the UV and optical light of many LRDs remains spatially unresolved even in the \jwst/Near-Infrared Camera (NIRCam) images (see, however, \citealp{rinaldi_lrd_2024} for cases where the UV images of LRDs are resolved), suggesting physical sizes smaller than 100\,-\,300 pc for the general population \citep{akins_lrd_2024} and even within 30 pc for particular cases \citep{furtak_abell2744_2023,furtak2023}.
To date, the physical nature of LRDs remains highly debated, and the explanations for their peculiar UV-optical turnover include stellar continua of massive and compact stellar populations \citep{perez-gonzalez_lrd_2024,baggen_lrd_2024,wang_break_2024,ma_lrd_2024,Nandal_smslrd_2025}, gas-obscured accretion disc emission of active galactic nuclei \citep[AGN,][]{Inayoshi_maiolino_2025,ji_lrdbreak_2025,degraaff_lrd_2025,naidu_lrd_2025,Begelman_2025,Liu_speddlrd_2025}, accretion disc emission of AGN attenuated by special dust attenuation curves \citep{Lizhengrong_lrd_2024}, or emission from extended and gravitationally unstable accretion discs of AGN \citep{Zhang_lrddisk_2025}.

One of the consequences of the stellar interpretation of the UV-optical turnover is the high stellar masses ($>\!10^{9-10}\,M_\odot$) for LRDs, which are mainly derived from the strong Balmer breaks in the turnover region.
These large masses, together with the high comoving-volume density of LRDs, lead to a stellar-mass function potentially incompatible with the standard $\rm \Lambda CDM$ cosmological model \citep[e.g.,][]{akins_lrd_2024,inayoshi_ombh_2024}.
Additionally, the compactness of LRDs implies extremely high stellar-mass densities reaching $10^{5-6}~M_\odot ~\text{pc}^{-2}$ within their half-light radii \citep{baggen_lrd_2024,labbe_monster,ma_lrd_2024}.
Such a high stellar density is only seen in the nuclear star clusters (NSCs) of local galaxies \citep{Neumayer_2020}, whose stellar masses are several orders of magnitude lower than those inferred for LRDs.
Furthermore, the high stellar masses of LRDs are incompatible with the low dynamical masses inferred from the widths of their emission lines \citep{juodzbalis_rosetta_2024,wangbingjie_2024,ji_lrdbreak_2025,deugenio_qso1_2025,Akins_ci_2025}.

Recently, it has been proposed that the stellar-mass problem of LRDs can be alleviated if their observed continua including the strong Balmer breaks are dominated by AGN accretion discs rather than stellar populations \citep{Inayoshi_maiolino_2025,ji_lrdbreak_2025}.
In this picture, the Balmer breaks are driven by dense nuclear gas potentially close to the broad-line region (BLR) that is optically thick to the Balmer continuum, without invoking massive stellar populations.
Recent models also explore the potential origin of the obscuring gas, including the ``quasi-star'' idea where the black hole formed through direct collapses and left a gas envelope \citep{Begelman_2006,Begelman_2025,naidu_lrd_2025} or the turbulent accretion flows surrounding the black hole \citep{Liu_speddlrd_2025}.
\redtxt{Such gas envelopes around early black holes might eventually disappear due to the drop of gas inflow rates due to the gas removal by supernova feedback in galaxies \citep{Inayoshi_gasremoval_2025}.}

It has been speculated that the presence of high-column density and dense gas is ubiquitous in early AGN at $z>2$ discovered by \jwst \citep{Maiolino_jadesagn_2024,juodzbalis_rosetta_2024}, with LRD AGN being a subpopulation that makes up roughly 30\% of the whole AGN population \citep{hainline_lrd_2024}.
Indeed, many of the spectroscopically confirmed LRDs show prominent broad Balmer emission lines with full widths at half maximum (FWHM) $>1000$ \kms, tracing the broad-line regions (BLRs) of AGN \citep{matthee2024,greene2024,rinaldi_lrd_2024}.
However, even the AGN interpretation is not without problems. Notably, these LRD AGN have peculiar properties not commonly seen in local AGN or bright quasars: extremely weak X-ray emission, 
the widespread absence of high-ionization emission lines (see, however, \citealp{Tang_nv_2025} for an exception), weak or no continuum variability, and $\sim$20\% of them exhibit strong Balmer-line (sometimes also helium-line) absorption with a non-stellar origin \citep{ananna_2024,yue_lrd_2024,Maiolino2024_Xrays,matthee2024,akins_lrd_2024,juodzbalis_rosetta_2024,taylor_agn_2024,lin_aspire_2024,labbe_monster,ji_lrdbreak_2025,deugenio_qso1_2025,kokubo_harikane2024,zhang_var_2024}.
The dense gas obscuration might explain the above peculiarities through Compton down-scattering of X-ray photons by high-column density gas (with $N_{\rm H}>1.5\times10^{24}~{\rm cm^{-2}}$) and absorption of hydrogen Balmer and helium transitions \citep{juodzbalis_rosetta_2024,wangbingjie_2024}.
Meanwhile, super-Eddington accretion could serve as an alternative or a complementary mechanism to produce intrinsically weak X-ray emission, weak high-ionization emission lines, weak variability \citep{king2024b,madau_superedd_2024,pacucci_superedd_2024,lambrides_superedd_2024,Inayoshi_agnvar_2024}, and potentially the red optical continuum without dust attenuation \citep{Liu_speddlrd_2025}.
Full investigation of the AGN scenario is usually limited by the lack of constraining multiwavelength data at high redshift \redtxt{\citep{degraaff_lrd_2025,greene_lbol_2025}} and  of multi-epoch observations \citep{kokubo_harikane2024}\footnote{The variability, however, can be well constrained with single-epoch observations of gravitationally lensed sources with multiple images \citep[see e.g.,][]{ji_lrdbreak_2025,Furtak_qso1var_2025}.}.

The main obstacle to studying LRDs is the rapid decline in their population at $z<4$ \citep{Kocevski_lrd_2024}, which could be evidence for a connection with cosmic evolution \citetext{\citealp{Loeb2024,inayoshi_firstacc_2025,Pacucci_2025}; however, note
that large-area \textit{Euclid} and \textit{Spitzer} photometry suggests an increase in comoving volume density of photometric LRD candidates 
with increasing cosmic time, \citealp{euclid_lrd_2025}}.
Thus far, there are only three spectroscopically confirmed LRDs at $z\approx 2$, which are the Rosetta Stone at $z=2.26$ \citep{juodzbalis_rosetta_2024}, DEEP23-z2LRD1 at $z=2.26$ \citep{ma_lrd2d26_2025}, and the Big Red Dot at $z=2.33$ \citep{Loiacono_2025}, and there are no credible spectroscopically selected LRD candidates with \jwst at $z<2$ \citep[although recent results from \textit{Euclid} have pushed the color-selected LRD candidates to $z\approx 0.33$,][]{euclid_lrd_2025}. 
{Establishing a truly local analog would be a huge leap forward in our ability to study these systems in detail}.
First of all, a local object can put stringent constraints on current theories that link LRDs to cosmic evolution, such as the first accretion event on supermassive black holes \citep{inayoshi_firstacc_2025}. 
More importantly, the time required for multi-wavelength observations would be greatly reduced, enabling
a panchromatic view of the spectral energy distribution (SED) of the LRD with stringent constraints on X-ray and radio emission, which is usually not detected in this kind of objects \citep[e.g.,][]{ananna_2024,yue_lrd_2024,akins_lrd_2024}.
Finally, detailed time-domain science would become possible, as cosmic time dilation is negligible. This will allow for variability studies of the AGN properties, which is extremely challenging for high-$z$ LRDs \citep{zhang_var_2024}.

Recently, \citet{Lin_lrdanalog_2025} presented a catalog of 19 broad-line AGN in green pea (GP) galaxies at $z<0.4$ as local spectroscopic analogs of high-redshift LRDs, \redtxt{among which 6 objects have V-shaped UV-optical continua}.
Many of these sources show strong \nii$\lambda 6583$ lines, which imply chemically enriched environments unlike the very metal-poor broad-line AGN selected by \jwst \citep{trefoloni_feii_2024}.
None of these candidates shows clear Balmer absorption as seen in LRDs spectroscopically confirmed by \jwst.
Also, some candidates show strong high-ionization lines such as \heii$\lambda 4686$, which
are, perhaps surprisingly, weak in \jwst-selected broad-line AGN, potentially due to their high-accretion rates (e.g., \citealp{lambrides_superedd_2024,juodzbalis_jadesagn_2025})\footnote{Sometimes strong \heii$\lambda 4686$ emission is also found in star-forming galaxies in the local Universe \citep{Shirazi_2012}.}.
Finally, it remains unclear whether the SEDs of the local analogs outside the optical regime actually match that of the average LRD selected by \jwst.
To find a local LRD, 
it is vital to match all the characteristics of the high-$z$ counterparts mentioned above in observations.
It is thus useful to start the search by looking into spectral features rarely seen in local sources, such as the V-shaped spectrum and the strong Balmer absorption.

More recently, \citet{linxiaojing_locallrd_2025} presented three local broad-line LRDs with V-shaped spectra at $z\approx0.1-0.2$.
These LRDs have low metallicities similar to high-$z$ counterparts, have weak high-ionization lines, and their V-shaped turnover points are close to the Balmer limit.
In addition, two of these LRDs show clear Balmer absorptions, making them the best local LRD candidates thus far.

We have independently identified the brightest and also the lowest redshift LRD present in \citet{linxiaojing_locallrd_2025}'s sample.
In this work,
we present a \redtxt{detailed} analysis of this local LRD, SDSS J102530.29+140207.3 (hereafter \target).
We provide a detailed panchromatic view on \target with the goal of demonstrating its similarities to high-redshift LRDs and how it provides new insights into the nature of the LRD population.
The structure of the manuscript is as follows.
In Sections~\ref{sec:data} and \ref{sec:method}, we present the data we used for \target and our spectral measurements.
In Section~\ref{sec:lrd_com}, we compare the observations of \target with those of high-$z$ LRDs.
In Section~\ref{sec:bh_params}, we perform emission-line and absorption-line diagnostics of \target.
We discuss the physical interpretations of the panchromatic spectrophotometric data of \target in Section~\ref{sec:sed}, and discuss issues with the derived stellar mass and energy budget in Section~\ref{sec:discuss}.
We draw our conclusions in Section~\ref{sec:conclude}.
Throughout this work, we assume a flat $\rm \Lambda CDM$ cosmology with $h=0.674$ and $\Omega _{\rm m} = 0.315$ \citep{planck2020}.
All magnitudes are given in the AB system and wavelengths are given in air.
\redtxt{All observations are corrected for Galactic extinction using the \citet{Schlafly_2011} dust map and a \citet{Fitzpatrick_1999} extinction curve with $R_{\rm V}=3.1$.
}

\section{Spectroscopic and photometric data}
\label{sec:data}

\begin{figure*}
    \centering
    \includegraphics[width=\textwidth]{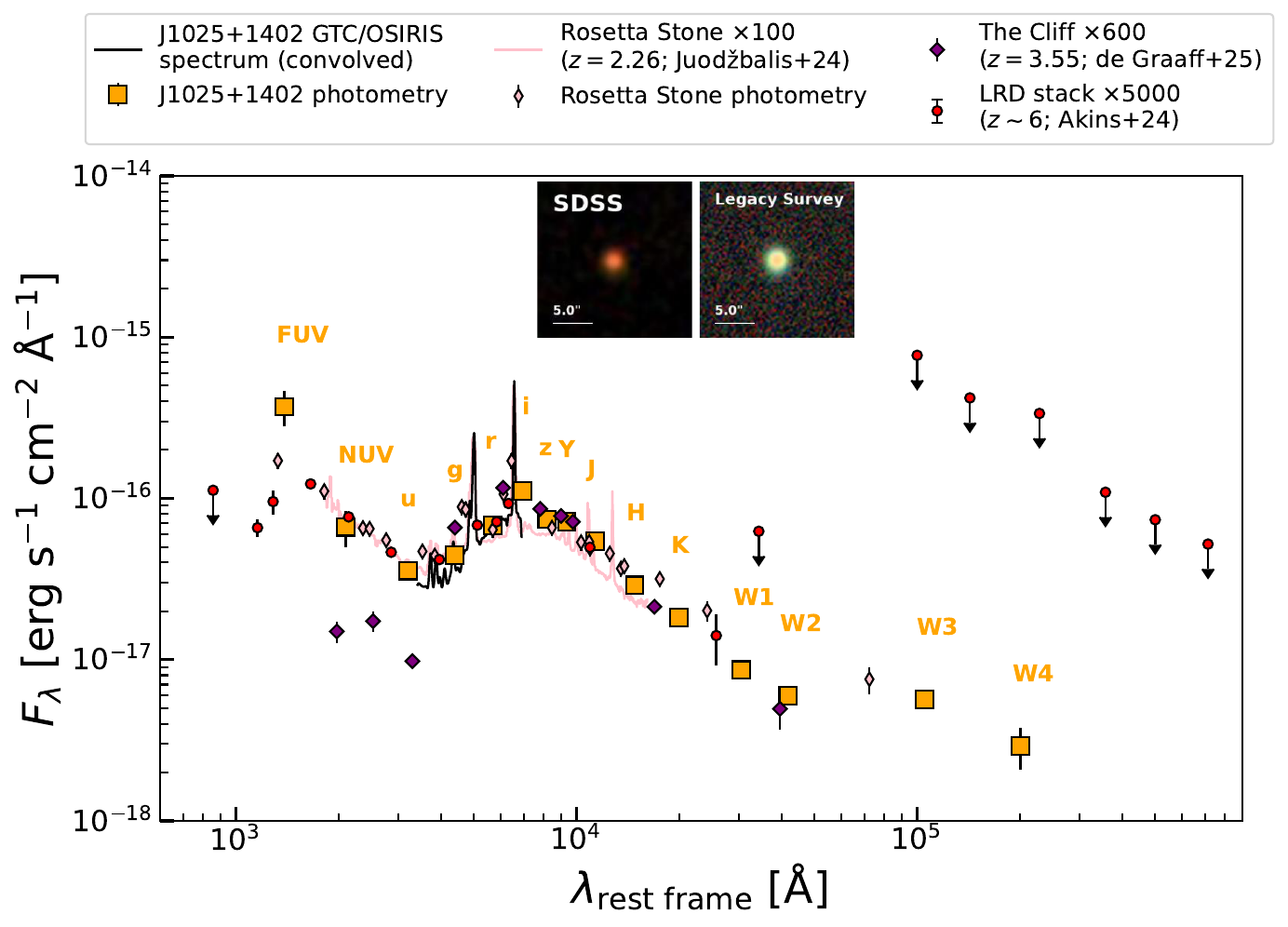}
    \caption{
    Comparison between the spectrophotometric SEDs of \target, the Rosetta Stone (one of the lowest-$z$ LRDs discovered by \jwst at $z=2.26$, \citealp{juodzbalis_rosetta_2024}), the Cliff (one of the LRDs showing the strongest Balmer break at $z=3.55$, \citealp{degraaff_lrd_2025}), and the median stack of the color-selected LRDs at $z\sim6$ from the COSMOS field \citep{Scoville_cosmos_2007} provided by \citet{akins_lrd_2024} (where non detections are plotted as $5\sigma$-upper limits following \citealp{akins_lrd_2024}).
    {The GTC spectrum (solid black) of \target is convolved to the resolution of the PRISM spectrum ($R\sim 100$) of the Rosetta Stone (solid pink) for illustrative purposes.}
    There is a close resemblance among the SEDs of \target, the Rosetta Stone, and the stacked LRDs, once normalized to similar flux density levels, with a common V-shaped turnover in the NUV-optical regime, a blue UV slope, a red optical slope, a peak in the NIR, and weak MIR emission.
    While the Cliff exhibits a much stronger break near the Balmer limit compared to all the other LRDs, its NIR SED is similar to that of \target.
    On the top, we show the SDSS \citep{york2000} $gri$-composite image and the Legacy Survey \citep{Dey_2019} $grz$-composite image of \target, where \target is unresolved.
    }
    \label{fig:full_sed}
\end{figure*}

\redtxt{We started our search of the local LRD candidates using the dwarf galaxy catalog of \citet{Izotov_2007} with broad-line detections in 35 sources at $z\lesssim 0.3$, which might host low-metallicity AGN similar to those found at high redshift \citep{Izotov_2008}.
We used archival photometric and spectroscopic data 
from the Sloan Digital Sky Survey (SDSS, \citealp{york2000}) data release 7 \citep[DR7,][]{Abazajian_dr7_2009}, and the Galaxy Evolution Explorer \citep[\textit{GALEX},][]{galex}
to select sources satisfying the following criteria}


{
\begin{enumerate}
    \item \redtxt{Following the photometric selections with \jwst/NIRCam bands by \citet{Kocevski_lrd_2024},}
    an ambiguous V-shape turnover indicated by photometric bands at 3000\,-\,4000, with the UV slope $\beta_{\rm UV}<-0.37$ \redtxt{(determined by \galex NUV and SDSS $u$ bands)} and the optical slope $\beta _{\rm optical}>0$ \redtxt{(determined by both SDSS $g$ and $r$, and $g$ and $i$ bands)}.
    \item Significant absorption in \ha revealed by archival spectroscopic data \redtxt{from SDSS (see, e.g., \citealp{burke_abs_2021})}.
    \item Point-source like morphology from the \redtxt{SDSS} imaging, where model photometric magnitudes are consistent with the point spread function (PSF) magnitudes within $2\sigma$ uncertainties.
    \item A metal-poor gaseous environment with $Z/Z_\odot \lesssim 0.1$ similar to \jwst-selected broad-line AGN \citep{trefoloni_feii_2024}.
    \redtxt{Here we took the metallicity measurements by \citet{Izotov_2007}.}
    \item High-ionization lines, such as \heii$\lambda 4686$, weaker than typical Seyferts \redtxt{\citep[with the flux ratio of \heii/\hb$\lesssim$0.06, e.g.,][]{juodzbalis_jadesagn_2025}}.
\end{enumerate}

\redtxt{Within the catalog, \target is the only source satisfying all criteria.}
Furthermore, \target has a stable light curve with tentative evidence of small variability in \ha and a high broad \ha luminosity $>10^{41}$ \ergs, which largely rule out the possibility of it being dominated by blue variable stars or supernovae \citep{Izotov_2008,burke_abs_2021}.
}

In this work, we \redtxt{focus on the single source, \target, and examine its physical properties leveraging both spectroscopic and photometric data.
For spectroscopic data, besides SDSS, we obtained archival Gemini-North observations described in \citet{burke_abs_2021}.
}
The publicly available Gemini observations were obtained within the programme GN-2020A-FT-204 (PI C. Burke) using the $1''$ slit and we performed our own reduction as detailed in Appendix~\ref{appendix:gemini_reduction}.
The Gemini spectrum has a wavelength coverage of 6050\,-\,8410 \AA and has an effective spectral resolution of {$R\approx 3400$}.

In addition to the archival spectroscopic data, we obtained new spectroscopic data in the optical using the 10.4 m-Gran Telescopio CANARIAS (GTC) with the Optical System for Imaging and low-Intermediate-Resolution Integrated Spectroscopy (OSIRIS) from the programme GTC03-25ADDT
(PI C. Ramos Almeida).
The observation was performed using the $0.\!\!^{\prime\prime}6$-slit, oriented along the parallactic angle, and R1000B grism with a total exposure time of 2 hr on source, which produces an effective spectral resolution of $R\approx 1000$ in the optical.
We describe the data reduction in detail in Appendix~\ref{appendix:gtc_reduction}.
Due to the smaller size of the slit ($0.\!\!^{\prime\prime}6$) compared to the seeing ($0.\!\!^{\prime\prime}8$\,-\,$1.\!\!^{\prime\prime}2$), there is significant aperture loss in the final spectrum, which we corrected by normalizing the continuum level to that of the SDSS spectrum.
The average normalization factor between the GTC and SDSS spectra is $2.40\pm 0.01$ and the derivation is detailed in Appendix~\ref{appendix:slit_loss}.
Figure~\ref{fig:full_sed} shows the full GTC spectrum covering 3380 \AA - 7140 \AA in the rest frame {(convolved to a low resolution of $R\sim 100$ to compare with the SEDs of high-$z$ LRDs)}.
The signal-to-noise (S/N) of the GTC spectrum is significantly higher than that of the SDSS spectrum, which helps reveal a series of emission lines as we describe later in Section~\ref{sec:method}.

We also collected archival and new photometric observations for \target from X-ray to Mid-infrared (MIR).
The archival X-ray data are from a small \chandra programme (PI T. Thuan) presented by \citet{Simmonds_xraydwarf_2016}, who reported no detection.
We obtained new X-ray observations with a \textit{NuSTAR} Director's Discretionary Time programme (PI A. Fabian) for 40 ks on-source that reaches high energies up to 50 keV and still had no detection.
The rest of the photometric data are queried through the photometric tool \textsc{VizieR} \citep{vizier}.
Specifically, we extracted FUV and NUV data from \textit{GALEX} \citep{galex}, $u,g,r,i,z$-band data from SDSS \citep{york2000}, 
Y, J, H, K-band data from UKIDSS \citep{ukidss}, and W1, W2, W3, W4-band data from \textit{WISE} \citep{wise}.
In Figure~\ref{fig:full_sed}, we plot the archival photometric points of \target, which show the characteristic V-shape in the UV-optical regime with the turnover point close to the location of SDSS $u$ band also revealed by the GTC/OSIRIS spectrum.
Next, we describe measurements of spectral features for \target.

\section{Spectral measurements}
\label{sec:method}

\begin{figure*}
{\phantomsubcaption\label{f.gmos.gaussfit.a}
 \phantomsubcaption\label{f.gmos.gaussfit.b}
 \phantomsubcaption\label{f.gmos.gaussfit.c}
 \phantomsubcaption\label{f.gmos.gaussfit.d}
 \phantomsubcaption\label{f.gmos.gaussfit.e}
 \phantomsubcaption\label{f.gmos.gaussfit.f}
 \phantomsubcaption\label{f.gmos.gaussfit.g} 
 \phantomsubcaption\label{f.gmos.gaussfit.h}}
\includegraphics[width=2\columnwidth]{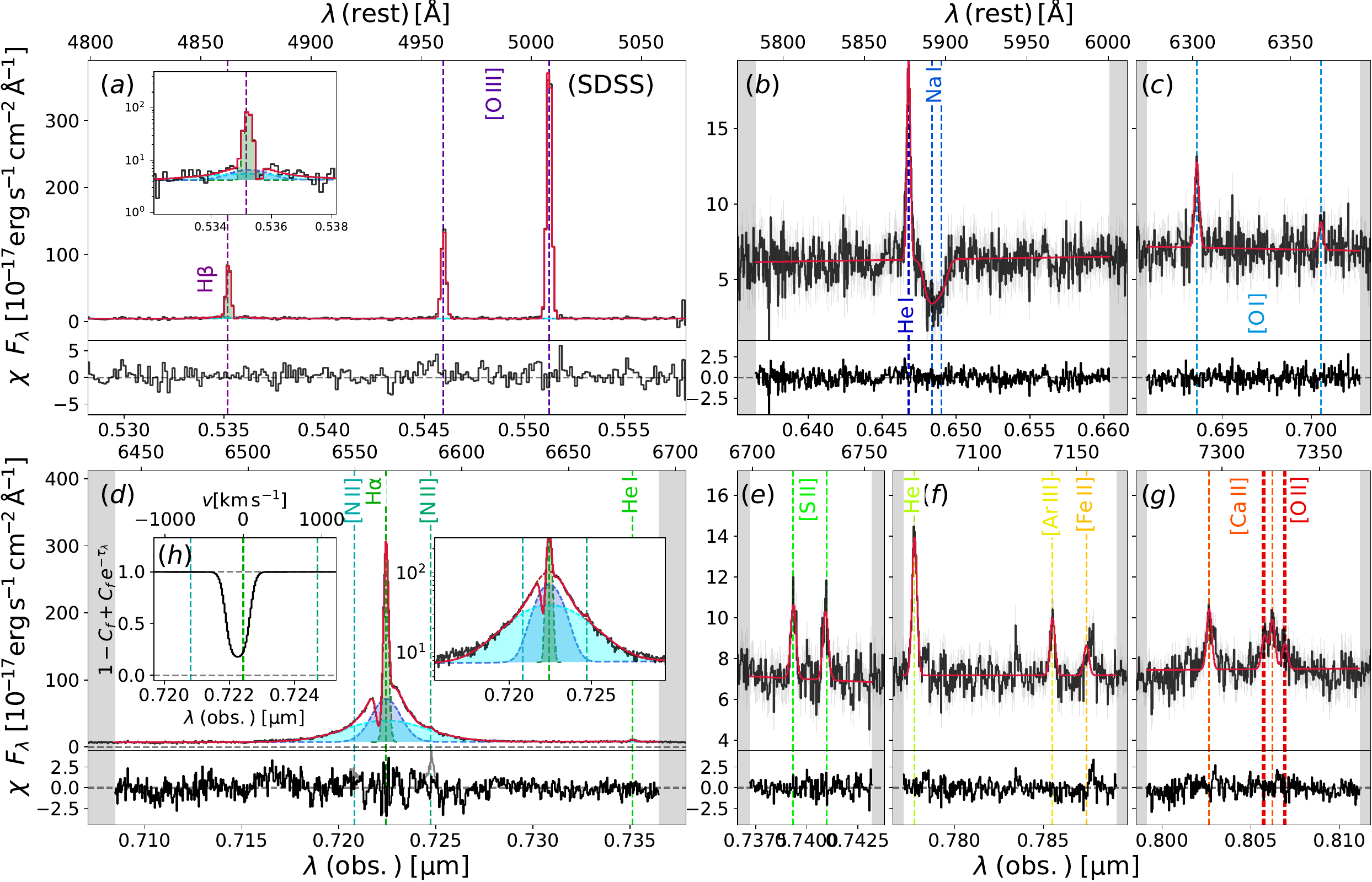}
\caption{Best-fit spectral models for part of the observed spectra of \target.
\textit{Panel (a)} SDSS spectrum around \hb and \oiii, where \hb has a narrow component, a broad component, and a redshifted absorption (modelled with Equation~\ref{eq:abs}) shown in the zoomed-in panel.
{Panels (b) and (c)} Gemini/GMOS spectrum around \hei, NaD, and \oi, where NaD is modelled as an absorber with kinematics independent of narrow emission lines.
\textit{Panel (d)} Gemini/GMOS spectrum around \ha, \nii, and \hei, where \ha has a narrow component, a broad component (modelled as a double-Gaussian function), and a blueshifted absorption shown in the zoomed-in panel.
\textit{Panels (e), (f), and (g)} Gemini/GMOS spectrum around \sii, \hei, [Ar\,{\sc iii}], [\feii], [\caii], and \oii.
[\feii] and [Ca\,{\sc ii}] show different line profiles compared to other narrow lines, suggesting that they come from a different region.
}\label{f.gmos.gaussfit}
\end{figure*}

\begin{figure*}
    \centering
    \includegraphics[width=\textwidth]{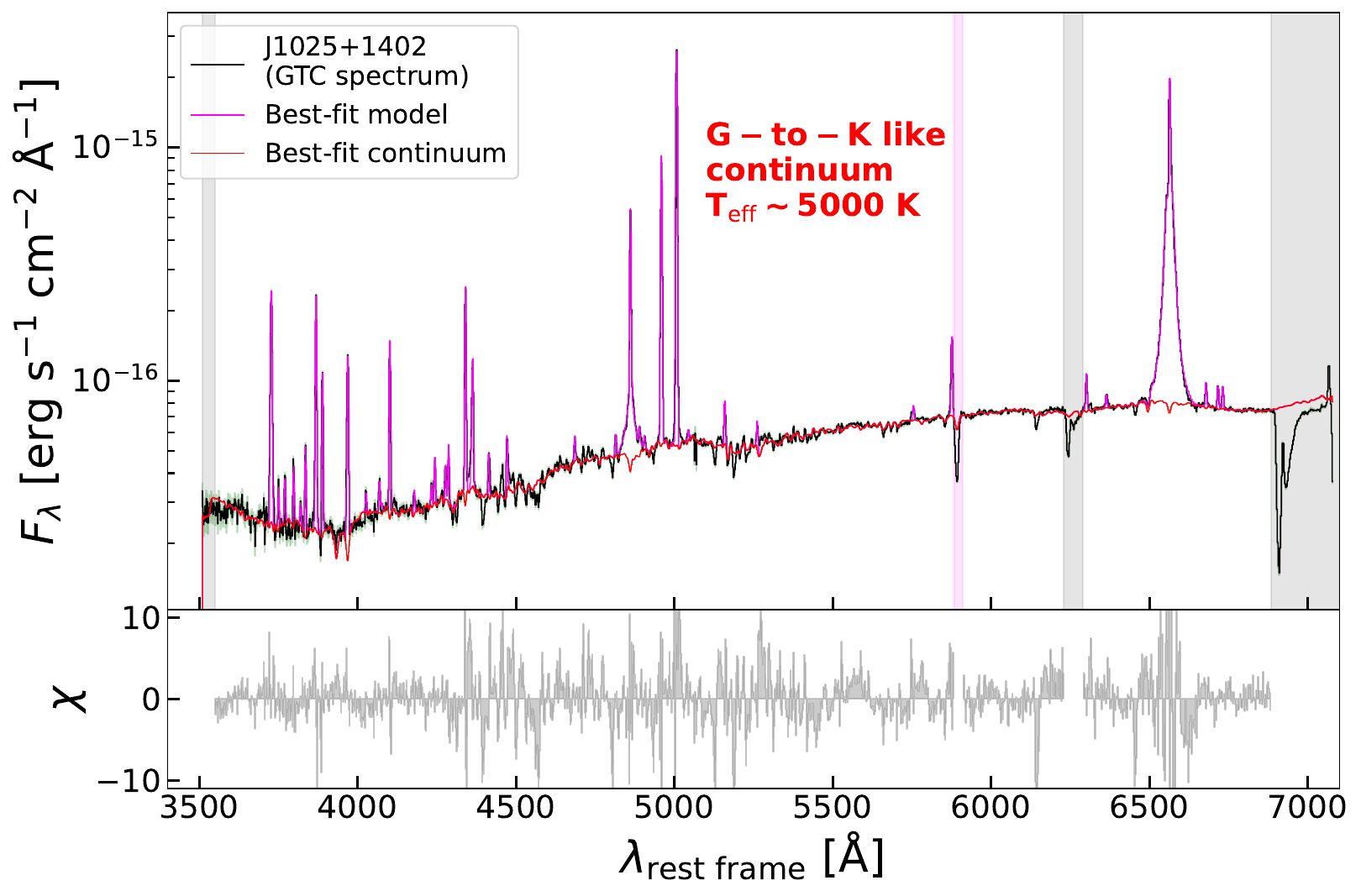}
    \includegraphics[width=\columnwidth]{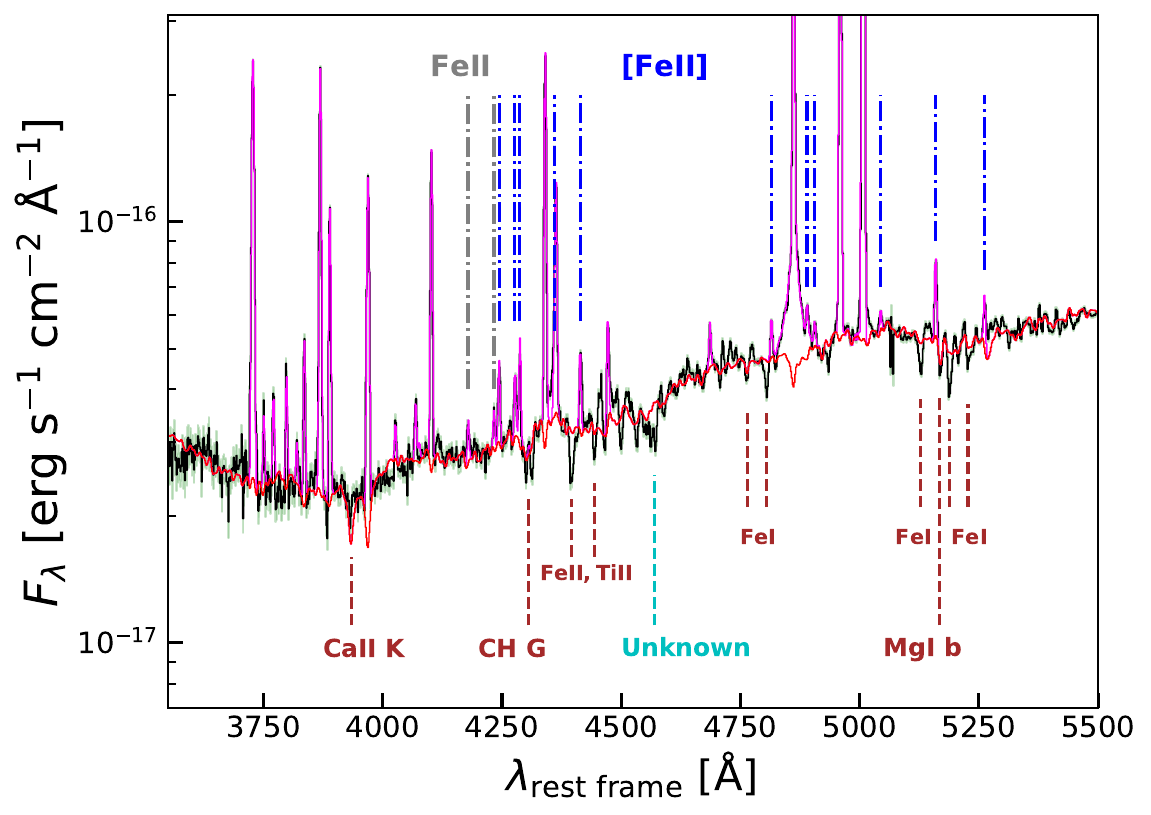}
    \includegraphics[width=\columnwidth]{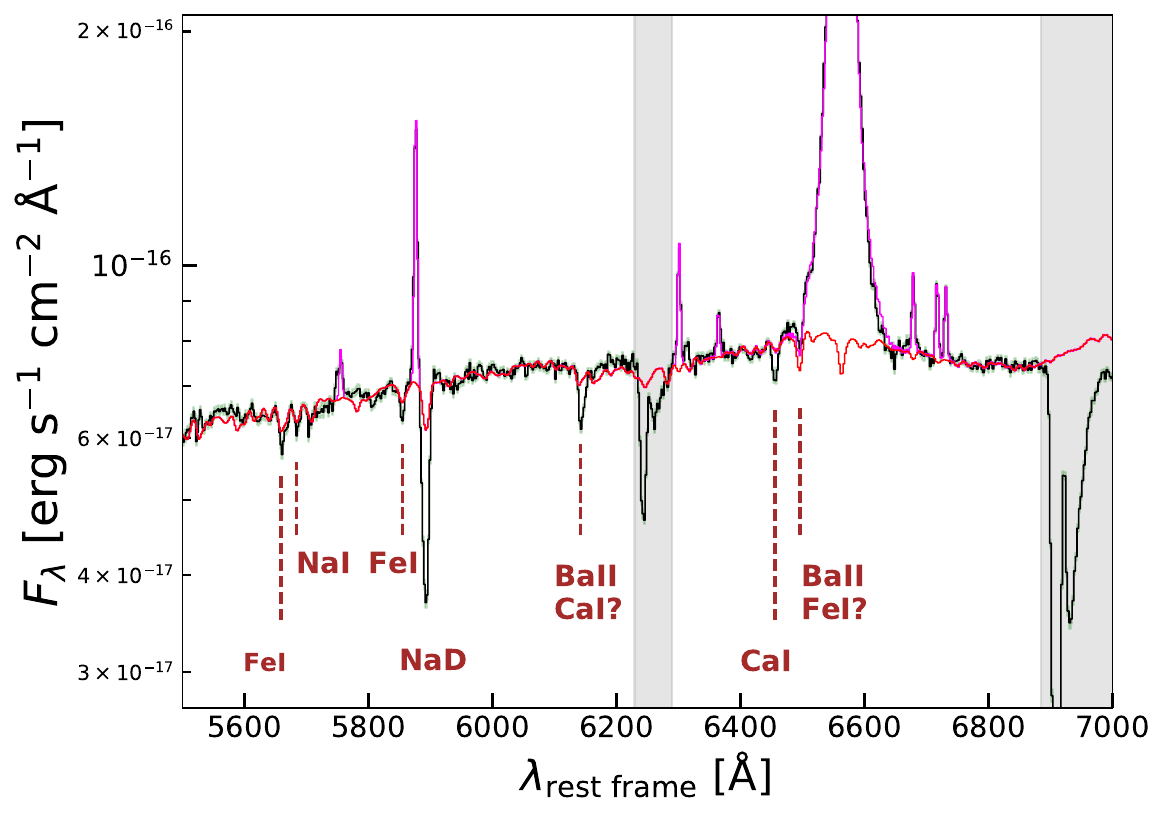}
    \caption{
    Best-fit spectral model for the GTC/OSIRIS spectrum of \target based on MILES empirical stellar templates \citep{miles} and additive polynomials with \textsc{pPXF} \citep{cappellari2004,cappellari2017}.
    We plot both the best-fit emission line + continuum model (magenta) and the best-fit continuum model (red) and compare them with the GTC spectrum (black with $1\sigma$ uncertainty in shaded green).
    Regions overlapping with the vertical shaded bands are excluded during the fit.
    We also plot the residual of the fit divided by the uncertainty (i.e., $\chi$).
    The best-fit models primarily consist of cool G-to-K type supergiant stars with $T_{\rm eff}\sim 5000$ K.
    The bottom panels show zoom-in views of the spectrum, where we mark tentative identifications of absorption lines and \feii\ emission lines that are rarely seen in galaxies.
    Despite adding the flexible polynomials, the strengths of many absorption lines are underfitted.
    }
    \label{fig:cool_star_fit}
\end{figure*}

\begin{table*}
    \centering
    \caption{
    Best-fit fluxes for major emission lines in different optical spectra of the local LRD \target at $z=0.1006$. The red end of the Gemini/GMOS spectrum has been normalized to that of the SDSS spectrum and the GTC/OSIRIS spectrum has been corrected for slit losses using a linear correction.
    See Appendices~\ref{appendix:gemini_reduction} and \ref{appendix:slit_loss} for details. 
    }
    \begin{tabular}{l c c c c}
    \hline
    Line & Flux [$10^{-17}$ \ergscm] & Flux [$10^{-17}$ \ergscm] & Flux [$10^{-17}$ \ergscm] \\
     & (SDSS, $R\approx 2000$) & (Gemini/GMOS, {$R\approx 3400$}) & (GTC/OSIRIS, $R\approx 1000$) \\
    \hline
    \ha (narrow) & 
    $952\pm 14$ 
    & $946\pm 8$ & $898\pm 2$  \\
    \ha (broad 1; ``NB'') & $(1.35\pm 0.04)\times 10^3$ & $(1.22\pm 0.02)\times 10^3$ & $(1.274\pm 0.007)\times10^3$ \\
    \ha (broad 2; ``BB'') & $(1.71\pm 0.03)\times 10^3$ & $(1.70\pm 0.02)\times 10^3$ & $(1.318\pm 0.004)\times 10^3$ \\
    \hb (narrow) & $260\pm 8$ & - & $270.9\pm 0.8$ \\
    \hb (broad 1; ``NB'') & $45\pm 14$ & - & $62\pm 2$ \\
    \hb (broad 2; ``BB'') & $58\pm 17$ & - & $63\pm 3$ \\
    \hg & $118\pm 3$ & - & $108.8\pm 0.6$ \\
    \hd & $57.7\pm 2.9$ & - & $58.3\pm 0.5$ \\
    H$\epsilon$ & $33.6\pm 3.3$ & - & $32.6\pm 0.5$ \\
    H8+\hei$\lambda 3889$ & $47.5\pm 3.0$ & - & $40.6\pm 0.5$ \\
    H9 & $21.4\pm 2.9$ & - & $13.8\pm 0.5$ \\ 
    H10 & $<9.4$ & - & $10.7\pm 0.5$ \\ 
    H11 & $<8.8$ & - & $8.6\pm 0.5$ \\  
    H12 & $<9.4$ & - & $5.6\pm 0.5$ \\
    \oii$\lambda 3726$ & $66.9\pm 3.8$ & - & $54.8\pm 1.0$ \\
    \oii$\lambda 3729$ & $73.9\pm 4.0$ & - & $81.6\pm 1.0$ \\
    \hei$\lambda 3820$ & $<8.9$ & - & $3.6\pm 0.5$ \\
    \neiii$\lambda 3869$ & $96.2\pm 3.3$  & - & $97.2\pm 0.6$ \\
    \hei$\lambda 4026$ & $<7.8$ & - & $4.0\pm 0.4$ \\
    \sii$\lambda 4069$ & $<8.2$ & - & $4.7\pm 0.4$ \\
    \oiii$\lambda 4363$ & $46.2\pm 2.6$ & - & $44.3\pm 0.5$ \\
    \heii$\lambda 4686$ & $9.4\pm 2.1$ & - & $6.0\pm 0.4$ \\
    \oiii$\lambda 5007$ & $(1.17\pm 0.01)\times10^3$ & - & $(1.31\pm 0.01)\times10^3$ \\
    \nii$\lambda 5755$ & $5.5\pm 1.7$ & - & $5.4\pm 0.3$ \\
    \hei$\lambda 5876$ & $45.7\pm 3.3$ & $39\pm 2$ & $47.9\pm 0.5$  \\
    \oi$\lambda 6300$ & $15.6\pm 1.6$ & $17\pm 1$ & $17.4\pm 0.3$  \\
    \nii$\lambda 6583$ & $11.8\pm 3.5$ & $13\pm 1$ & $9.9\pm 0.7$ \\
    \hei$\lambda 6678$ & $11.7\pm 1.9$ & $11\pm 1$ & $10.4\pm 0.3$ \\
    \sii$\lambda 6716$ & $12.1\pm 2.2$ & $12\pm 1$ & $11.9\pm 0.3$ \\
    \sii$\lambda 6731$ & $12.6\pm 2.0$ & $11\pm 1$ & $10.9\pm 0.3$ \\
    \hei$\lambda 7065$ & $23.0\pm 2.0$ & $24\pm 1$ & - \\
    $[$Ar\,{\sc iii}$]$$\lambda 7136$ & $<12$ & $9.2\pm 1.0$ & - \\
    $[$Fe\,{\sc ii}$]$$\lambda 7155$ & $<6.8$ & $13\pm 1$ & - \\
    $[$Ca\,{\sc ii}$]$$\lambda 7291$ & $25.4\pm 3.9$ & $19\pm1$ & - \\
    \oii$\lambda 7320$ & $5.7\pm 2.1$ & $5.1\pm 0.7$ & - \\
    \oii$\lambda 7331$ & $6.5 \pm 2.0$ & $2.6\pm 0.3$ & - \\
    $[$Ca\,{\sc ii}$]$$\lambda 7324$ & $12.6\pm 2.9$ & $15\pm 1$ & - \\
    \hline
    \end{tabular}
    \begin{tablenotes}
        \small
        \item $\bf Notes.$
        \item All upper limits listed are $3\sigma$-upper limits. For doublets with fixed flux ratios we only list the stronger lines.
    \end{tablenotes}
    \label{tab:lines}
\end{table*}

\begin{table}
    \centering
    \caption{Best-fit kinematics for emission lines and absorption lines measured from the observed spectra of \target, where the velocity of narrow lines is set as the zero point.}
    \begin{tabular}{l c c}
    \hline
    Line & Velocity [\kms] & FWHM [\kms]\\
    \hline
    \multicolumn{3}{|c|}{Gemini/GMOS spectrum ($R\approx 3400$)}\\
    \hline
    \ha, \hei, \oi, \nii, & 0 & $92\pm 1$ \\
    \sii, \ariii\ (narrow) & & \\
    $[$\feii$]$, [\caii] & $19\pm 7$ & {$240\pm 40$} \\
    \ha (broad 1; ``NB'') & $3\pm 1$ & $727\pm 10$ \\
    \ha (broad 2; ``BB'') & $3\pm 1$ & $2050\pm 20$ \\
    \ha (absorption)$^\dagger$ & $-62\pm 1$ & $188\pm 5$ \\
    NaD (absorption)$^\dagger$ & {$-50\pm 25$} & $260\pm 50$ \\
    \hline
    \multicolumn{3}{|c|}{SDSS spectrum ($R\approx 2000$)}\\
    \hline
    \hb (broad 1; ``NB'')$\ddag$ & $25\pm4$   & $790\pm30$ \\
    \hb (broad 2; ``BB'')$\ddag$ & $25\pm4$   & $2120\pm60$\\
    \hb (absorption)$^\dagger$             & $100\pm20$ & 
    {$24_{-19}^{+82}$} \\
    \hline
    \end{tabular}
    \label{tab:kinematics}

\raggedright{
$^\dagger$ For absorption the nominal FWHM is calculated as $\sqrt{4\ln{4}}~\sigma$, where $\sigma$ is the velocity dispersion.
}\\
\raggedright{
$\ddag$ In the SDSS model, the velocity, FWHM and flux ratio of the two broad components of \hb are tied to the equivalent components of \ha.
}
\end{table}

\begin{table}
    \centering
    \caption{
    Identifications and flux measurements of weak lines associated with \feii\ in the local LRD \target at $z=0.1006$. All measurements are made in the GTC spectrum corrected for slit losses.}
    \begin{tabular}{l c}
    \hline
    Line & Flux [$10^{-17}$ \ergscm]\\
    \hline
    $[$\feii$]$$\lambda 4178.96^{\rm a}$ & $3.48\pm 0.35$ \\
    \feii$\lambda 4233.17$ & $3.31\pm 0.35$ \\
    $[$\feii$]$$\lambda \lambda 4243.97,4244.81$ & $8.25\pm 0.36$ \\
    $[$\feii$]$$\lambda 4276.83$ & $6.40\pm 0.32$ \\
    $[$\feii$]$$\lambda 4287.39^{\rm b}$ & $11.8\pm 0.3$ \\
    $[$\feii$]$$\lambda 4305.89^{\rm c}$ & $<0.8^{\rm d}$ \\
    $[$\feii$]$$\lambda 4359.33^{\rm b}$ & $9.40\pm 0.20$ \\
    $[$\feii$]$$\lambda 4413.78$ & $9.05\pm 0.34$ \\
    $[$\feii$]$$\lambda 4814.53$ & $5.48\pm 0.30$ \\
    $[$\feii$]$$\lambda 4889.62$ & $5.93\pm 0.36$ \\
    $[$\feii$]$$\lambda 4905.34$ & $3.12\pm 0.31$ \\
    $[$\feii$]$$\lambda 5043.52$ & $3.43\pm 0.29$ \\
    $[$\feii$]$$\lambda \lambda 5158.00,5158.78$ & $15.4\pm 0.3$ \\
    $[$\feii$]$$\lambda 5261.62^{\rm c}$ & $8.79\pm 0.28$ \\
    \hline
    \end{tabular}
    \begin{tablenotes}
        \small
        \item $\bf Notes.$
        \item $\rm ^{a}$ Tentative identification. Blended with the permitted transition \feii$\lambda 4178.86$.
        \item $\rm ^{b}$ Fixed doublet flux ratio with [\feii]$\lambda 4287$/[\feii]$\lambda 4359$ = 1.25.
        \item $\rm ^{c}$ Blended with absorptions.
        \item $\rm ^{d}$ $3\sigma$ upper limit.
    \end{tablenotes}
    \label{tab:fe_forest}
\end{table}

We measured emission as well as absorption lines present in the archival SDSS and Gemini spectra and the new GTC spectrum of \target.
Due to the different spectral coverages, instrumental resolutions, and S/N of these spectra, we adopted the following strategies for line measurements.
First of all, we removed the relative flux calibration difference (mainly due to the flux calibration uncertainty of the Gemini spectrum) by renormalizing the continuum of the Gemini spectrum to that of the SDSS spectrum (see Appendix~\ref{appendix:gemini_reduction}).
While this step prohibits studies of the continuum variability for \target (as the Gemini observations were taken $\sim 16$ yr after the SDSS observations, see \citealp{burke_abs_2021}), the spectral variability can still be constrained through the equivalent width of broad emission lines, which we discuss later in Section~\ref{sec:bh_params}.
For the GTC spectrum, as we have mentioned, there are significant \redtxt{slit} losses, and we applied corrections also based on the SDSS spectrum (see Appendix~\ref{appendix:slit_loss}).

To fit the lines in the SDSS, we used the Penalized PiXel-Fitting code \citep[{\sc pPXF},][]{cappellari2004,cappellari2017} and assumed the continuum is a superposition of three power laws (with slopes of $\alpha=-2.33,~-1,~{\rm and}~0.5$) multiplied by a fourth-order Legendre polynomial.
By doing so, we remain agnostic about the physical nature of the continuum, \redtxt{given that the relatively low S/N of the continuum does not allow detailed modeling with stellar or AGN templates}.
During the fit, we fixed the flux ratios of lines in certain doublets, forcing \oiii$\lambda 4959$/\oiii$\lambda 5007$ = 0.347, \oi$\lambda 6363$/\oi$\lambda 6300$ = 0.33, \nii$\lambda 6548$/\nii$\lambda 6583$ = 0.33, and \neiii$\lambda 3968$/\neiii$\lambda 3869$ = \redtxt{0.30 based on theoretical emissivity ratios \citep{chianti1}}.
We assumed a single narrow component for lines except \ha and \hb and tied the kinematics of all narrow lines.
For \ha and \hb, we include narrow, broad, and absorption components.
We used the standard parameterization of
\begin{equation}
    f_\lambda/f_{\lambda;0}=1-C_f+C_f\exp[-\tau_0e^{-(1-\lambda/\lambda_0)^2c^2/v^2}],
    \label{eq:abs}
\end{equation}
to model the \redtxt{Balmer} absorption,
which assumes a screen geometry for the absorber.
In the equation above, $f_\lambda/f_{\lambda;0}$ is the ratio between the transmitted flux and the initial flux, $C_f$ is the covering fraction of the absorber (varying between 0 and 1 assuming the source is partially and isotropically obscured), $\tau_0$ is the optical depth at the line center, $\lambda_0$ is the central wavelength, $c$ is the speed of light, and $v$ is the effective velocity broadening of the absorbing gas.
\redtxt{The reason for modeling these absorptions separately is due to the fact that they have large equivalent widths and are offset from the narrow lines, suggesting they do not have a stellar origin \citep[see, e.g.,][]{burke_abs_2021,juodzbalis_jadesagn_2025,ji_lrdbreak_2025,deugenio_qso1_2025}.
}
{During the fit, we assume both broad lines and the continuum are absorbed. Due to the relatively low continuum level compared to emission lines, whether the continuum is absorbed or not only has a small impact on narrow line fluxes and does not change our conclusions.}
The same formalism was adopted to fit the NaD absorption present in the spectrum of \target, where we allowed the parameters to be independent of those describing the \ha absorption.

A similar fitting strategy is adopted for the Gemini/GMOS spectrum.
The only difference is that the Gemini/GMOS spectrum shows that profiles of [\feii]$\lambda 7155$, [\caii]$\lambda 7291$, and possibly [\caii]$\lambda 7324$ are clearly broader than other narrow lines, and thus we untied the kinematics of these lines from those of other narrow lines.
Since the Gemini/GMOS data have higher spectral resolution, we used its fitting results as the fiducial values for different spectral components of \ha and NaD.
{However, the Gemini/GMOS spectrum does not cover \hb, which has an indication of a redshifted absorption in the SDSS spectrum and is confirmed in the new observations obtained by \citet{linxiaojing_locallrd_2025}.
Also, the GTC spectrum does not have enough spectral resolution to fit the \hb absorption.
Therefore, we used the SDSS spectrum to fit the \hb absorption and plotted the result in Figure~\ref{f.gmos.gaussfit.a}, where the redshifted absorption can be clearly seen in the zoomed-in panel.
}

For the \ha emission, we initiated our fit assuming both broad and narrow components have Gaussian intrinsic profiles.
Inspired by the recent work of \citet{deugenio_qso1_2025}, we examined the possibility that the broad component is described by two Gaussians, which are also found to be typical in a sample of $z\sim 2$ quasars \citep{Santos_gravity_2025}.
Based on the Bayesian Inference Criterion \citep[BIC,][]{liddle_2007}, we found that the double-Gaussian model provides a good description of the broad line profile and is strongly preferred over the single Gaussian profile ($\Delta\,\text{BIC}>200$).
The best-fit model for \ha\ as well as other spectral features in the Gemini/GMOS spectrum is shown in Figure~\ref{f.gmos.gaussfit.b}\,-\,\ref{f.gmos.gaussfit.g}.
One potential physical interpretation of the double-Gaussian broad component is that there are kinematically distinct regions within the BLR with the narrower broad component potentially originating in the larger radius \citep[e.g.,][]{Brotherton_1994,popvic_2004,Zhu_2009,zhang_2011,Nagoshi_2024}.

We also checked another possible model for the broad component recently proposed by \citet{Rusakov_escattering_2025} for high-$z$ LRDs, where the line profile is dominated by electron scattering and has exponential wings.
Such line profiles have been identified in some local AGN and \citet{Laor_2006} suggests the exponential profile becomes apparent in BLRs of low-mass BHs, where the virial broadening of the lines is $<1000$ \kms.
The exponential model, which also provides a reasonably good fit to the line profile, is
presented in Appendix~\ref{appendix:gemini_exp}.
Statistically, the exponential model is strongly preferred, with $\Delta\,\text{BIC}=199$.
This preference is not due to differences in the fainter, forbidden lines, since we find almost the same $\Delta\,\text{BIC}$ when restricting the analysis to \ha alone.
The preference for an exponential model based on \ha is in agreement with high-redshift LRDs
\citetext{\citealp{Rusakov_escattering_2025}; but see below for other hydrogen lines}.
\redtxt{In the case of the exponential model, \citet{Rusakov_escattering_2025} suggest that one can still use the virial theorem to derive the single-epoch BH mass, but the line width needs to be replaced by the intrinsic width of the broad-line profile before undergoing electron scattering (see Appendix~\ref{appendix:gemini_exp} for the derivation).
}
The derived mass of the black hole is lowered by roughly a factor of 3 (i.e., 0.5 dex) with the exponential model due to the intrinsically smaller virial broadening.
We note here that this simple electron scattering model also predicts an electron temperature for the scattering medium, but the inferred value is only $T_{\rm e}=2,500$ K, which is too low for photoionized gas of AGN or star-forming regions that can provide free electrons \citetext{\citealp{agn3_2006}; see Appendix~\ref{appendix:gemini_exp} for more details}.
Furthermore, the recent work of \citet{Brazzini_2025} has shown that simple electron scattering due to
a foreground ionized gas screen is unlikely to explain the Balmer and Paschen line profiles observed in the Rosetta Stone, an LRD at $z=2.26$, which is remarkably similar to \target as we demonstrate later in this manuscript.
For this work, we do not have the spectral coverage of Paschen lines to investigate the origin of the line broadening and we simply use the double-Gaussian fit as the fiducial fit.
For more discussions on the broad-line profiles of LRDs, see \redtxt{\citet{Rusakov_escattering_2025,juodzbalis_jadesagn_2025,juodzbalis_specast_2025,Brazzini_2025}.}

For \hb, the absorption feature appears clearly redshifted, unlike for \ha, where the
absorption is blueshifted. For this reason, and supported by similar 
observations in other LRDs \citetext{cf.~\citealp{ji_lrdbreak_2025} versus \citealp{deugenio_qso1_2025}; \citealp{deugenio_lrdoutflow_2025}}, we allowed the \hb absorption to have independent kinematics from the \ha absorption.
We present our best-fit profile in Figure~\ref{f.gmos.gaussfit.a}.
Comparing Figure~\ref{f.gmos.gaussfit.a} with ~\ref{f.gmos.gaussfit.d}, one can clearly see the different kinematics of \ha and \hb absorptions.

To fit the GTC spectrum, which has a much higher S/N throughout the optical range compared to the SDSS spectrum, we invoked empirical stellar templates to match the absorption features in detail.
{Although, as argued by \citet{ji_lrdbreak_2025,degraaff_lrd_2025,naidu_lrd_2025} for high-$z$ LRDs and \citet{linxiaojing_locallrd_2025} for \target, the optical continua of LRDs can originate from AGN emission modified by gas envelopes, our goal here is to identify key absorption features based on stellar templates and see if stellar populations are capable of producing the observed strengths of the absorptions.}
To do the fit, we again used the \textsc{pPXF} code.
We used empirical stellar templates from MILES \citep{miles} that cover the optical range and assumed Gaussian profiles for emission lines as input models, and the continuum and the lines are fitted simultaneously with their kinematics being independent.
The templates are convolved with the instrumental resolution of the GTC spectrum before fitting so that we can recover the intrinsic width of lines, although the intrinsic broadening dominates the profiles of narrow lines.
{Since the Balmer absorption lines are barely resolved in the GTC spectrum, we fixed the parameters of the absorber to those measured from the SDSS and Gemini spectra.}
To fit the other absorption lines, we adopted a flexible $\rm 20^{th}$-order additive polynomial to adjust the depths of absorption, which is not physical but provides a significantly improved fit compared to the one without the polynomial or the ones with low-order polynomials (e.g., 8 or 4) typically adopted for fitting stellar kinematics.
During the fit, we masked strong telluric features as well as the strong NaD absorption, as it likely has an outflow origin.
Our best-fit model has $\chi^2_{\nu}=14.9$ over the full spectral range and is plotted in Figure~\ref{fig:cool_star_fit}.
This model roughly reproduces the overall continuum shape and a few absorption features.
Still, many of the absorption lines, which cannot come from telluric absorption, are not reproduced and we emphasize again that the continuum model is not physical.
We made tentative identifications of absorption lines in the bottom panels of Figure~\ref{fig:cool_star_fit} based on known stellar absorption lines \citep{stellar_spec}.
All absorption lines are likely from low-ionization ions, atoms, or even molecules.
We further discuss the interpretation of the continuum in Section~\ref{sec:sed}.

{In addition to the major emission lines typically seen in the interstellar medium (ISM) of galaxies, we identified a series of weak and narrow lines associated with \feii\ in our newly obtained GTC spectrum.
We marked the location of the \feii\ lines we identified in the bottom left panel of Figure~\ref{fig:cool_star_fit}.
These \feii\ lines are mainly forbidden transitions and can be produced by collisional excitation and fluorescence of UV photons \citep{Baldwin_1996}.
In the local Universe, these lines are seen in the dense gas regions of the Orion nebula, some Seyfert galaxies, supernova remnants, envelopes of blue variable stars, and $\eta$ Carinae-like objects \citep{Bautista_1998,Gull_etacar_2001,Choe_etacar_2025}.
Intriguingly, similar \feii\ transitions are identified in a $z=2.26$ LRD as we show later, as well as in a $z \sim 6.7$ LRD \redtxt{\citep{irony}} and a $z\sim 5$ LRD \citep{tripodi2025deepdivebroadlineregion}.
Many of these weak \feii\ lines are close to absorption features in the spectrum, making the recovery of their intrinsic strengths subject to the continuum model.
}

We summarize our measurements of fluxes of major emission lines in Table~\ref{tab:lines} and the kinematics of different components in Table~\ref{tab:kinematics}.
For weak forbidden and permitted lines of \feii, we list the measurements separately in Table~\ref{tab:fe_forest}.
The $1\sigma$ uncertainties of all line measurements are extracted from a Markov chain Monte Carlo (MCMC) method with 1000 steps using the \textsc{Python} package \textsc{emcee} \citep{emcee}.
Next, we show why \target is considered a local LRD by comparing its spectrophotometric measurements with high-redshift sources.

\section{Comparison with high-redshift LRDs}
\label{sec:lrd_com}

In this section, we
check the SED characteristics of \redtxt{\target} based on typical selection criteria of high-$z$ LRDs
We further compare some observed properties of \target with those of high-$z$ LRDs selected from \jwst observations.

\subsection{Photometric characteristics}

The definition of the LRD varies in the literature.
Still, there is a general consensus on criteria including a flat or blue UV slope \citep[e.g., with a UV slope $\beta _{\rm UV}<-0.37$,][]{Kocevski_lrd_2024}, a red optical color \citep[e.g., with an optical slope $\beta _{\rm optical}>0$,][]{Kocevski_lrd_2024}, and a compact morphology \citep[with a physical effective radius typically $\lesssim 300$ pc,][]{akins_lrd_2024}.


\redtxt{The UV slope computed from the NUV and $u$ bands of \target is $\beta _{\rm UV}=-1.5\pm 0.6$.
While the uncertainty is large, the FUV-$u$ slope is $-2.8 \pm 0.3$, suggesting the overall UV slope should be significantly below $-0.37$.
Meanwhile, the optical slope from the $g$ and $r$ bands are $\beta _{\rm optical}=1.6\pm 0.1>0$.
The computed slopes confirm the SED of \target satisfies the V-shaped criteria of \citet{Kocevski_lrd_2024}.
}

In Figure~\ref{fig:full_sed}, we have compared the spectrophotometric SED of \target with those of \jwst-selected LRDs at $z>2$ from the rest-frame FUV to MIR.
As a representative case of high-$z$ LRDs, we show the UV-to-NIR spectrum and photometric data of GN-28074, which is one of the closest LRDs observed spectroscopically by \jwst at $z=2.26$ \citep[the ``Rosetta Stone'',][]{juodzbalis_rosetta_2024}.
The spectrum of GN-28074 shown here is obtained with the low-resolution PRISM ($R\sim 100$) from the JWST Advanced Deep Extragalactic Survey \citep[JADES,][]{rieke_jades_2023,eisenstein2023,jades_jof,bunker_dr1_2024} data release 3 \citep[DR3,][]{deugenio_dr3_2024}, which is upscaled by 100 times to compare with the SED of \target.
The photometric data are compiled from \jwst, \textit{HST}, and \textit{Spitzer} as described in \citet{juodzbalis_rosetta_2024} and are upscaled also by 100 times.
Additionally, we show the median stacked SED of LRDs covering the rest-frame FUV to FIR at $z\sim 6$ \citep{akins_lrd_2024} selected in the COSMOS field \citep{Scoville_cosmos_2007}.
Intriguingly, the three SEDs show remarkable consistency in the UV-optical regime, with blue shapes in the UV, red shapes in the optical peaking at the NIR, and V-shaped turnovers around 3000\,-\,4000 \AA in the rest frame.
In addition to the typical LRDs, we plot the \jwst/NIRCam + MIRI photometric data of a special LRD, the Cliff, at $z=3.55$, which shows an extreme change in the flux density near the Balmer limit \citep{degraaff_lrd_2025}.
Although the UV part of the Cliff is significantly fainter compared to typical high-$z$ LRDs and \target, their optical-to-NIR spectra show excellent agreement, suggesting that (part of) the UV of the LRDs could be a physically distinct component compared to the optical \citep{rinaldi_lrd_2024,ji_lrdbreak_2025,naidu_lrd_2025,torralba+2025}.
Thanks to the low redshift of \target, the IR detection is pushed to the MIR regime and is significantly lower than the upper limits set by high-$z$ observations.
This new constraint, together with the curious UV component is further discussed in Section~\ref{sec:sed}.

In Figure~\ref{fig:full_sed}, we also show color composite images of \target.
While \target is unresolved in the SDSS and Legacy Survey images, the large FWHM of the PSFs of these images ($\sim 1.\!\!^{\prime\prime}1$\,-\,$1.\!\!^{\prime\prime}3$) set very loose constraints on the physical sizes.
A better size constraint comes from the Gemini observation with the best seeing, where the source image remains unresolved as described in Appendix~\ref{appendix:gemini_reduction}.
This gives an effective (half-light) radius of $R_{\rm e}<0.\!\!^{\prime\prime}32$ in the optical, corresponding to $R_{\rm e}<620$ pc.
While the limit is still loose compared to the limit for typical LRDs selected by \jwst (optical $R_{\rm e}\lesssim100$\,-\,$300$ pc; \citealp{akins_lrd_2024}), this is clearly a compact source in the local Universe.

Taking the nominal value of the total stellar mass of $M_\star = 10^{9.9\pm 0.1}~M_\odot$ reported by \citet{burke_abs_2021} obtained with the SED fitting code \textsc{cigale} \citep{cigale}, one can calculate the nominal stellar mass surface density limit, which is $\Sigma _{\rm \star,e}=M_\star /2 \pi R_{\rm e}^2>10^{3.5}~M_\odot~{\rm pc^{-2}}$. Such a high stellar mass surface density is comparable to those of individual star clusters (which are also candidates of progenitors of current-day globular clusters) in the gravitationally lensed galaxy, the Firefly Sparkle, at $z=8.3$ \citep{Mowla_2024}, yet \target is $10^4$ times more massive compared to individual star clusters and $10^3$ times more massive compared to the whole Firefly Sparkle.
Such a ``too dense and too massive'' problem has also been noted for high-$z$ LRDs \citep{baggen_lrd_2024,ma_lrd_2024,akins_lrd_2024}, and it has been suggested that the stellar mass is overestimated due to the overlooked AGN contribution to the optical continuum \citep{ji_lrdbreak_2025}, which we discuss in Section~\ref{sec:sed}.

\begin{figure}
    \centering
    \includegraphics[width=\columnwidth]{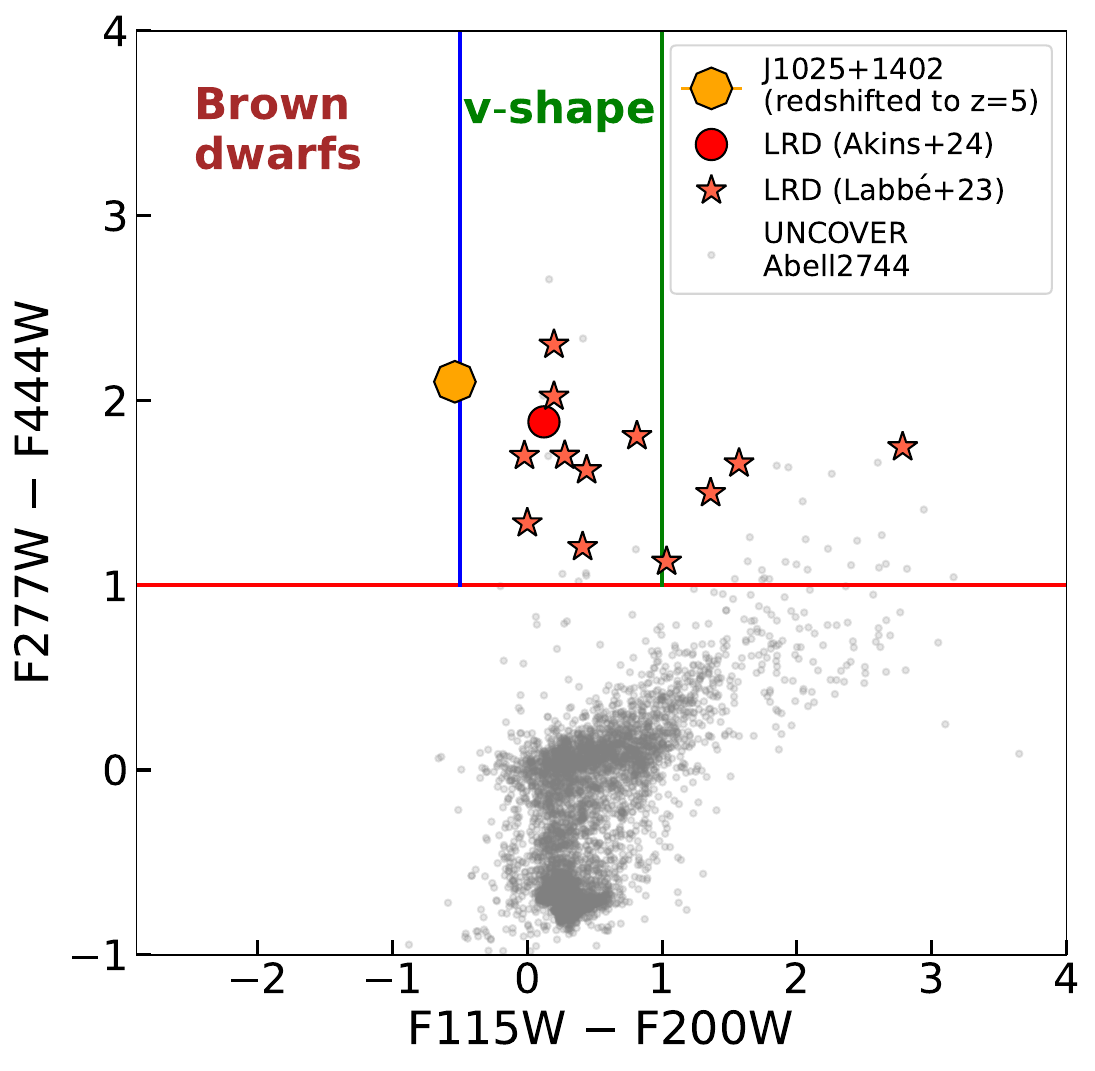}
    \caption{\jwst/NIRCam color-color diagram used for selecting high-$z$ LRDs.
    The mock NIRCam colors for \target is generated by redshifting its SED to $z=5$.
    The V-shape zone in the diagram defined in \citet{greene2024} is indicated by solid demarcation lines, which also set a zone for excluding brown dwarfs that contaminate the selection of the high-$z$ sample but not in our case.
    For comparison, we show the locations of normal high-$z$ galaxies from the UNCOVER survey \citep{uncover,uncoverdr_2024} as well as LRDs selected by \citet{labbe_2023}.
    Some of \citet{labbe_2023}'s sources are outside the V-shaped zone due to a slightly different selection based on F150W-F200W.
    The location of the stacked LRD of \citet{akins_lrd_2024} are also shown.
    }
    \label{fig:cc_diagram}
\end{figure}

To further illustrate that \target would be selected as an LRD if it were at high redshift, in Figure~\ref{fig:cc_diagram}, we show the \jwst/NIRCam color-color diagram frequently used for selecting high-$z$ LRDs.
The F277W-F444W color and F115W-F200W color trace rest-frame optical and UV of galaxies at $z\sim 4-6$.
The V-shape zone defined in \citet{greene2024} sets a rough cut for optically red and UV flat objects.
In the upper left corner where the UV is blue, high-$z$ photometric samples have the risk of being contaminated by local brown dwarfs, which is not a concern for spectroscopic samples and our local source.
In this diagram, we plot a sample of galaxies observed within the Ultradeep NIRSpec and NIRCam ObserVations before the Epoch of Reionization (UNCOVER) survey \citep{uncover,uncoverdr_2024} as well as the stacked LRD of \citet{akins_lrd_2024}.
To compare \target's color with high-z sources, we redshifted its spectrophotometric data to a typical redshift of $z=5$ for high-$z$ LRDs.
The UV color is measured by setting a power law connecting the \galex FUV flux and SDSS $u$ band flux.
The location of \target is within the brown dwarf zone and close to the V-shaped zone defined in \citet{greene2024}, where most LRDs lie, and the spectral data rule out the brown dwarf possibility.
Overall, the photometric properties of \target are very similar to those of high-$z$ LRDs, and we discuss the spectroscopic properties next.

\subsection{Spectroscopic characteristics}

One of the key characteristics of LRDs is the V-shaped spectral turnover.
As shown by \citet{setton_lrd_2024} \redtxt{and recently updated by \citealp{degraaff_2025}}, the V-shaped turnovers in low-resolution NIRSpec PRISM spectra of high-$z$ LRDs appear ubiquitously close to the Balmer limit of 3645.1 \AA in the rest frame \redtxt{(with $\sim 50\%$ of the LRDs showing the turnovers at the Balmer limit)}, implying physical processes associated with atomic hydrogen.
Whether this conclusion holds for all photometrically selected LRDs, which is a much larger sample, remains unclear\footnote{Such a question seems to be a Hempel's Ravens problem \citep{hempel_ravens}.}, but the general sample of high-$z$ LRDs hinted a turnover point at 3000\,-\,4000 \AA (see e.g., Figure~\ref{fig:full_sed}).

\begin{figure*}
    \centering
    \includegraphics[width=\columnwidth]{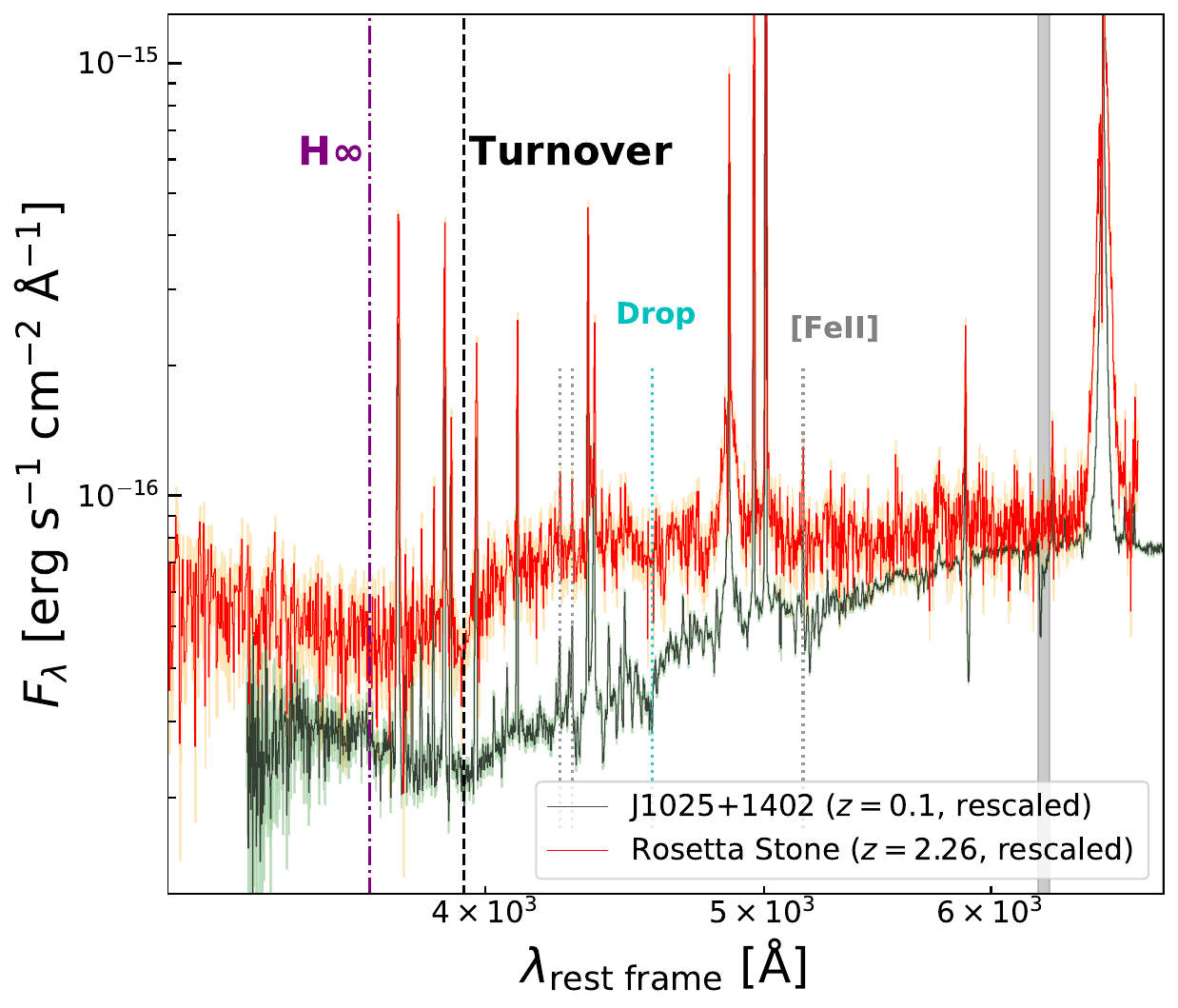}
    \includegraphics[width=\columnwidth]{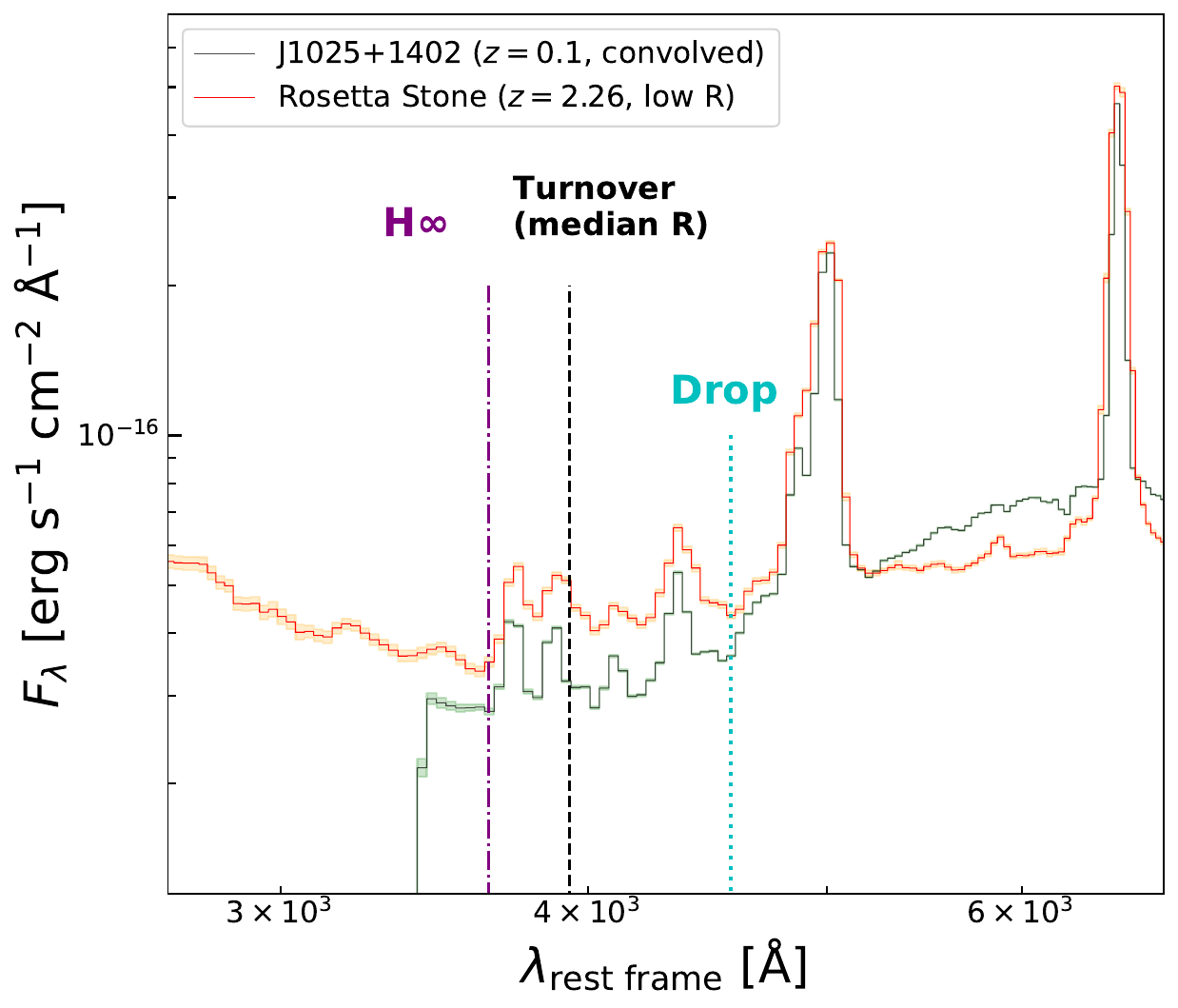}
    \caption{
    Comparison between the UV-optical spectra of \target and the high-$z$ LRD, the Rosetta Stone, at $z=2.26$ \citep{juodzbalis_rosetta_2024}.
    \textit{Left:} the median-resolution (G140M, $R\sim 1000$) spectrum of the Rosetta Stone (red) is compared with the GTC spectrum of \target.
    Fluxes of both spectra are rescaled for illustrative purposes.
    The shaded regions correspond to $1\sigma$ flux density uncertainties.
    The vertical grey band masks the telluric absorption in the GTC spectrum.
    The vertical dash-dotted purple line marks the Balmer limit.
    The vertical dashed black line marks the turnover point seen in the GTC spectrum.
    The vertical dotted grey lines mark several narrow lines corresponding to the forbidden transitions of \feii\ seen in both spectra.
    The vertical dotted cyan line marks a potential absorption feature in both spectra.
    The two sources have overall similar spectral shapes, although \target \redtxt{appears} redder in the optical.
    \textit{Right:} the low-resolution (PRISM, $R\sim 100$) spectrum of the Rosetta Stone is compared with the GTC spectrum convolved to the same LSF.
    At low resolutions, it becomes difficult to determine whether the turnover point is at the Balmer limit or at redder wavelengths due to the blended lines, although the spectrum of the Rosetta Stone seems to favor a turnover at $\rm H_\infty$.
    }
    \label{fig:uvo_sed}
\end{figure*}

In the GTC spectrum of \target, the turnover point, if defined as the lowest spectral point, appears to be at $\sim 3940$ \AA, roughly the location of the \caii\ K absorption.
In the left panel of Figure~\ref{fig:uvo_sed}, we show a zoomed-in view of the spectrum of \target around the turnover region and compare it with the NIRSpec median-resolution (G140M, $R\sim 1000$) spectrum of the Rosetta Stone from JADES \citep{juodzbalis_rosetta_2024,deugenio_dr3_2024}.
In both spectra, there appears to be a drop in the flux density around 3900 \AA, which is also roughly the turnover point in the spectrum of \target.
In comparison, the Balmer limit, $\lambda _{\rm H\infty}$, is blueward of the turnover point and matches the edge of a tentative Balmer continuum.
For the Rosetta Stone, due to the nosier spectrum blueward of 3900 \AA, it is unclear whether the turnover point is different from $\lambda _{\rm H\infty}$.
In the right panel of Figure~\ref{fig:uvo_sed}, we plot the high-S/N low-resolution ($R\sim 100$) spectra for both sources.
The R100 spectrum of the Rosetta Stone was obtained with NIRSpec PRISM and from the JADES DR3 \citep{deugenio_dr3_2024}, whereas the R100 spectrum of \target was constructed by convolving the GTC spectrum to the \redtxt{instrumental line spread function (LSF)} of the PRISM spectrum \citep[assuming it is point-source like,][]{degraaff_lsf_2024} and then resampling to the wavelength grid of the Rosetta Stone using \textsc{spectres} \citep{spectres}.
In the R100 spectrum of the Rosetta Stone, the turnover point matches $\lambda _{\rm H\infty}$.
For \target, it is unclear whether the turnover point is shifted from $\lambda _{\rm H\infty}$ in its R100 spectrum due to the flat spectral shape.
This comparison shows that for LRDs with strong nebular emission, the visual identification of a turnover point at $\lambda _{\rm H\infty}$ in the low-resolution spectrum does not guarantee it is a true spectral turnover, as blended emission lines can shift the apparent turnover wavelength.

\begin{figure*}
    \centering
    \includegraphics[width=0.85\textwidth]{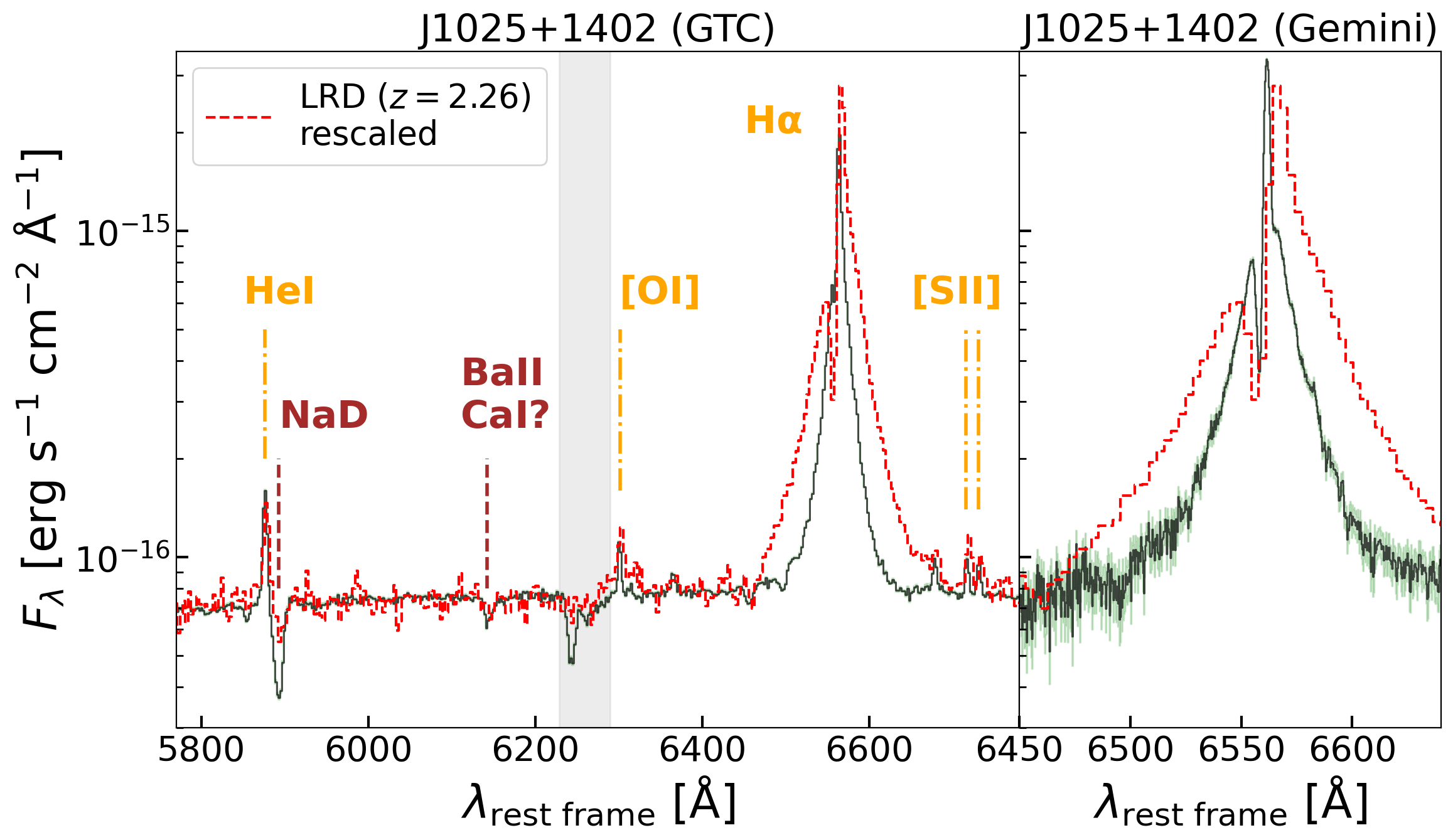}
    
    \caption{
    Comparison between the spectra of \target and the rescaled \jwst/NIRSpec G235M spectrum ($R\sim 1000$) of the Rosetta Stone, a bright LRD at $z=2.26$ \citep{juodzbalis_rosetta_2024}, in the rest frame around \ha.
    \textit{Left:} comparison between the GTC spectrum of \target and the rescaled NIRSpec R1000 spectrum of the Rosetta Stone.
    The telluric absorption is marked by the gray shaded region.
    Both spectra show a clear broad component in \ha.
    In addition, there are clear indications of \ha absorption and NaD absorption in both spectra.
    There is a clear 6142 \AA-absorption in \target, which is tentatively present in the Rosetta Stone.
    \textit{Right:} show case of the high-resolution ({$R\sim 3400$}) Gemini spectrum of \target zoomed in around \ha.
    The \ha absorption is narrow, deep, and slightly blueshifted with respect to the centroids of narrow \ha and broad \ha, as the case in the Rosetta Stone.
    }
    \label{fig:nadhaabs}
\end{figure*}

As another characteristic of LRDs, \citet{Hviding_2025} recently showed that broad Balmer lines ($\rm FWHM>1000$ \kms) are present in at least $80\%$ of the spectroscopically observed LRDs with V-shaped SEDs and compact optical morphologies, implying the presence of accreting black holes (see also \citealp{greene2024}).
In Figure~\ref{fig:nadhaabs}, we plot the spectral region around \ha for both the Rosetta Stone and \target.
In the left panel, both spectra with median resolutions ($R\sim 1000$) are compared and clear absorption in \ha as well as NaD can be seen.
The strong Balmer absorption is a key feature present in many LRDs with spectroscopic observations. Such absorption cannot be explained by typical stellar atmospheric models and instead could originate from dense and/or high-column density gas around the accreting black holes \citep[e.g.,][]{matthee2024,juodzbalis_rosetta_2024,Inayoshi_maiolino_2025,ji_lrdbreak_2025,degraaff_lrd_2025,naidu_lrd_2025,Rusakov_escattering_2025}.
The detection rate of the Balmer absorption in \jwst-selected broad-line AGN is $\sim 20\%$ but it could be significantly underestimated due to the S/N and spectral resolutions of \jwst observations \citep{deugenio_lrdoutflow_2025}.
There is an indication of a common weak absorption feature at 6142 \AA, which is not identified as a telluric line and could originate from Ba\,{\sc ii} or Ca\,{\sc i}.
There is another weak absorption at the blue wing of the broad \ha around 6497 \AA in \target, potentially matching the 6497 \AA blend seen in some stellar atmospheres and is present in the best-fit stellar model in Figure~\ref{fig:cool_star_fit}.
It is unclear whether the 6497 \AA blend is also present in the Rosetta Stone due to the much broader \ha profile.

In summary, \target \redtxt{shows} spectral features similar to high-$z$ LRDs including the V-shaped turnover, broad Balmer lines, and strong Balmer absorption.
It also reveals additional absorption from neutral sodium and potentially other low-ionization ions and atoms, which could also be present in high-$z$ LRDs.
Next, we describe the derived physical properties based on spectral lines observed in \target.



\section{Emission-line and absorption-line diagnostics}
\label{sec:bh_params}

\begin{table}
    \centering
    \caption{Derived properties for the local LRD \target.}
    \begin{tabular}{l c}
    \hline
    Parameter & Value \\
    \hline
    \multicolumn{2}{|c|}{Black hole/BLR}\\
    \hline
    Maximum $A_{\rm V}$ [mag] & \redtxt{$5.1\pm 0.1$} [assuming $\rm (H\alpha/H\beta)_{intrinsic}=3.06$] \\
    $L_{\rm bol}$ [\ergs] & \redtxt{$(1.039\pm 0.004)\times 10^{44}$} (w/o dust) \\
    & {$(3.7\pm0.3)\times 10^{45}$ (w/ dust)} \\
    $\log (\mbh/M_\odot)$ & \redtxt{$6.49\pm 0.01$} (w/o dust) \\
     & $7.22\pm 0.04$ (w/ dust) \\
    $\lambda _{\rm Edd}$ & \redtxt{$0.26\pm 0.01$} (w/o dust) \\
    & $1.40^{+0.06}_{-0.05}$ (w/ dust) \\
    \hline
    \multicolumn{2}{|c|}{Outflowing neutral gas (probed by \ha and NaD)}\\
    \hline
    $C_f$ & $0.86^{+0.03}_{-0.02}$ (\ha) \\
     & $0.48^{+0.07}_{-0.05}$ (\nai) \\
    $\tau _0$ & $3.2\pm 0.3$ (\ha) \\
     & $2.1^{+1.5}_{-1.0}$ (\nai) \\
    $\log N_{\rm H~(n=2)}$ [$\rm cm^{-2}$] & $13.77\pm 0.04$ \\
     $\log N_{\rm Na^0}$ [$\rm cm^{-2}$] & $14.04\pm 0.24$ (w/ $\sigma _{\rm NaD}$)  \\
      & $13.90^{+0.17}_{-0.30}$ (w/ $\sigma _{\rm H\alpha}$) \\
    $v_{\rm centroid}$ [\kms] & $-62\pm 1$ (\ha) \\
     & $-50\pm 25$ (\nai) \\
    $\max~v_{\rm turb}$ [\kms] & $112\pm 1$ (\ha) \\
     & $156^{+42}_{-28}$ (\nai) \\
    \hline
    \multicolumn{2}{|c|}{NLR/ISM}\\
    \hline
    $A_{\rm V}$ [mag] & \redtxt{$0.38\pm 0.01$} \\
    $n_{\rm e}$ [$\rm cm^{-3}$] & $380_{-220}^{+360}$ (\sii) \\
     & $1.41^{+0.61}_{-0.49}\times 10^3$ (\oii) \\
     & $\sim 10^{5-7}$ ([\feii]) \\
    $T_{\rm e}~({\rm O^{2+}})$ [K] & $2.17^{+0.08}_{-0.09}\times 10^4$ \\
    $T_{\rm e}~({\rm O^{+}})$ [K] & $(1.02\pm 0.15)\times 10^4$ \\
    $\rm 12+\log(O^{2+}/H^+)$ & $7.36\pm 0.03$ \\
    $\rm 12+\log(O^{+}/H^+)$ & $7.46^{+0.31}_{-0.25}$ \\
    $\rm ICF(O^{++}+O^+)$ & $1.04\pm 0.02$ \\
    $\rm 12+\log(O/H)$ & $7.73_{-0.14}^{+0.21}$ \\
    \hline
    \multicolumn{2}{|c|}{Host galaxy}\\
    \hline
    $A_{\rm V}$ [mag] & \redtxt{1.7\,-\,1.9} (\textsc{cigale}) \\
    $\log(M_\star/M\odot)$ & 10\,-\,11 (\textsc{cigale}) \\
    $R_{\rm e}$ & $<620$ pc (Gemini slit image)\\
    $\log (M_{\rm dyn}/M_\odot)$ & {$<9.5$} \\
    \hline
    \end{tabular}
    \begin{tablenotes}
        \small
        \item $\bf Notes.$
        \item We used the bolometric conversion from \citet{sternlaor_2012} and the single-epoch virial black hole mass relation from \citet{rv_bhmass_2015}.
        The dust attenuation curve is from \citet{calzetti2000} and we set $R_{\rm V}=3.1$.
        The atomic data set for relevant calculations are from CHIANTI \citep[v10,][]{chianti0,chianti1}.
    \end{tablenotes}
    \label{tab:properties}
\end{table}

Based on our spectral fitting results for \target, we discuss the properties of the supermassive black hole in \target as well as its gaseous environment probed by emission and absorption lines.
In addition, we compare these properties with those of \jwst-selected AGN, thereby placing \target within the broader context of AGN rather than an LRD, which is potentially a subcategory of AGN based on the UV-optical colors \citep{hainline_lrd_2024}.


\subsection{Black hole parameters}

We started by investigating the black hole parameters based on the observed spectra of \target.
While \citet{burke_abs_2021} have already derived black hole parameters for \target based on the same Gemini observation, we show below that some of these values might need to be reevaluated.

From the high-resolution Gemini spectrum, the luminosity of the broad \ha is \redtxt{$L_{\rm b}({\rm H\alpha})=(7.99\pm 0.03)\times10^{41}$} \ergs\ in total, with the narrower broad (hereafter NB) component (which dominates the core of the profile) being 70\% of the broader broad (hereafter BB) component (which dominates the wing of the profile).
The luminosity of the BB component in our fit is consistent with the total luminosity of the broad \ha fitted by \citet{burke_abs_2021} within $2\sigma$.
We note that \citet{burke_abs_2021} also used a double-Gaussian model, as evident from their Fig. 1. However, \citet{burke_abs_2021} ascribed the NB component to part of the narrow component in their fit and thus exclude it from the BLR.
As a result, 
the FWHM of the total broad component in our fit is largely set by the NB component since it dominates the core, and we have $\rm FWHM_b(H\alpha)=934\pm 10$ \kms, which is only $\sim 1/2$ of what is inferred by \citet{burke_abs_2021} for the total broad \ha.

One might wonder whether the NB component should be considered as a narrow component outside the BLR.
We argue that this scenario is less likely due to the following reason.
The NB component in \ha is significantly brighter and broader compared to the narrowest component ($\rm FWHM_n=92\pm1$ \kms) marginally resolved in the Gemini spectrum.
There is no smooth transition between the NB component and the narrowest component, implying that the NB component likely originates in a spatially distinct region, such as an outflow.
However, no NB component has been found in other emission lines, even in \oiii$\lambda 5007$ that is similarly bright as \ha.
Also, for \oiii$\lambda 4363$ that is detected at $\rm S/N\approx18$ in the SDSS spectrum and at $\rm S/N\approx 90$ in the GTC spectrum, there is no NB component.
This means the gas giving rise to the NB component needs to have a gas density higher than the critical densities of these metal lines, in other words, $n_{\rm e}>10^7~{\rm cm^{-3}}$.
Such a high density is more typical for BLR clouds rather than the ISM in galaxies \citep{netzer_1990}.
One could still argue that the NB component in other lines is simply extinguished by dust, meaning the NB component sees a significantly larger amount of dust compared to the narrowest component.
While plausible, this dust reddened scenario for \oiii\ might be somewhat contrived since there is significant detection of a broad component in \hb.
In addition, recent observations show that double-Gaussian profiles are common in the BLR components of AGN at high $z$ if sufficiently high spectral S/N and resolution are achieved \citep{deugenio_qso1_2025,Santos_gravity_2025}.
Potential physical pictures for complex BLR profiles include a stratified BLR structure \citep[e.g.,][]{Storchi-Bergmann_2017}, turbulence caused by gravitational instability in the accretion disc at high accretion rates \citep[e.g.,][]{Collin_2006}, electron scattering \citep{Baldwin_1975,Laor_2006,Rusakov_escattering_2025}, or radiation pressure-dominated BLR clouds \citep{Blumenthal_1975}.
In what follows, we adopt the assumption that both the NB and BB components originate in the BLR (see \citealp{Rusakov_escattering_2025,Brazzini_2025} for further discussions on broad-line profiles in LRDs).

With $\rm FWHM_b(H\alpha)=934\pm 10$ \kms and \redtxt{$L_{\rm b}({\rm H\alpha})=(7.99\pm 0.03)\times10^{41}$} \ergs, we derived \redtxt{$\log (\mbh/M_\odot)=6.49\pm 0.01$} using the relation of \citet{rv_bhmass_2015}.
The 0.01 dex uncertainty only includes measurement uncertainties, and we caution that the single-epoch method based on \ha can have a systematic uncertainty as large as 0.5 dex \citep{Shen_qsomass_2013}.
To estimate the Eddington ratio, $\lambda _{\rm Edd}\equiv L_{\rm  bol}/L_{\rm Edd}$, we used the bolometric conversion of $L_{\rm bol} = 130~L_{\rm b}({\rm H\alpha})$ from \citet{sternlaor_2012}, and calculated $L_{\rm Edd}\approx 1.26\times 10^{38}~(\mbh/M_\odot)$ \ergs.
We obtained \redtxt{$\lambda _{\rm Edd}=0.26\pm 0.01$}. Again, the uncertainty only includes the measurement uncertainty, and we caution that the bolometric conversion for LRD AGN can be highly uncertain \citep{lambrides_superedd_2024}.
The Eddington ratio we derived for \target is comparable to high-$z$ LRDs with broad-line measurements, despite having a considerably lower black hole mass \citep[e.g.,][]{juodzbalis_rosetta_2024,furtak2023,ji_lrdbreak_2025,deugenio_qso1_2025}.
We summarize our derived values in Table~\ref{tab:properties}.

The black hole mass and the Eddington ratio we derived above are not corrected for dust attenuation.
The Balmer decrement measured from the broad line ratio in the GTC spectrum is \redtxt{$F_{\rm b}({\rm H\alpha})/F_{\rm b}({\rm H\beta})=21\pm 1$}.
This is a very large ratio compared to the typical Case B value of 2.86 in the ISM.
Even using an average intrinsic Balmer decrement of 3.06 obtained by \citet{Dong_2008} for low-$z$ AGN, the implied attenuation is \redtxt{$A_{\rm V}=5.1\pm 0.1$} assuming a \citet{calzetti2000} attenuation curve with $R_{\rm V}=3.1$.
The reddening correction would make $\log (\mbh/M_\odot)=7.22\pm 0.04$ and $\lambda _{\rm Edd}=1.40^{+0.06}_{-0.05}$, suggesting \target is accreting above the Eddington limit.
However, we note that the intrinsic Balmer decrement in the BLR can be enhanced by collisional excitation, especially when the ionizing photon intensity is low, which would mimic extreme dust attenuation \citep{ilic_2012}.
Such an effect has also been proposed for \jwst-selected AGN and LRDs to explain their high broad-line Balmer decrement \citep[e.g.,][]{lambrides_superedd_2024,degraaff_lrd_2025}.
Therefore, we considered the derived $A_{\rm V}$ for the BLR only as a maximum value and we further discuss it in the context of energy budget in Section~\ref{sec:discuss}.

Combining the above black hole mass with the nominal stellar mass of $M_\star=10^{9.9\pm 0.1}~M_\odot$ derived by \citet{burke_abs_2021}, \target exhibits a black hole mass to stellar mass ratio of $\sim 10^{-3}$, not overmassive compared to many \jwst-selected high-$z$ AGN that have $\mbh/M_\star \gtrsim 10^{-2}$ \citep{harikane2023,Maiolino_jadesagn_2024,juodzbalis_jadesagn_2025,ubler2023a,ubler2024}.
This result is subject to the systematics in the SED fitting. As already shown implicitly by \citet{burke_abs_2021}, their best-fit SED model fails to fit the UV continuum and also overpredicts the strength of the Balmer break in the optical.
We discuss SED fitting with stellar populations later in this manuscript.

Overall, \target is similar to the majority of \jwst-selected AGN given its low black hole mass and high accretion rate \citep{harikane2023,matthee2024,Maiolino_jadesagn_2024,juodzbalis_jadesagn_2025}.
Next, we investigate the gaseous environment in \target using the observed emission and absorption features.

\subsection{Gaseous environment}

There is rich nebular emission in the spectra of \target, as already shown in the SDSS spectrum.
Our new GTC observation further reveals a series of weak emission and absorption lines, implying complex gaseous environments in \target.
In this section, we discuss in detail nebular diagnostics associated with emission and absorption lines.

\subsubsection{Excitation of narrow lines}

\begin{figure*}
    \centering
    \includegraphics[width=\textwidth]{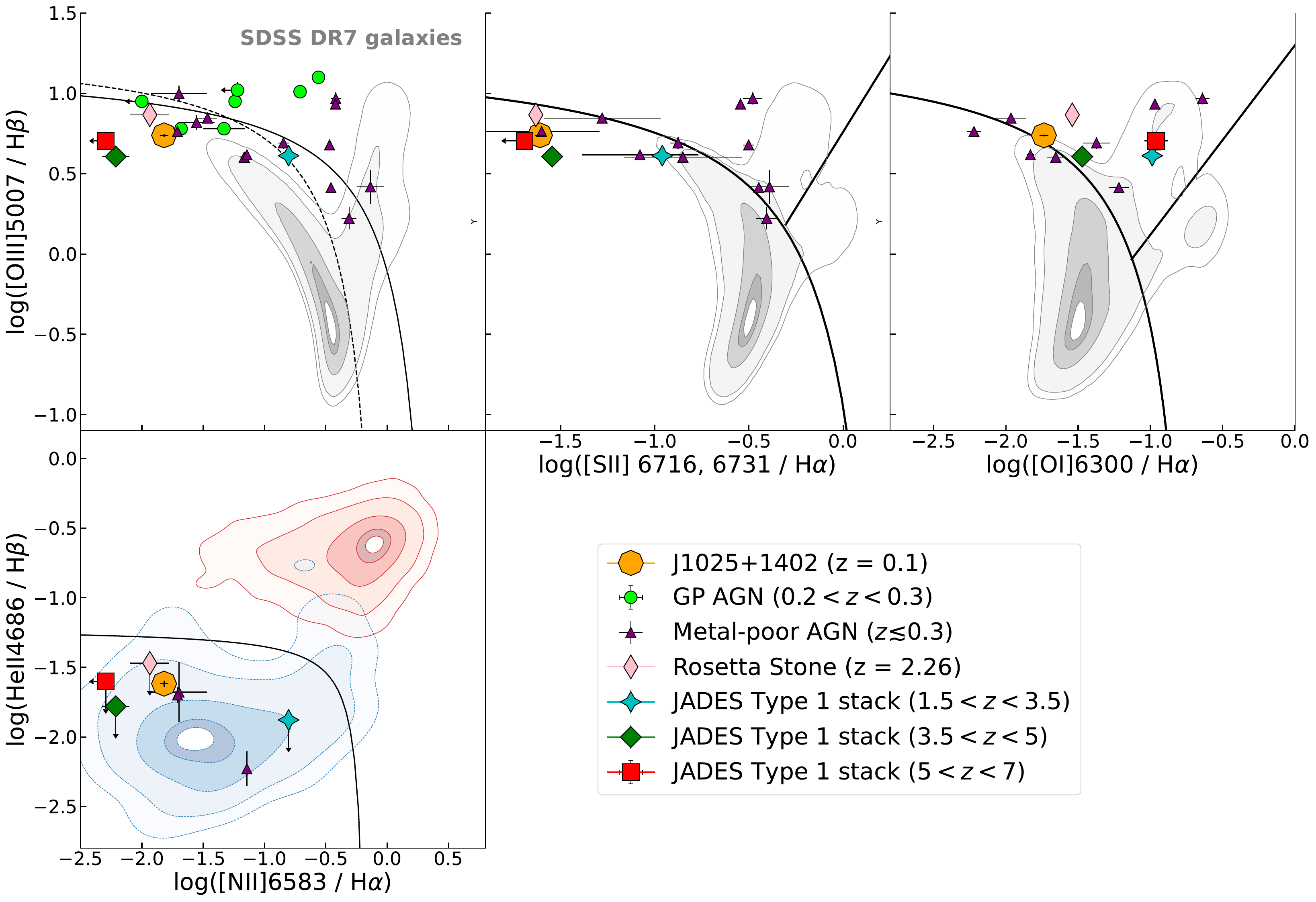}
    \caption{
    Locations of \target in four optical diagnostic diagrams. In comparison, we show SDSS DR7 galaxies with $\rm S/N > 5$ for all lines involved as contours, which correspond to their probability density distributions and enclose 5\%, 16\%, 50\%, 84\%, and 95\% of the galaxies from the innermost contour to the outer most contour.
    We also show broad-line GP galaxies selected by \citet{Lin_lrdanalog_2025} as local analogs for LRDs, 
    \redtxt{metal-poor AGN candidates with broad-line detections and X-ray constraints at $z\lesssim 0.3$ present by \citet{Simmonds_xraydwarf_2016} and \citet{Baldassare_2017} (originally from \citealp{Izotov_2007} and \citealp{reines_2013}, respectively)},
    \redtxt{the Rosetta Stone LRD at $z=2.26$ \citep{juodzbalis_rosetta_2024},}
    and stacked broad-line AGN selected from JADES over $1.5<z<7$ divided into three redshift bins \citep{juodzbalis_jadesagn_2025}.
    \textit{Top panels:} BPT/VO diagrams \citep{bpt,vo} with demarcation lines from \citet{Kewley_2001,Kewley_2006,Kauffmann_2003}.
    \textit{Bottom panel:} \heii-diagram with the demarcation line from \citet{Shirazi_2012}.
    The solid red contours and dashed blue contours correspond to \heii-emitting AGN and SF galaxies compiled by \citet{Shirazi_2012}.
    Overall, \target lies within or close to SF regions similar to high-$z$ broad-line AGN in all diagrams.
    }
    \label{fig:bpt}
\end{figure*}

\begin{figure}
    \centering
    \includegraphics[width=\columnwidth]{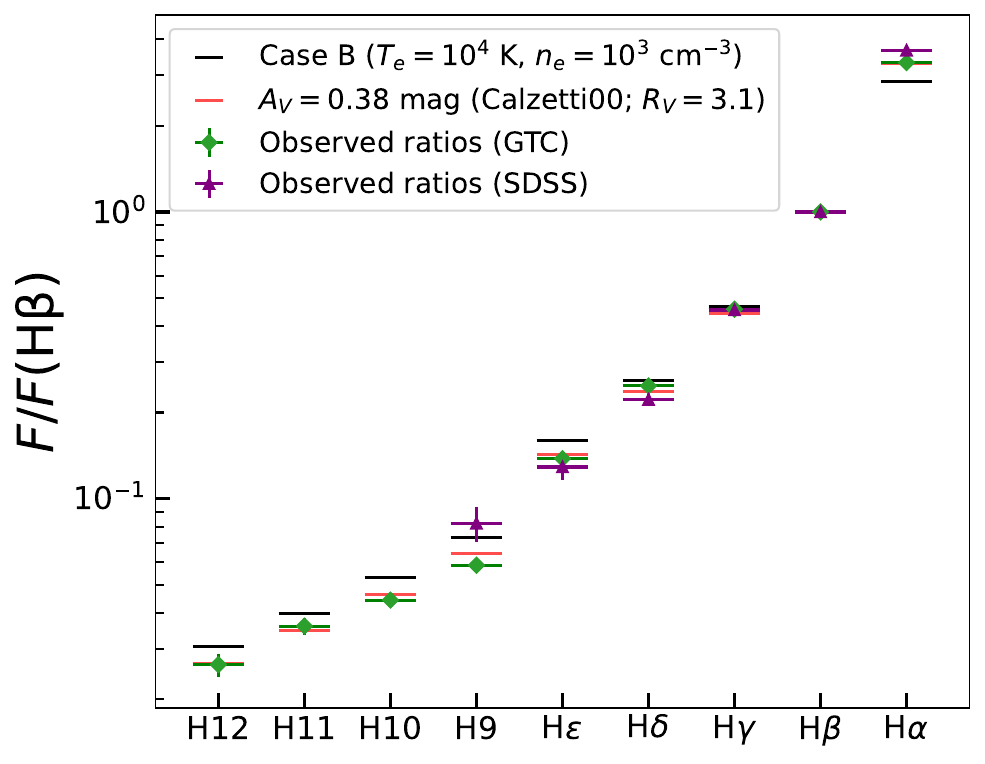}
    \caption{
    Balmer decrements measured from the optical GTC and SDSS spectra of \target, which are consistent with \redtxt{$A_{\rm V}=0.38\pm 0.01$} mag assuming a \citet{calzetti2000} attenuation curve.
    The Balmer line H8 is not included as it is blended with \hei$\lambda 3889$.
    A similarly moderate narrow-line attenuation is also found in the Rosetta Stone \citep{juodzbalis_rosetta_2024}, in contrast to the red optical continuum and extreme broad-line Balmer decrements in both cases.
    }
    \label{fig:bal_decrement}
\end{figure}

We examined the excitation diagnostics of narrow emission lines in \target.
In Figure~\ref{fig:bpt}, we plot \target\ in BPT/VO diagrams \citep{bpt,vo} as well as the \heii/\hb versus \nii/\ha diagram \citep{Shirazi_2012}.
We also plot SDSS galaxies from DR7, broad-line GP galaxies at $0.2<z<0.3$ as LRD analogs selected by \citet{Lin_lrdanalog_2025}, 
\redtxt{metal-poor broad-line AGN candidates at $z\lesssim 0.3$ present by \citet{Simmonds_xraydwarf_2016} and \citet{Baldassare_2017}, the Rosetta Stone LRD at $z=2.26$ \citep{juodzbalis_jadesagn_2025},}
and stacked broad-line AGN at $1.5<z<7$ selected from JADES \citep{rieke_jades_2023,eisenstein2023,jades_jof,bunker_dr1_2024,deugenio_dr3_2024} by \citet{juodzbalis_jadesagn_2025}.

In both \nii- and \sii-based BPT/VO diagrams, \target lies below the SF demarcation lines of \citet{Kewley_2001} and \citet{Kauffmann_2003}, similar to high-$z$ Type 1 AGN selected with \jwst.
In contrast, local GP AGN \redtxt{and metal-poor AGN candidates} show a wide distribution in \nii/\ha with \redtxt{several} sources lying above the theoretical extreme starburst line of \citet{Kewley_2001} and inside the AGN region in the BPT diagram.
The generally low \nii/\ha and \sii/\ha in high-$z$ AGN and \target are likely caused by the low metallicities in these systems \citep{ubler2023a,Maiolino_jadesagn_2024,juodzbalis_jadesagn_2025,scholtz2023}, where N/O and S/O are well below solar values.
Notably, these AGN are relocated in the Seyfert region in the \oi-based VO diagram.
This could be in part due to the lower sensitivity of \oi$\lambda 6300$ to the metallicity compared to \nii$\lambda 6583$ and \sii$\lambda \lambda 6716,6731$ \citep{ji_2020}.

In the \heii-based diagram, both JADES Type 1 AGN and \target are again more consistent with local SF galaxies selected from SDSS.
The only difference is that in \target, \heii$\lambda 4686$ is well detected in the GTC spectrum, whereas this line is not detected even in the stacked spectra of JADES AGN \citep{juodzbalis_jadesagn_2025}.
In principle, \heii$\lambda 4686$/\hb is sensitive to the hardness of the \redtxt{extreme ultraviolet (EUV)} SED due to the high ionization potential (IP) of $\rm He^+$ (54.4 eV).
This seems to imply the EUV SEDs are intrinsically soft in JADES AGN and \target, unlike those of typical local AGN (e.g., \citealp{oh_2017}; see, however, \citealp{ramosalmeida_2025} for a local Type 2 quasar with comparably low \heii$\lambda 4686$/\hb).
The softening of the ionizing SEDs might be an indication of high accretion rates for the supermassive black holes \citep{lambrides_superedd_2024}.
Alternatively, the narrow lines in these regions might be predominantly ionized by stellar populations, meaning there are no classical NLRs in these AGN.
The latter explanation would be in conflict with the high \oi/\ha, as stellar populations generally do not have enough soft X-ray photons to produce extended partially ionized zones that enhance low-ionization lines \citep{ho_2008}, unless other ionization and excitation mechanisms such as shocks are involved.
Besides \heii, we found no detection of any other high-ionization lines with $\rm IP\gtrsim50~eV$ in \target, again suggesting a soft EUV spectrum \citep[see][]{ramosalmeida_2025}.

Finally, for the general population of broad-line AGN selected by \jwst, the lack of high-ionization lines is often found \redtxt{\citep{juodzbalis_jadesagn_2025,Zucchi_2025}}, but not always \citep[e.g.,][]{ubler2023a,ji2024,Tang_nv_2025}, which might imply different evolutionary stages of AGN and their host galaxies.
The optical diagnostics suggest \target is not only similar to high-$z$ LRDs, but more generally to most of high-$z$ AGN discovered by \jwst.
If \target does represent an early evolutionary stage of an AGN in the low-redshift Universe, one might expect a low metallicity as another evidence, which we discuss next.

\subsubsection{Gas-phase metallicity}

Based on the SDSS spectrum, \citet{Izotov_2008} inferred a very low metallicity of $\rm 12+\log(O/H)=7.36\pm 0.08$ for \target, which assumes that the dominant species of oxygen is $\rm O^{2+}$.
This clearly places \target significantly below the mass-metallicity relation (MZR) of $z\sim 0$ galaxies given its nominal stellar mass \citep{Tremonti_2004} but tentatively consistent with the extrapolation of the MZR of $z>3$ galaxies observed by \jwst \citep{curti_mzr_2024}.
With the high-S/N Gemini and GTC spectra, we reexamine the above conclusion.

First of all, the derivation of $\rm 12+\log(O^{2+}/H^+)$ is relatively insensitive to any dust reddening as all the lines involved (i.e., \oiii$\lambda 4363$, \oiii$\lambda 5007$, and \hb) are close in the wavelength space.
The only caveat is the contamination of [\feii]$\lambda 4359$ to \oiii$\lambda 4363$, which is clearly identified and fitted in the GTC spectrum.
Using \oiii$\lambda 4363$/\oiii$\lambda 5007$, we obtained an electron temperature of $T_{\rm e}({\rm O^{2+}})=2.17^{+0.08}_{-0.09}\times 10^4$ K with \textsc{PyNeb} (\citealp{luridiana2015}; where the atomic data are from \citealp{oiii_as_tz17,chianti0,chianti1}).
The electron density is determined through \sii$\lambda \lambda 6716,6731$ and \oii$\lambda \lambda 3726,3729$ and are all in the low-density regime with $n_{\rm e}<10^4~{\rm cm^{-3}}$, thus having little impact on the \oiii\ temperature derivation.
We then derived $\rm 12+\log(O^{2+}/H^+)=7.36\pm 0.03$ based on the narrow-line ratio of \oiii$\lambda 5007$/\hb, similar to the \textit{total} metallicity derived by \citet{Izotov_2008}.
We list all derived values in Table~\ref{tab:properties}.

Next, we derived $\rm 12+\log(O^{+}/H^+)$, which relies on the low-ionization zone temperature derived from \oii$\lambda \lambda 3726,3729$/\oii$\lambda \lambda 7320,7330$ and thus is very sensitive to dust attenuation.
The narrow-line Balmer decrement we measured for \target significantly deviates from the Case B value.
For \ha and \hb, for example, the ratio from the GTC spectrum is
\redtxt{$\rm H\alpha/H\beta = 3.31\pm 0.01>2.86$}.
In Figure~\ref{fig:bal_decrement}, we compare the Balmer decrement measured from the GTC spectrum up to H12 with Case B values computed at $T_{\rm e}=10^4$ K and $n_{\rm e}=10^3~{\rm cm^{-3}}$ with \textsc{PyNeb}.
Overall the Balmer decrement can be fitted with \redtxt{$A_{\rm V}=0.38\pm 0.02$} mag using MCMC assuming a \citet{calzetti2000} attenuation curve, implying a small amount of dust attenuation \citep[see also,][]{Izotov_2008}.
We also tried the \redtxt{Small Magellanic Cloud (SMC)} extinction curve of \citet{gordon2003}, which produces a slightly worse fit but does not have a significant impact on our results.
{Considering that the Balmer decrement measured from the GTC spectrum might be affected by the slit-loss correction, we performed the same analysis using the SDSS spectrum, where the highest order Balmer line that can be measured is H9.
This time we obtained a best-fit attenuation magnitude of \redtxt{$A_{\rm V}=0.6\pm 0.1$} mag when including \ha to H9, or \redtxt{$A_{\rm V}=0.4\pm 0.1$} mag if we exclude \ha that might be more affected by absorption, again suggesting that \target has a moderate amount of dust.
}

With the dust attenuation corrected \oii$\lambda \lambda 3726,3729$/\oii$\lambda \lambda 7320,7330$, we obtained $T_{\rm e}({\rm O^+})=(1.02\pm 0.15)\times10^4$ K, much lower than $T_{\rm e}({\rm O^{2+}})$, but such a difference is typical for Seyferts \citep{dors2020}.
The low low-ionization zone temperature, combined with the strong \oii$\lambda \lambda 3726,3729$, implies $\rm 12+\log(O^{+}/H^+)=7.46^{+0.31}_{-0.25}$ comparable to $\rm 12+\log(O^{2+}/H^+)$.
We further calculated the ionization correction factor (ICF) to correct for higher ionization species of oxygen, by approximating $\rm ICF(O^{++}+O^+)\approx (He^{++}+He^{+})/He^+$ and using the observed fluxes of \heii$\lambda 4686$ and \hei$\lambda 5876$ and their emissivities calculated with \textsc{PyNeb}.
The derived ICF has a modest value of $1.04\pm 0.02$, and the final metallicity is $\rm 12+\log(O/H)=7.73^{+0.21}_{-0.14}$, higher than the value derived by \citet{Izotov_2008} yet still low compared to the local MZR \citep{Tremonti_2004}.
The metallicity of \target is comparable to the narrow-line metallicity of \jwst-selected broad-line AGN and LRDs \citep{trefoloni_feii_2024,Isobe_2025,juodzbalis_rosetta_2024,ji_noeg_2025}.

With the above derived metallicity of $\sim 11\%$ solar, one expects little enrichment of Fe in the ISM due to the low Fe/O from the core-collapse supernova ejecta in the early phase of galaxy evolution \citep{Kobayashi_2020}.
Intriguingly, as can be seen in Figures~\ref{fig:uvo_sed} and \ref{fig:hbfeii}, there is rich [\feii] emission in \target, perhaps also in high-$z$ LRDs.
We discuss these transitions next.

\subsubsection{Fe forest}

\begin{figure*}
    \centering
    \includegraphics[width=0.85\textwidth]{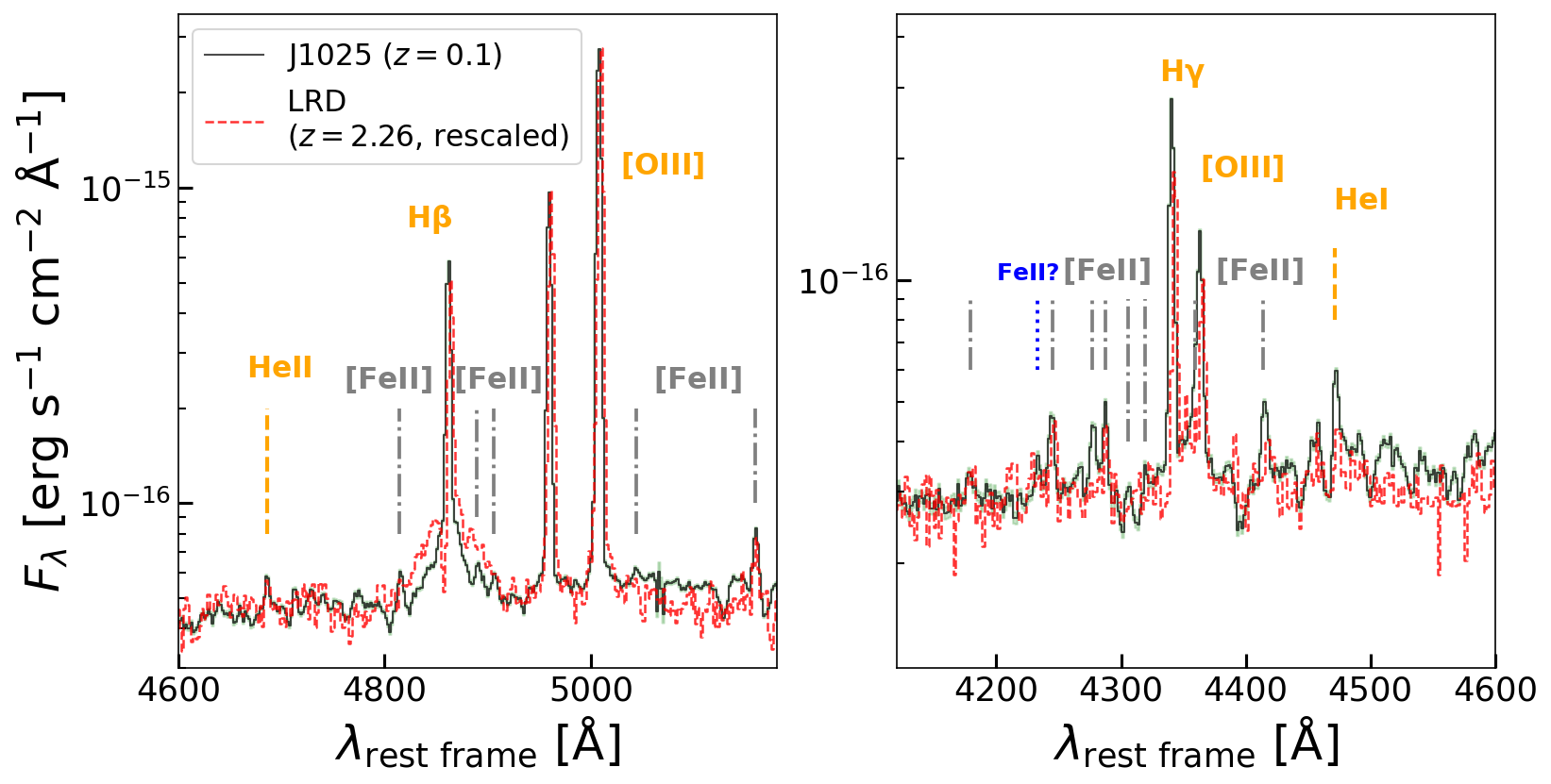}
    
    \caption{
    Comparison between the GTC/OSIRIS spectrum of \target and the rescaled \jwst/NIRSpec G235M spectrum ($R\sim 1000$) of the Rosetta Stone, a bright LRD at $z=2.26$ \citep{juodzbalis_rosetta_2024} around \hb and \hg.
    \textit{Left:} comparison of the two spectra around \hb.
    There are clear broad \hb in both spectra.
    While the Rosetta Stone has a blueshifted \hb absorption, the existence of the \hb absorption in the GTC spectrum of \target is not clear and it is only present in the higher resolution SDSS spectrum shown in Figure~\ref{f.gmos.gaussfit}.
    There is a weak \heii\ line visible in both spectra.
    The GTC spectrum of \target further reveals several [\feii] lines, with the strongest one, [\feii]$\lambda 5159$ also visible in the NIRSpec spectrum of the Rosetta Stone.
    \textit{Right:} comparison of the two spectra around \hg.
    There are a series of [\feii] lines significantly detected in \target, among which [\feii]$\lambda 4245$, [\feii]$\lambda 4287$, and potentially [\feii]$\lambda 4414$ are also present in the Rosetta Stone.
    The existence of the ``\feii\ forest'' implies very low-ionization gas in the LRDs.
    }
    \label{fig:hbfeii}
\end{figure*}

\begin{figure*}
    \centering
    \includegraphics[width=2\columnwidth]{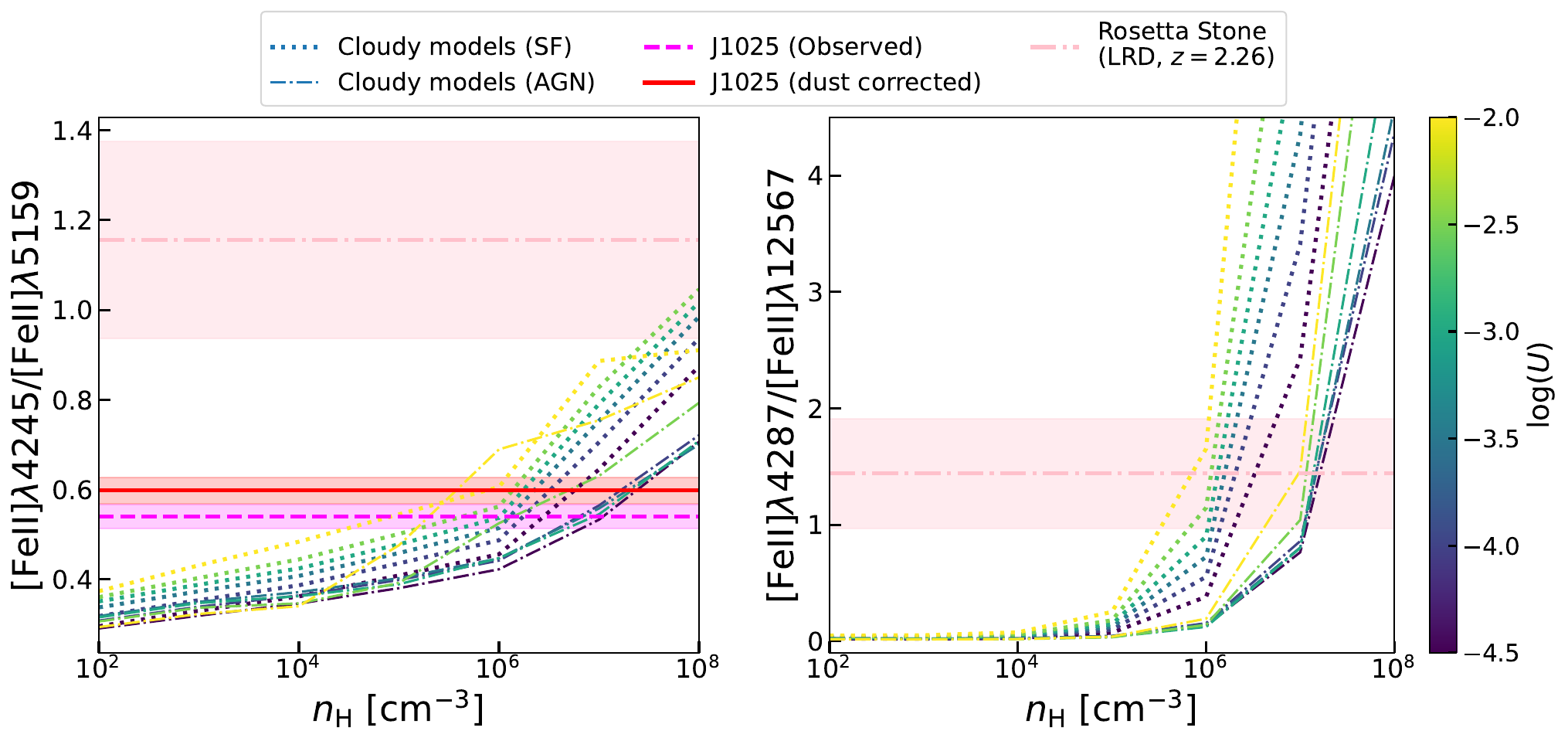}
    \caption{
    Density diagnostic of \citet{Bautista_1998} with [\feii] lines.
    \textit{Left:} comparison between the density-sensitive line line ratio, [\feii]$\lambda 4245$/[\feii]$\lambda 5159$ (each line is actually a blend of two lines), observed in \target and those predicted by \cloudy photoionization models.
    Both the observed line ratios and the dust attenuation corrected line ratios together with the $1\sigma$ uncertainties are shown.
    Two sets of models with different ionizing SEDs and $-4.5\leq \log U \leq -2$ are plotted.
    Overall, the dust attenuation corrected line ratios of \target favor high-density models with $n_{\rm H}\sim 10^{5-7}~{\rm cm^{-3}}$. {For comparison, the line ratio measured from the Rosetta Stone, an LRD at $z=2.26$, is plotted, which also implies a high density despite the larger uncertainty.}
    \textit{Right:} density diagnostic with [\feii]$\lambda 4287$/[\feii]$\lambda 12567$. The measured value from the Rosetta Stone, which has the NIR coverage from NIRSpec, is compared with the same \cloudy models.
    Again, the observed line ratio indicates a high density for the [\feii]-emitting region.
    }
    \label{fig:feii_diag}
\end{figure*}

\begin{figure}
    \centering
    \includegraphics[width=\columnwidth]{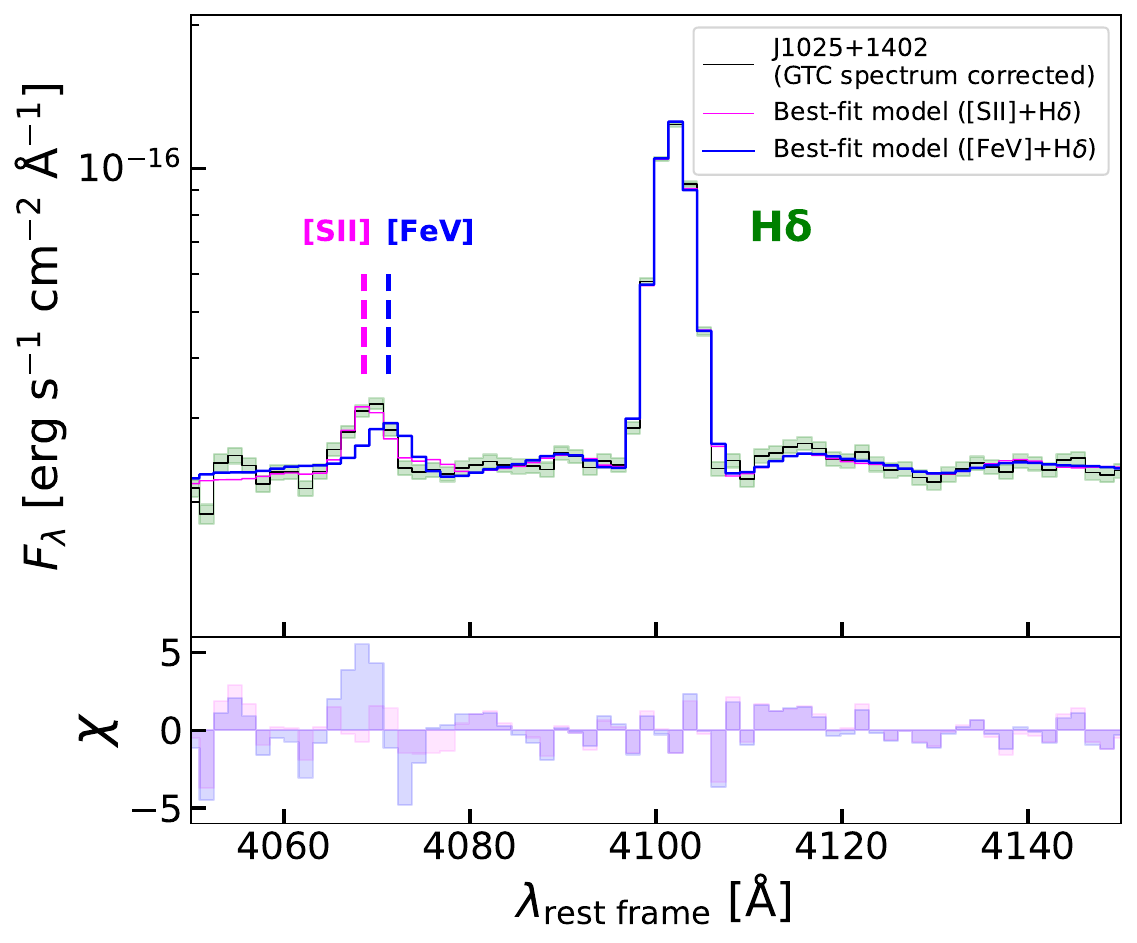}
    \caption{
    Comparison between two spectral fits with \textsc{pPXF} within a wavelength window around \hd, where we tied the kinematics of the weak line around 4070 \AA to those of \hd.
    The two fits assume the weak line is \sii$\lambda \lambda 4069,4076$ doublet and \fev$\lambda 4071$, respectively, and their residuals normalized by $1\sigma$ uncertainties are shown in the bottom.
    If the weak line has the same kinematics as those of \hd, the fit with \sii\ is significantly better compared to the one with \fev.
    The reduced $\chi^2$ of the two fits are $\chi^2_\nu[{\rm SII}]=1.8$ and $\chi^2_\nu[{\rm FeV}]=3.4$, respectively.
    }
    \label{fig:fev_fit}
\end{figure}

\begin{table}
        \centering
        \caption{Input parameters for \cloudy photoionization models to predict [\feii] line fluxes.}
        \label{tab:models}
        \begin{tabular}{l c}
            \hline
            Parameter & Values \\
            \hline
            $Z/Z_\odot $ & {0.1}\\
            \hline
            $\log U$& $-4.5$, $-4$, $-3.5$, $-3$, $-2.5$, $-2$ \\
            \hline
            $\log (n_{\rm H}/{\rm cm^{-3}})$& 2, 3, 4, 5, 6, 7, 8 \\
            \hline
            Stopping criterion & $n_{\rm e}/n_{\rm H}\leq 0.01$ or $T_{\rm e}\leq100$ K \\
            \hline
            SED & Black hole accretion disc and SSP (see text) \\
            \hline
            Solar abundance set & \citet{grevesse2010} abundance set\\
            \hline
            Dust & No dust\\
            \hline
            \feii\ atomic data& \citet{Bautista_2015}, \citet{Tayal_2018},\\
            & \citet{Smyth_2019}\\
            \hline
        \end{tabular}
\end{table}

Based on the high-S/N Gemini spectrum, we identified 14 emission lines associated with \feii\ within $4100~\AA < \lambda < 5300~\AA$.
To identify these transitions, we checked the line list of \citet{Bautista_1998}, which includes [\feii] emission lines observed in the Orion nebula.
In addition, we checked the built-in line list in the photoionizaiton code \cloudy \citep{ferland2017,Chatzikos_2023}, which uses \feii\ atomic data sets from \citet{Bautista_2015,Tayal_2018,Smyth_2019} as detailed in \citet{sarkar2021}.
We also checked transitions from Fe\,{\sc iii}, which generally do not match the lines we identified.
The absence of Fe\,{\sc iii} suggests a peculiar ionizing condition characterized by low-energy photons.

The fluxes we measured for the identified Fe lines are listed in Table~\ref{tab:fe_forest}, among which 9 lines are comparable to or stronger than \heii$\lambda 4686$ measured in the GTC spectrum.
From Figure~\ref{fig:uvo_sed}, some of these [\feii] lines are potentially present in the Rosetta Stone (LRD at $z=2.26$) as well, with the strongest one being [\feii]$\lambda 5159$. {In Figure~\ref{fig:hbfeii}, the R1000 spectrum of the Rosetta Stone is rescaled to match the continuum level of \target near \hb and \hg, where there are many [\feii] emission lines present in \target.
By comparing the two LRDs, one can see the Rosetta Stone has indications of [\feii]$\lambda 4245$, [\feii]$\lambda 4287$, [\feii]$\lambda 5159$, and potentially [\feii]$\lambda 4414$.
}
In fact, as shown by {\citet{tripodi2025deepdivebroadlineregion} and \redtxt{\citet{irony}}}, similar \feii\ transitions are identified in high-$z$ LRDs once enough S/N is achieved.
This suggests similarly low-ionization environments in these high-$z$ LRDs.

The [\feii] lines can be produced through either fluorescence by UV photons or collisional excitation \citep{Baldwin_1996,Bautista_1998}.
Due to the low IP of $\sim 7.6$ eV of Fe, [\feii] lines are presumably from the transition zones between ionized gas and neutral gas.
The presence of strong nebular absorption in the spectrum of \target, including the NaD and Balmer absorption, also implies the presence of a large column of neutral gas along the line-of-sight (LOS).

With the \feii\ transitions, one can perform diagnostics of the gas properties in the transition zone.
For example, \citet{Bautista_1998} show the flux ratio between the two [\feii] blends, [\feii]$\lambda 4245$/[\feii]$\lambda 5159$ is a good tracer of the gas density.
These two blends are also not affected by any obvious underlying absorption.
In Figure~\ref{fig:feii_diag}, we plot the flux ratio of [\feii]$\lambda 4245$/[\feii]$\lambda 5159$ and compare it with the model predicted ratios as a function of the gas density using the photoionization code \cloudy \citep[v17.03,][]{ferland2017}.
For the model, we used two sets of ionizing SEDs, one is
a theoretical AGN accretion disc with $M_{\rm BH}=10^6~M_\odot$ and $\lambda _{\rm Edd}=0.1$ from \citet{Pezzulli_2017}, the other is a young simple stellar population (SSP) with an age of 1 Myr from the Binary Population and Spectral Synthesis models \citep[BPASS;][]{stanway2018,byrne2022}.
We considered a range of ionization parameters and fixed the metallicity to about 10\% solar.
There is a secondary dependence of the flux ratio on $\log U$, although it appears unlikely that $\log U$ could reach a high value due to the absence of Fe\,{\sc iii} transitions.
We summarize our model parameters in Table~\ref{tab:models}.
Based on the models, we inferred a high gas density of roughly $n_{\rm H}\sim 10^{5-7}~{\rm cm^{-3}}$ for the [\feii]-emitting region, which is broadly consistent with the model-based result of \citet{linxiaojing_locallrd_2025}, who fitted a density of $n_{\rm H}\approx 10^{7.5}~{\rm cm^{-3}}$.
This density is much higher compared to those inferred from the low-ionization lines of \sii\ and \oii\ ($n_{\rm e}\sim10^{2-3}~{\rm cm^{-3}}$, see Table~\ref{tab:properties}), and these lines are likely saturated in the [\feii]-emitting region.
Our derived [\feii]-density suggests a stratified NLR, where the [\feii]-emitting clouds are in a denser environment.
The stratified emission is not uncommon in NLRs of AGN \citep[see, e.g.,][]{ji2024}, and in the case of \target, the high-density zone is characterized by very low-ionization emission.
Intriguingly, from the line kinematics we measured from the Gemini/GMOS spectrum, the [\feii]$\lambda 7155$ line has a broader line profile compared to other narrow lines and might be slightly redshifted (see Table~\ref{tab:kinematics}), suggesting the high-density zone is also kinematically distinct.
This is in contrast to the previous observations of local AGN where high-ionization coronal lines are broader compared to lower ionization narrow lines \citep[e.g.,][]{rodriguez-ardila_2004,rodriguez-ardila_2011,muller-sanchez_2011,ramosalmeida_2025}.

{For comparison, we also plotted the line ratio measured from the NIRSpec G140M ($R\sim 1000$) spectrum of the Rosetta Stone, which show clear detection of [\feii] lines at $\lambda = 4245$, 4287, and 5159 \AA (see Figure~\ref{fig:hbfeii}).
Despite the much larger uncertainty, the [\feii] emission in this high-$z$ LRD also matches high-density models with $n_{\rm H}\gtrsim 10^{7}~{\rm cm^{-3}}$.
In addition, the NIRSpec G395M spectrum of the Rosetta Stone covers [\feii]$\lambda 12567$ in the NIR, allowing us to perform another density diagnostic illustrated by \citet{Bautista_1998}.
In the right panel of Figure~\ref{fig:feii_diag}, we compared the flux ratio of [\feii]$\lambda 4287$/[\feii]$\lambda 12567$ between the measured value from the Rosetta Stone and those predicted by \cloudy models. Again, the observed ratio is consistent with $n_{\rm H}\sim 10^{6-7}~{\rm cm^{-3}}$ for the Rosetta Stone.
This result is subject to the relative flux calibrations between different NIRSpec gratings {(which can introduce a $\sim 10$\% systematic, \citealp{deugenio_dr3_2024})}.
We also note that, while we did not correct the line ratio for dust attenuation, the dust-corrected ratio would only be higher and correspond to an even higher density.
}
As another caveat, we note that the optical [\feii] lines can have strong fluoresence contributions from the UV photons \citep{Baldwin_1996,Bautista_1998}, which might not be precisely modeled given the unknown intrinsic UV fluxes.
The inclusion of other weaker NIR [\feii] lines ($\lambda =8617$, 8892) would be useful as they are predominantly produced by collisional excitation \citep{Baldwin_1996}, which requires deep observations in the NIR.

{The recent work of \citet{linxiaojing_locallrd_2025} claims the detection of the high-ionization line, \fev$\lambda 4071$.}
In our GTC spectrum, our best-fit model prefers the auroral line \sii$\lambda 4069$ near this location.
The presence of this auroral line is expected as auroral lines of \oii$\lambda \lambda 7320,7330$ and \nii$\lambda 5755$ are both detected in the GTC spectrum.
In fact, if we jointly fit \sii$\lambda 4069$ and \fev$\lambda 4071$ in the GTC spectrum, the latter is not detected.
To avoid having the spectral fit of \sii\ and \fev\ dominated by other spectral regions, we performed a test by isolating a spectral window around \hd.
We performed two separate fits assuming only \sii\ or \fev\ is present in addition to \hd.
The results are shown in Figure~\ref{fig:fev_fit}.
If we tied the kinematics of \sii\ or \fev\ to those of \hd, then the fit with \sii\ is significantly better.
Of course, this does not rule out the possibility that \fev$\lambda 4071$ is blueshifted with respect to other narrow lines by $\sim 150$ \kms.
In the optical, \fev$\lambda 4227$ is usually much stronger than \fev$\lambda 4071$ but it is also not detected in our GTC spectrum.
\fev$\lambda 4227$ has a critical density of $n_{\rm crit.}\sim 10^6~{\rm cm^{-3}}$, which is lower than that of \fev$\lambda 4071$ ($n_{\rm crit.}\sim 10^8~{\rm cm^{-3}}$).
Thus, if one believes the \fev$\lambda 4071$ detection, a high density of $n_{\rm e}\gtrsim 10^6~{\rm cm^{-3}}$ needs to be present to not detect \fev$\lambda 4227$.
Still, there could be a small contribution from \fev$\lambda 4071$ that is blended with \sii$\lambda 4069$, as the ionization potential of $\rm Fe^{3+}$ is very similar to that of $\rm He^+$ and we detect a weak \heii$\lambda 4686$.
Overall, the optical spectrum of \target is dominated by low-ionization species.

\subsubsection{Absorbing medium}

One of the most intriguing features of \target is the deep absorption line near \ha, as already noted by \citet{burke_abs_2021}.
As can be seen in Figure~\ref{fig:nadhaabs}, this absorption is clearly deeper than the underlying continuum, indicating that it must be absorbing the broad \ha at the same time.
A very similar absorption line is observed in the spectrum of the LRD GN-28074 as shown in Figure~\ref{fig:uvo_sed}.
To date, about 10\,-\,20\% of \jwst-selected broad-line AGN show similar Balmer absorption \citep{matthee2024,juodzbalis_rosetta_2024,lin_aspire_2024,labbe_monster,ji_lrdbreak_2025,deugenio_qso1_2025,naidu_lrd_2025,taylor_agn_2024} and more could have been missed due to the insufficient spectral resolutions and S/N \citep{deugenio_lrdoutflow_2025}.
In comparison, the occurrence of narrow Balmer absorption in the local AGN is very rare \citep{Izotov_2008,wangxu_2015,burke_abs_2021,linxiaojing_locallrd_2025}.
As shown by \citet{juodzbalis_rosetta_2024}, such deep Balmer absorption cannot have a stellar atmospheric origin and is originating in gas with high densities capable of collisionally exciting hydrogen to $n=2$ and/or high column densities capable of trapping Ly$\alpha$ photons.
The implied gas conditions support a picture of high-covering fraction but slowly moving clouds close to the BLR \citep{juodzbalis_rosetta_2024}.

The strength of the absorption in \ha would give the column density of hydrogen in the excited state at $n=2$, $N_{\rm H~(n=2)}$.
\citet{burke_abs_2021} derived $\log N_{\rm H~(n=2)}=14.2~{\rm cm^{-2}}$ using the EW of \ha absorption following the method of \citet{wangxu_2015}.
This approach is only appropriate when the core of the line is not strongly saturated.
By doing so, \citet{burke_abs_2021} implicitly assumed the absorption only applies to the observed continuum below \ha, which is not exact as the absorption profile is clearly deeper than the continuum.
We derived $N_{\rm H~(n=2)}$ based on a different approach, where we modeled the absorption profile explicitly and obtained the integrated optical depth, $\tau$, over the entire line profile.
The column density is given by
\begin{equation}
    N_{\rm H~(n=2)} = \frac{m_{\rm e}c}{4\pi e^2 f_0 \lambda _0}\tau,
\end{equation}
where $m_{\rm e}$ is the electron mass, $c$ is the speed of light, $e\equiv q_{\rm e}/\sqrt{4\pi \epsilon _0}$ is the ``Gaussian electron charge'' ($q_{\rm e}$ is the electron charge and $\epsilon_0$ is the vacuum permittivity), $f_0$ is the oscillator strength, $\lambda _0$ is the central wavelength, and $\tau$ is the integrated optical depth with a dimension of velocity.
We obtained $\log N_{\rm H~(n=2)}=13.77\pm 0.04$ much lower than \citet{burke_abs_2021}'s result.
This is because, as shown in Section~\ref{sec:method}, our best-fit model implies absorption in \textit{both} the continuum and the broad \ha rather than just in the continuum.
With $N_{\rm H~(n=2)}$, \citet{burke_abs_2021} then inferred a total column density of $N_{\rm H}=1.6\times 10^{17}~{\rm cm^{-2}}$ assuming that the gas cloud is homogeneous in all physical conditions with $T_{\rm e}=7500$ K, which again is not exact.
As shown by the photoionization models of \citet{juodzbalis_rosetta_2024}, most of $N_{\rm H~(n=2)}$ is accumulated in the transition zone between the ionized gas and the neutral gas, meaning that the homogeneous gas assumption tends to bias $N_{\rm H}$ low.
Based on the analysis of \citet{juodzbalis_rosetta_2024}, one generally needs $n_{\rm H}>10^{8}~{\rm cm^{-3}}$ and $N_{\rm H}>10^{21.5}~{\rm cm^{-2}}$ to accumulate $N_{\rm H~(n=2)}\gtrsim 10^{14}~{\rm cm^{-2}}$ in a cloud photoionized by a supermassive black hole accretion disc.

{In addition to the \ha absorption, there is detection of the \hb absorption in the SDSS spectrum as shown in Figure~\ref{f.gmos.gaussfit.a}.
This \hb\ absorption has also been confirmed in the \redtxt{Large Binocular Telescope (LBT) Multi-Object Double Spectrograph (MODS)} spectrum obtained by \citet{linxiaojing_locallrd_2025}.
Interestingly, and perhaps surprisingly, the \hb absorption shows a different kinematic compared to the \ha absorption and they have opposite signs of velocities with respect to narrow lines (see Table~\ref{tab:kinematics}).
Such a difference in the kinematics is also seen in the high-resolution ($R\sim 2700$) \jwst spectrum of the triply imaged LRD, Abell2744-QSO1 at $z=7.04$, although the \hb absorption in this high-$z$ LRD is detected only at $4\sigma$ significance \citep{ji_lrdbreak_2025,deugenio_qso1_2025}.
As another example, \redtxt{\citet{irony}} find significant detection of a redshifted \hb\ absorption and a blueshifted \ha\ absorption in an LRD at $z\sim 6.7$.}
Similar inverse P-Cygni-like features in \hb and \ha can be seen in the spectrum of the supernova SN 2010jl interacting with the circumstellar medium \citep{Fransson_sne_2014}.
The difference implies that the \ha\ absorption comes from outflowing gas whereas the \hb\ absorption is dominated by infalling gas, and their optical depths are not fixed by their oscillator strength ratio.
Whether this kinematic inconsistency is ubiquitous in LRDs \redtxt{remains} unclear.
At least the Rosetta Stone analyzed by \citet{juodzbalis_rosetta_2024} shows consistent \ha and \hb absorptions.
We will present a detailed comparison between the absorption features of different LRDs in future work.

In the spectrum of \target, the second strongest absorption feature is the NaD absorption.
The \nai absorption is usually found in the spectra of cool stars, but it can also trace ISM absorption and neutral gas outflows.
The observed strength of the NaD absorption is much deeper than that reproduced by stellar templates.
We measured the kinematics of NaD as an independent absorption line.
Despite relatively large uncertainties, we obtained a velocity of $-50\pm 25$ \kms consistent with the \ha\ absorption velocity and a width of {$\rm FWHM=260\pm 50$ \kms} marginally higher than that of the \ha absorption.
This implies the NaD absorption can have an outflowing component similar to that of the Balmer absorption.
We derived a very high column density of $\log N_{\rm Na^0}[{\rm cm^{-2}}]=14.04\pm 0.24$, or $\log N_{\rm Na^0}[{\rm cm^{-2}}]=13.9^{+0.17}_{-0.30}$ if we assume the intrinsic width should be the same as \ha.
It is difficult to reconcile the comparably high column densities of neutral Na and excited H(n=2) in a single low-metallicity cloud, and one possibility is that NaD has an additional contribution from stellar atmospheres or the ISM.
Finally, we calculated the maximally allowed turbulence in the absorbing gas from the absorption line width and obtained $v_{\rm turb}<\sqrt{2}\sigma=112$ \kms for \ha.
This value is similar to the measurement made in high-$z$ LRDs \citep[e.g.,][]{ji_lrdbreak_2025,degraaff_lrd_2025} and is compatible with the microturbulence value required for BLRs \citep{bottorff2000} or ensembles of outflowing clouds.

In addition to the Balmer and NaD absorptions, the GTC spectrum of \target shows a series of absorption lines potentially related to ionic, atomic, and even molecular transitions.
Whether any of these lines can be contributed by the same slowly outflowing gas is unclear, but many of the absorption lines are clearly deeper than what can be reproduced by our stellar templates.
We discuss the identities and implications of these weaker absorption lines in Section~\ref{subsec:cool_optical}.

Interestingly, if one compares the \ha\ absorptions in SDSS and Gemini spectra that were obtained at different epochs, there is a remarkable similarity in the shapes of the absorption (see Appendix~\ref{appendix:gemini_reduction}), suggesting the absorbing medium as well as the BLR emission are rather stable.
This is in clear contrast to the luminous local Seyfert, NGC 4151, where the Balmer absorption is varying drastically with the broad lines in a timescale of roughly 1 yr \citep{hutchings2002}.
Next, we discuss the spectral variability of \target based on line measurements.

\subsection{Variability}
\label{subsec:var}

\begin{figure}
    \centering
    \includegraphics[width=\columnwidth]{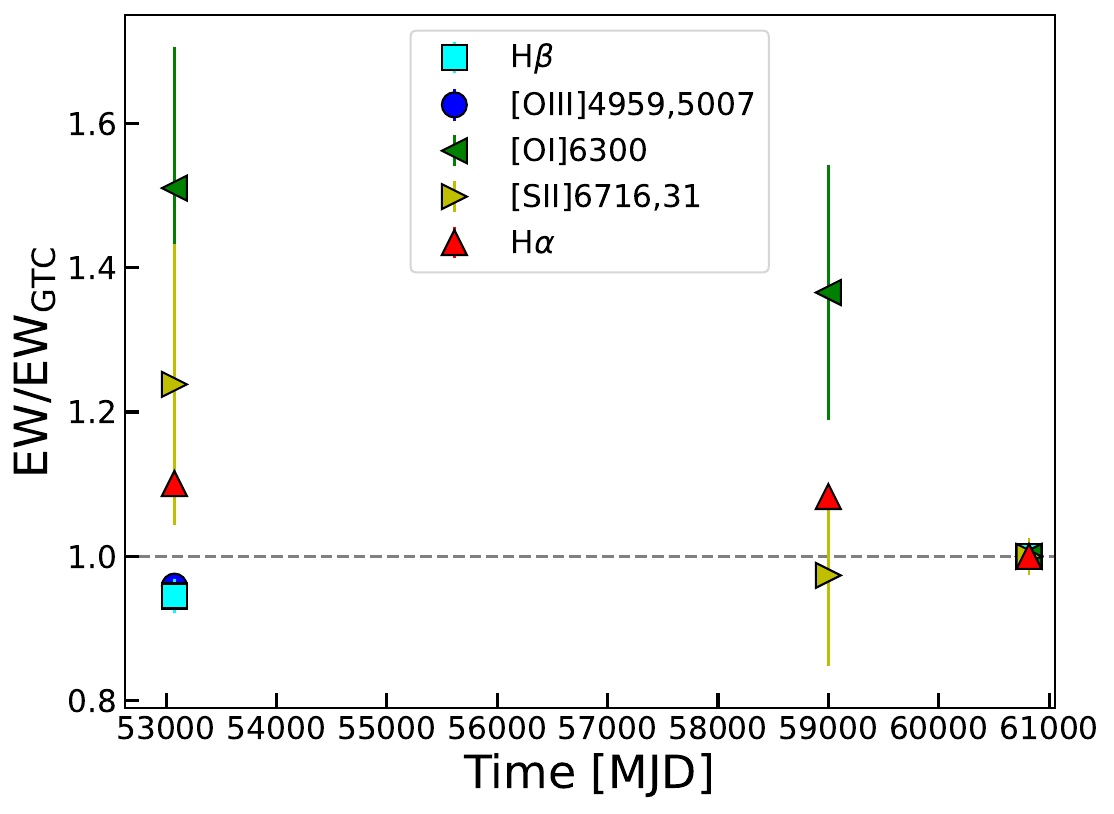}
    \caption{Relative equivalent widths of emission lines measured in the SDSS spectrum ($\rm MJD=53075$), the Gemini spectrum ($\rm MJD=58999$), and the GTC spectrum ($\rm MJD=60817$).
    All line EWs are normalized to the EWs of corresponding lines measured in the GTC.
    Only the spectral region around \ha\ show significant variation at $>5\sigma$, while the EW differences between the SDSS and GTC spectra in \hb\ and \oiii\ are marginal ($\lesssim 3\sigma$).
    Considering the slit loss effect is unlikely to reduce the EWs in the GTC spectrum if the continuum is more spatially extended compared to lines, the $\sim10\%$ variability in \ha\ implies tentative continuum variation.
    }
    \label{fig:line_var}
\end{figure}

As shown by \citet{kokubo_harikane2024} and \citet{zhang_var_2024} for high-$z$ LRDs and by \citet{Maiolino2024_Xrays} for general \jwst-selected broad-line AGN, there is a clear lack of continuum variability in these sources compared to local galaxies \citep{burke2021}.
However, not all LRDs lack variability.
For example, \citet{ji_lrdbreak_2025} and \citet{Furtak_qso1var_2025} recently showed that the triply imaged LRD, Abell2744-QSO1, at $z=7.04$ \citep{furtak_abell2744_2023,furtak2023} exhibits significant variability in terms of the \hb EWs measured from different gravitationally lensed images.
There is also an indication of variability reported by \citet{naidu_lrd_2025} for an LRD at $z=7.8$.

For \target, \citet{burke_abs_2021} show by comparing the SDSS spectrum taken at Mar 2004 and the Gemini spectrum taken at May 2020, there is only 10-20\% variation in the luminosity of broad \ha, consistent with low variability seen in the observations taken with the 3.5-m Sloan Telescope during these two observations \citep{Izotov_2008}.
Due to the uncertainty in the flux calibration of the Gemini spectrum, and given that we now have a new spectrum from GTC taken at May 2025, we performed a new variability analysis below combining the SDSS, Gemini, and GTC spectra.

We started by examining the spectral variability around \ha.
We normalized the continua of all three spectra (SDSS, Gemini, and GTC) to the same level in the wavelength range of $5950~\AA < \lambda < 6850~\AA$.
This step is to remove any potential systematics induced by relative flux calibrations between the continua bluewards and redwards of \ha.
However, we emphasize that this step actually has little impact on the variability analysis we made based on line EWs.
To normalize the spectra, we used the approach described in Appendix~\ref{appendix:slit_loss}.
Specifically, we defined seven 50 \AA-wide and non-overlapping spectral windows within the above wavelength range and calculated median flux densities and uncertainties within the windows for individual spectra.
We then fit the relative flux densities between the three spectra as a linear function of wavelength and applied the corrections to the Gemini and GTC spectra.
For the GTC spectrum this leads to a correction factor of 2.45 at 6563 \AA.
The final results are plotted in Figure~\ref{fig:line_var}, where $\rm EW(H\alpha)_{SDSS}=456.5\pm 2.3~\AA$ and $\rm EW(H\alpha)_{GTC}=414.7\pm 0.3~\AA$, implying a significant change of 9\% ($18\sigma$) over 19 years in the rest frame of \target.
In the Gemini spectrum, we found $\rm EW(H\alpha)_{SDSS}=449.1\pm 1.1~\AA$, suggesting 8\% variation occurred between the Gemini and GTC observations ($\sim 4$ years in the rest frame).
We also measured the EWs of weaker narrow lines near \ha, including \sii$\lambda\lambda 6716,6731$ and \oi$\lambda 6300$ and plotted their values in Figure~\ref{fig:line_var}.
Both lines show tentative enhancement in their EWs in the SDSS spectrum, although this is statistically insignificant due to the large measurement uncertainties.
If EWs of all the above lines are actually varying, the plausible explanation is the continuum itself is varying.
Alternatively, if only \ha\ is varying, the variability could come from a response to the previous variation in the underlying continuum.

To understand whether the EW variations could be caused by any systematics, we performed another measurement for \hb\ and \oiii\ (not covered by the Gemini spectrum).
This time, the EWs of both lines are tentatively lower by 5\% in the SDSS spectrum, although insignificant again ($\lesssim 3\sigma$).
We also checked GTC spectra coadded from two separate groups of science exposures with a total exposure time of 1 hr each, which have significantly different seeing conditions ($0.\!\!''8$ versus $1.\!\!''2$), and we found consistent results supporting low but significant variation in the EW of \ha.

Since the GTC spectrum was obtained with a $0.\!\!''6$-slit, one might wonder whether the small change in the EWs might be due to differential slit lossess, affecting more the more spatially extended component.
For example, any continuum due to stellar populations should be more
extended than the BLR component.
However, while the loss of the continuum could explain the tentatively higher EWs of \hb and \oiii\ in the GTC spectrum, it cannot explain the significantly lower EWs of \ha in the GTC spectrum.
{Also, the lowering \ha EW is unlikely to be explained by the possibility that the narrow \ha\ is more spatially extended.
This is because, based on the analysis of the Gemini slit image, the core 
of \ha\ remains unresolved, with $\rm FWHM<0.\!\!''64$ and no detectable
difference from either the continuum and the broad wings of the line.
In addition, if the narrow lines are more spatially extended, \oiii\ and \hb EWs should be decreasing more significantly compared to \ha, instead of the marginal increase we observed.
}
Thus, a more natural explanation for the change in the \ha\ EWs, and possibly the \sii\ and \oi\ EWs, is a small variation in the accreting black hole.
Overall, the variability in \target is modest and consistent with the small variability reported by \citet{Izotov_2008,burke_abs_2021}.
The stable broad \ha\ EW rules out the possibility of \target being a supernova, and its luminosity is too high to be explained by any variable star.

Finally, we comment on the photometric variability of \target.
Based on the photometric data we used (see Section~\ref{sec:data}), there is no clear indication of any variability beyond $1\sigma$ from the FUV to MIR, which has been confirmed by \citet{linxiaojing_locallrd_2025} as well.
We note that \citet{burke_abs_2021} have shown a comparison between the light curve constructed from the Catalina Real-Time Transient Survey \citep[CRTS,][]{crts} $V$-band data and the Zwicky Transient Facility \citep[ZTF,][]{ztf} $r$-band data and the prediction from the damped random walk-like model for QSO variability \citep{Butler_2011}.
\citet{burke_abs_2021} found that the light curve of \target
has a variability significance of $2.3\sigma$ and the significance of the variability being
consistent with the QSO variability pattern is $3.7\sigma$.
Thus, there is evidence that the optical light of \target is varying QSO-like, but clearly not significant.
\redtxt{More recently, \citet{burke_var_2025} updated the photometric variability analysis for \target leveraging new epochs of ZTF observations, confirming a low intrinsic variability of $<3-4$ \% at $3\sigma$.
}
To better understand the contribution from the AGN to the observed spectrum, it is vital to investigate the panchromatic SED of \target, which we describe in the next section.




\section{Interpretation of the SED}
\label{sec:sed}

In this section, we discuss the physical interpretation of the different parts of the spectrophotometric energy distribution of \target.
We also discuss whether any of these interpretations can be applied to high-$z$ LRDs discovered by \jwst.

\begin{figure}
    \centering
    \includegraphics[width=\columnwidth]{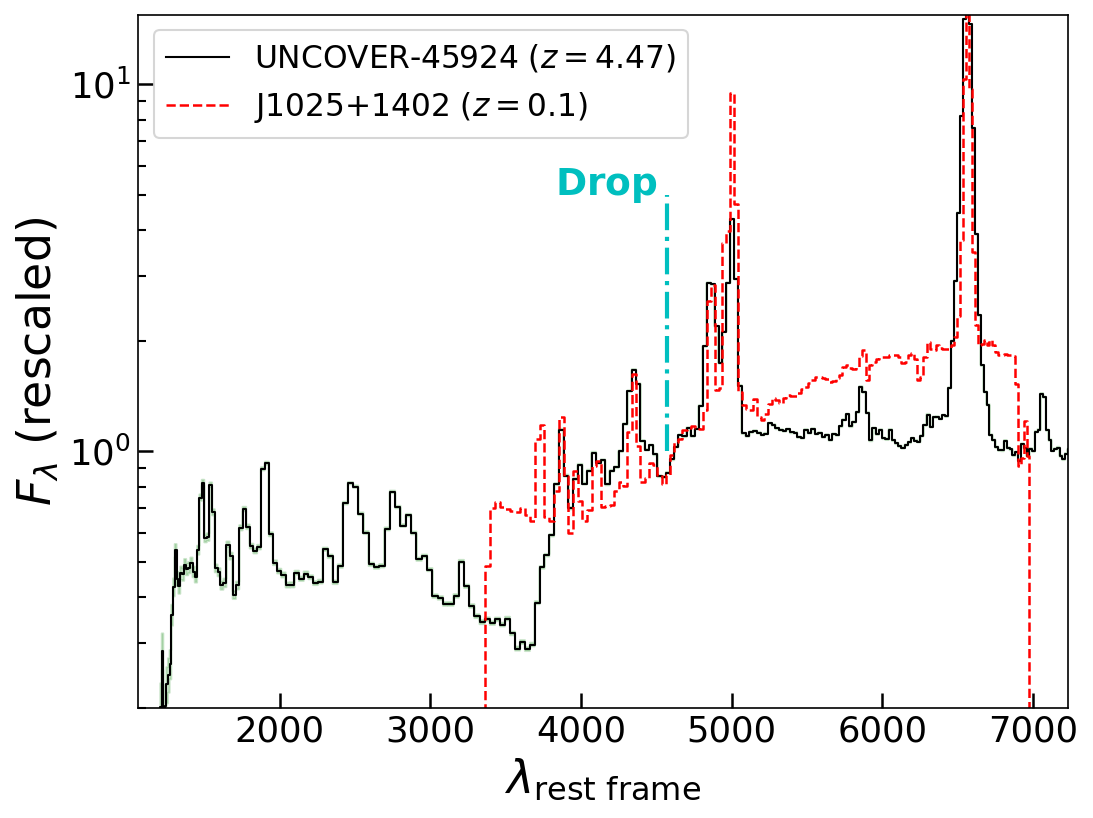}
    \caption{
    Comparison between the \jwst/NIRSpec PRISM ($R\sim 100$) spectrum of the LRD, UNCOVER-45924 at $z=4.47$, and the GTC spectrum of \target convolved to the PRISM resolution.
    The flux densities of both spectra are rescaled to be compared.
    Despite being redder in the optical, \target shows an exact same drop and a continuum bump redwards of the drop as in UNCOVER-45924.
    This implies the drop feature is a real absorption in the continuum, rather than an artifact made by broad emission lines on both sides as recently proposed by \citet{labbe_monster}.
    }
    \label{fig:monster_drop}
\end{figure}

\subsection{Cool optical}
\label{subsec:cool_optical}

One of the debating issues for \jwst-selected LRDs is the origin of the optical continuum, which appears ubiquitously red and peaks around the location of \ha\ in terms of $F_\lambda$ (see e.g., Figure~\ref{fig:full_sed}).
The high-S/N GTC spectrum reveals rich absorption features in \target, whose identities can help to understand the ionization and temperature conditions of the medium that produces them.

With \textsc{pPXF} we used a semi-empirical approach to fit the optical continuum of \target, as shown in Figure~\ref{fig:cool_star_fit}.
Based on the best-fit weights of MILES stellar templates, we found that the dominant stellar templates are a G0Ia yellow supergiant, HD 18391, with $T_{\rm eff}=5500$ K, and a K0 chemically peculiar star, BD+06\_0648, with $T_{\rm eff}=4400$ K.
These two stars make up 80\% of the total weight.
\redtxt{However, we note that it does not mean that the spectrum of \target is made up of two stars only, but rather the continuum shape is best described by cool supergiant-like stellar populations.
}
Indeed, for a spectral peak around \ha, one expects cool stars with $T_{\rm eff}\sim 4000-5000$ K.
In addition, there is no TiO band present in the spectrum, indicating a not too low temperature consistent with G to K giants.
While dust attenuation would allow for hotter stars, this would be inconsistent with the small nebular attenuation.
\redtxt{It is worth noting that we have included additive polynomials in our \textsc{pPXF} fit, which modifies the strengths of the absorptions in the original stellar templates to better fit the observed continuum.
However, even with the unphysical modifications by the additive polynomials, many of the observed absorption lines are still deeper compared to the model.
}
This might suggest a non-stellar origin for the absorptions, similar to the Balmer absorption and NaD absorption, as well as the NIR \caii\ T absorptions shown by \citet{linxiaojing_locallrd_2025}, or at least a non-standard stellar population.

We also identified typical stellar absorptions including the \caii\ K absorption (although the strength of \caii\ K is clearly over-fit by the stellar templates), G-band absorption, and the Mgb absorption.
The strong G-band absorption, which is produced by the CH (methylidyne) molecule, suggests a cool atmosphere where molecules are not photo-dissociated.
Compared to other ionic and atomic absorptions, CH absorption revealed by the GTC spectrum is an important feature.
This is because, as a typical G-to-K type star feature, the CH absorption is primarily produced in atmospheric conditions with $T_{\rm eff}\sim 5000$ K \textit{and} extremely high gas densities of $n_{\rm H}\gtrsim 10^{14}~{\rm cm^{-3}}$ \citep{Gray_2008}.
The extreme conditions make the CH absorption much more difficult to be explained by an external origin in the ISM or outflows.

In addition to the typical stellar absorptions above, we present tentative identifications of absorption lines in Figure~\ref{fig:cool_star_fit} based on absorption lines mostly identified in K-to-G stars \citep{stellar_spec}. These lines are mostly associated with low-ionization ions and atoms, and their observed strengths are stronger than those in typical stars indicated by the stellar templates.
In the blue part of the spectrum, there is a clear drop near 4570 \AA with unknown origin, potentially resulted from a blend of absorption lines including Ti\,{\sc ii}$\lambda 4572$.
This drop has been unambiguously identified in high-$z$ LRDs, including the Rosetta Stone at $z=2.26$ as shown in Figure~\ref{fig:uvo_sed} and another bright LRD at $z=4.47$ present by \citet{labbe_monster}.
In Figure~\ref{fig:monster_drop}, we compare the spectrum of \target to the \jwst/NIRSpec PRISM spectrum of a luminous LRD, the UNCOVER ``Monster'' Abell2744-45924, at $z=4.47$ \citep{greene2024,labbe_monster}.
Once again, the 4570 \AA-drop is clearly visible in the PRISM spectrum of the LRD.
If we convolved the GTC spectrum to the PRISM resolution and rescaled the spectra, the 4570 \AA-drops and the small bumps redwards of the drop in both spectra match well.

\citet{labbe_monster} argue the 4570 \AA-drop is not an actual absorption, but a feature made by broad \feii\ and \heii$\lambda 4686$ emission from the BLR on both sides of the drop.
While this scenario is partly justified by the strong \feii\ emission seen in the UV of the Monster, such a shape of the optical \feii\ bump was not observed in typical AGN or quasars previously, and the general population of high-$z$ broad-line AGN selected by \jwst\ show very weak optical \feii\ bumps \citep{trefoloni_feii_2024}.
In fact, typical optical \feii\ bump is characterized by a peak around 4570 \AA rather than a drop, and UV-to-optical \feii\ strengths vary strongly with the gas-phase metallicity and other physical conditions in the outskirt of BLRs \citep{shields2010}.
With the medium-resolution ($R\sim 1000$) GTC spectrum, we found no clear indication of any broad-\feii\ emission lines, nor any strong broad \heii\ emission, and there is a true absorption (band) around 4570 \AA.
Still, the origin of the absorption is unclear and we will investigate it by exploring a more comprehensive stellar library in future work.

In the red part of the spectrum as shown in the bottom right panel of Figure~\ref{fig:cool_star_fit}, there are again multiple absorption lines associated with atomic transitions.
At $\lambda = 6142$ \AA, there is a clear absorption line potentially from Ba\,{\sc ii}$\lambda 6142$ or Ca\,{\sc i}$\lambda 6142$.
We note that \citet{linxiaojing_locallrd_2025} identify the line as \feii$\lambda 6148$, which, however, does not match the central wavelength of the absorption in our GTC spectrum.
Finally, at the blue wing of the broad \ha, there is a known absorption feature, the 6497 \AA-blend, typically seen in G-to-K type stars, which \citet{linxiaojing_locallrd_2025} identified instead as a diffuse interstellar band (DIB) absorption.
The 6497 \AA-feature could be a blend of Ba\,{\sc ii} and Fe\,{\sc i} and its strength seems recovered by the stellar template.
Given the Balmer absorption in \ha, another interesting possibility for this feature is a P-Cygni absorption from stellar winds, similar to the double-absorption profiles seen in $\eta$ Carinae systems \citep{Gull_etacar_2001}, which also exhibit abundant low-ionization emission lines typically associated with \feii\ \citep[e.g.,][]{Choe_etacar_2025}.
However, this shift would imply a wind velocity of $\sim 3000$ \kms, and no other absorptions around, for example, \hb or \hei$\lambda 5876$ can be interpreted as P-Cygni features with similar velocities.
Notably, the G-to-K giant-like optical, in combination with the blue UV, makes \target share similar spectral shapes as some of the symbiotic star systems or X-ray binaries, such as AG Dra and V404 Cyg, both of which have their optical spectra dominated by K (sub)giants and UV dominated by a hot star or an accreting black hole \citep{symbiotic_2003,v404}.

The low-ionization potentials of the identified ionic and atomic transitions imply a cool atmosphere.
As recently discussed in \citet{ji_lrdbreak_2025,degraaff_lrd_2025,naidu_lrd_2025}, the optical continua of high-$z$ LRDs can be explained by AGN accretion discs and BLRs obscured by dense neutral gas with $n_{\rm H}\sim 10^{10-11}~{\rm cm^{-3}}$, which produces strong Balmer absorption that is clearly deeper than the underlying continua and strong Balmer breaks deeper than what can be produced by typical stellar populations.
Alternatively, turbulent accretion flows surrounding the accretion disc can make a K-star like optical continuum with a Balmer break \citep{Liu_speddlrd_2025}.
Clearly, as we have shown, there is no strong Balmer break in \target, despite the presence of deep Balmer absorption.
This is not unexpected as the presence of Balmer absorption with a non-stellar atmospheric origin does not guarantee the presence of a Balmer break \citep{juodzbalis_rosetta_2024,Inayoshi_maiolino_2025,ji_lrdbreak_2025}.
The presence of the G-band absorption from the molecular CH adds more constraints to the AGN models, suggesting very dense and cool gas in the accretion flows or the vicinity of the BLR, if it has a non-stellar origin.
While densities as high as $n_{\rm H}\approx10^{14}~{\rm cm^{-3}}$ are not implausible for BLRs \citep{Netzer_2013,Moloney_2014,Temple_2020}, whether the rest of the gas conditions would allow for reproducing the observed absorption lines require further investigation.
Alternatively, if the optical absorptions are truly associated stars, one would need exotic stellar atmospheric conditions to explain many underpredicted absorption lines.

As a result, the current constraints based on the optical continuum alone still pose challenges for explaining LRDs with either AGN or stellar models.
Next, we explore observational evidence from other wavelength ranges.


\subsection{Missing X-rays}

\begin{figure*}
    \centering
    \includegraphics[width=1.75\columnwidth]{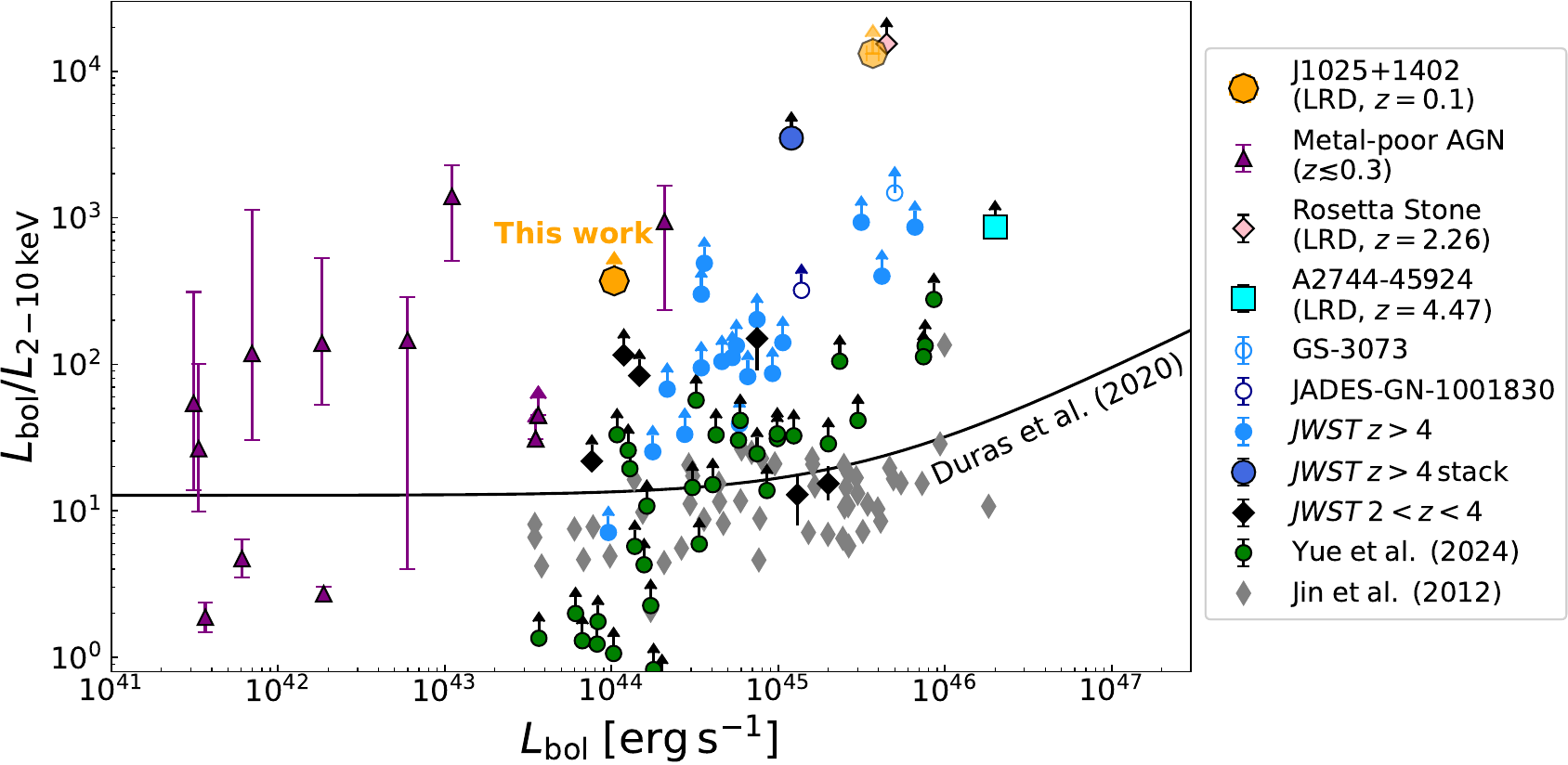}
    \caption{
    Comparison between the nominal bolometric luminosities and the bolometric-to-X-ray luminosity ratios for different samples of broad-line AGN. 
    For \target, the solid symbol represents the derived values without any dust attenuation correction and the transparent symbol represents the derived values with the maximum dust attenuation correction.
    In addition to the local LRD \target at $z=0.1$,
    \redtxt{we plot a sample of metal-poor AGN candidates in dwarf galaxies at $z\lesssim 0.3$ from \citet{Simmonds_xraydwarf_2016} and \citet{Baldassare_2017}.}
    We also plot two bright LRDs, the Rosetta Stone at $z=2.26$ \citep{juodzbalis_rosetta_2024} and Abell2744-45924 (the UNCOVER Monster) at $z=4.47$ \citep{labbe_monster}.
    We also plot two blue high-$z$ AGN, GS\_3073 at $z=5.55$ \citep{ubler2023a} and JADES-GN-1001830 at $z=6.68$ \citep{juodzbalis_dormantagn_2024}.
    Constraints for \jwst-selected broad-line AGN at $z>2$ are from \citet{Maiolino2024_Xrays}, and constraints for a sample of LRDs come from \citet{yue_lrd_2024}.
    We also plot a sample of local broad-line AGN from \citet{Jin_2012} at $z<0.4$ for comparison, as well as the best-fit relation based on a sample of optically selected AGN by \citet{Duras_2020}.
    \target stands out as the broad-line AGN with the most stringent constraint on $L_{\rm bol}/L_{\rm 2-10~keV}$ at the low luminosity end of $L_{\rm bol}\sim 10^{44}$ \ergs. All X-ray limits plotted are 90\% confidence upper limits. We also provide $3\sigma$ upper limit for \target in the text, which would give \redtxt{$L_{\rm bol}/L_{\rm 2-10~keV}>145$}.
    }
    \label{fig:xray_com_fde}
\end{figure*}

\begin{figure*}
    \centering
    \includegraphics[width=2\columnwidth]{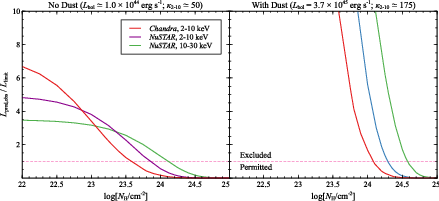}
    \caption{
    Ratio of the observed luminosity predicted assuming \target\ intrinsically exhibits a standard AGN SED to the limit implied by the X-ray non-detections from \textit{Chandra} and \textit{NuSTAR}, as a function of line-of-sight column density. Calculations are shown for both models for the BLR presented in Table \ref{tab:properties}: no dust -- left panel, dust obscured -- right panel. 
    Different curves correspond to the X-ray limits set by different telescopes and energy bands.
    The horizontal dashed line indicates the observational constraint and only the region below the line would be compatible with the current observations.
    The non-detection at high energies by \textit{NuSTAR} sets the most stringent constraint on the column density that would be required for a standard AGN SED to be consistent with the observational data, and would require a Compton-thick medium (i.e. $N_{\rm H}> 1.5 \times 10^{24}~{\rm cm^{-2}}$) along the line-of-sight in each case.
    }
    \label{fig:xray_column}
\end{figure*}


As noted above, there is no X-ray detection of \target\ to date, either in the archival \textit{Chandra} observation ($\sim$5\,ks exposure; \citealt{Simmonds_xraydwarf_2016}) or in our more recent \textit{NuSTAR} observation. We therefore analyzed the data from both observations to compute upper limits on the X-ray luminosity of \target\ for a couple of bands of interest, using the Bayesian approach outlined in \citet{Kraft91}. 
For the \textit{Chandra} data we consider a region 2$''$ in radius around the source position, and a region of 30$''$ radius for the \textit{NuSTAR} data when computing these limits; in both cases the background level was estimated from larger regions of blank sky on the same chip as the target position. For the \textit{NuSTAR} analysis we combine the data from FPMA and FPMB, and also consider both the `science' and `spacecraft science' data (\textit{NuSTAR} modes 1 and 6; see \citealt{Walton16cyg}), giving a total on-source exposure of $\sim$45\,ks. Initially we assume a simple, `unobscured' model consisting of a $\Gamma = 1.7$ powerlaw continuum modified only by the Galactic absorption column toward \target\ ($N_{\rm{H,Gal}} = 4 \times 10^{20}$\,cm$^{-2}$; \citealt{NH2016}) to allow for direct comparisons with relevant literature (e.g., \citealt{Maiolino2024_Xrays}), and use \textsc{pimms} (\citealt{PIMMS}) to convert between count rates and fluxes (adopting the appropriate cycle 10 responses for \textit{Chandra}).
For the 2--10\,keV band the 90\% and 99.73\% (3$\sigma$) limits we find for the \textit{Chandra} data with this model are $L_{\rm 2-10~keV} < 2.8 \times 10^{41}$\,\ergs\ and $< 7.5 \times 10^{41}$\,\ergs, respectively.\footnote{We note that \citet{Simmonds_xraydwarf_2016} reported a 99\% limit of $L_{\rm 2-10~keV}< 1.1 \times 10^{41}$\,\ergs\ based on the same \chandra observation, whereas our analysis yields a higher value of $L_{\rm 2-10~keV}< 5.5 \times 10^{41}$\,\ergs\ for the 99\% confidence limit. This is because the limit in \cite{Simmonds_xraydwarf_2016} is actually initially computed over a \textit{Chandra} bandpass that extends to lower energies (0.5--7\,keV), and is then converted into a 2--10\,keV limit assuming a similar spectral model to the unobscured case considered above. In contrast, the \textit{Chandra} analysis presented here focuses only on the 2--10\,keV band throughout, and therefore does not involve the softer X-ray data that would be most impacted by the presence of any further obscuration local to \target.}
For the \textit{NuSTAR} data the corresponding limits are $L_{\rm 2-10~keV} < 3.9 \times 10^{41}$\,\ergs\ and $< 7.7 \times 10^{41}$\,\ergs, respectively (initially calculated for the 3--10\,keV band and then converted into the 2--10\,keV band using the above spectral form). Both observatories thus provide comparable constraints on the luminosity of \target\ below 10\,keV.

However, should \target\ be heavily obscured, one might expect its X-ray spectrum to peak above 10\,keV, energies which can only be probed by \textit{NuSTAR}. 
While there is still no detection of \target\ above 10\,keV in the \textit{NuSTAR} data, we can compute a separate upper limit on its luminosity in the 10--30\,keV band. Here we adopt a more obscured spectral model, allowing reasonably heavy obscuration at the redshift of \target\ with $N_{\rm{H,int}} = 10^{23}$\,cm$^{-2}$. 
We find 90\% and 99.73\% limits of $L_{\rm 10-30~keV} < 5.9 \times 10^{41}$\,\ergs\ and $< 1.3 \times 10^{42}$\,\ergs, respectively.


The above results suggest \target is extremely X-ray weak compared to typical AGN in the local Universe.
In Figure~\ref{fig:xray_com_fde}, we compare the bolometric-to-X-ray luminosity ratio of \target with those of optically selected local AGN as well as high-$z$ broad-line AGN and LRDs selected by \jwst.
In this plot, all X-ray limits included represent 90\% confidence limits.
While \target lies at the low-luminosity end, it has \redtxt{$L_{\rm bol}/L_{\rm 2-10~keV}>145$} as the $3\sigma$ limit and \redtxt{$L_{\rm bol}/L_{\rm 2-10~keV}>371$} as the 90\% confidence limit (assuming a standard bolometric conversion from \ha\ from \citealp{sternlaor_2012}), which is significantly higher compared to the local AGN, and even the current limits for \jwst-selected Type 1 AGN and LRDs.
For comparison, we plot two bright LRDs, the Rosetta Stone and the Monster, which have more stringent limits on $L_{\rm bol}/L_{\rm 2-10~keV}$ but have $1-2$ magnitudes higher nominal bolometric luminosities.
\redtxt{Some metal-poor AGN candidates in dwarf galaxies selected through broad lines at $z\lesssim0.3$ exhibit certain levels of X-ray weakness, as measured by \citet{Simmonds_xraydwarf_2016} and \citet{Baldassare_2017} and shown in Figure~\ref{fig:xray_com_fde}, although the uncertainties are large.
A more systematic study with careful sample selection is needed to compare the X-ray properties of AGN candidates in local dwarfs and LRDs, which is beyond the scope of the current work.
}

The extreme apparent X-ray weakness of \target\ might potentially indicate similarly extreme levels of obscuration in this system. To investigate this further, we estimate the absorption columns that would be needed to result in the X-ray non-detections reported above, under the assumption that \target\ should intrinsically be similar to other AGN. 
To do so, we use the information provided in Table \ref{tab:properties} along with the $\lambda_{\rm{Edd}}$-dependent 2--10\,keV bolometric corrections ($\kappa_{2-10}$) presented for type-1 AGN in \cite{Lusso12} to predict what the intrinsic 2--10\,keV luminosity ($L_{\rm{pred,int}}$) of \target\ should be (and in the case of the 10-30\,keV band, subsequently determine the predicted intrinsic 10--30\,keV luminosity assuming a $\Gamma = 1.7$ powerlaw continuum). 
We then determine what the observed luminosity ($L_{\rm{pred,obs}}$) should be for a range of different line-of-sight column densities at the redshift of \target, accounting for losses due to both photoelectric absorption and scattering using models for both included in \textsc{xspec} (\citealt{xspec}).
These predicted luminosities are then compared to the corresponding limit on the luminosity provided by the X-ray data (which are also re-computed for each value of $N_{\rm{H}}$ considered; we again consider 90\% confidence limits here, as in Figure \ref{fig:xray_com_fde}). 

This process is repeated for both the models with and without dust, for which we find predicted intrinsic 2--10\,keV luminosities of $L_{\rm{pred,int}} \sim 1.3 \times 10^{43}$ and $\sim 2 \times 10^{42}$\,\ergs, respectively, based on the values presented in Table \ref{tab:properties}. 
The results are shown in Figure \ref{fig:xray_column}, where we plot the ratio of the predicted and limiting luminosities as a function of the hydrogen column density; ratios $>$ 1 would be excluded by the existing observations, while ratios $<$ 1 are still permitted by them.
The most stringent constraints here come from the high energy \textit{NuSTAR} data. Should \target\ intrinsically exhibit a relatively normal AGN SED, the level of obscuration required for a non-detection above 10\,keV by \textit{NuSTAR} would need to be in the Compton-thick regime, i.e., $N_{\rm{H}} > 1.5 \times 10^{24}$\,cm$^{-2}$. These limits can likely be viewed as conservative, as the presence of any reflected emission from the obscurer (predicted by all physical `torus' models, e.g. \citealt{mytorus, torus, borus}), and reducing the metallicity of the absorber to sub-solar values would both drive the limits on $N_{\rm{H}}$ even higher.

Alternatively, the extreme X-ray weakness of \target might indicate photon-trapping/a change in the structure of the corona in the super-Eddington regime, as recently proposed for high-$z$ AGN \citep{Maiolino2024_Xrays,lambrides_superedd_2024,madau_superedd_2024,Inayoshi_agnvar_2024}.
It is worth noting that \citet{Inayoshi_agnvar_2024} also connect the X-ray weakness to low optical variability, and the lack of high-ionization lines and the high broad-line Balmer decrement could be indications of softening ionizing radiation field from the accretion disc due to photon-trapping \citep{lambrides_superedd_2024}.
Without an actual detection or a stronger constraint from high-energy X-ray photons, we cannot further constrain the models.
Fortunately, given the relatively low-redshift of \target, future X-ray follow-ups will be valuable and efficient to understand the X-ray signatures of this LRD.
Next, we move to the UV regime where \target instead appears bright.



\subsection{Extreme FUV}
\label{subsec:fuv_crisis}

\begin{figure*}
    \centering
    \includegraphics[width=\textwidth]{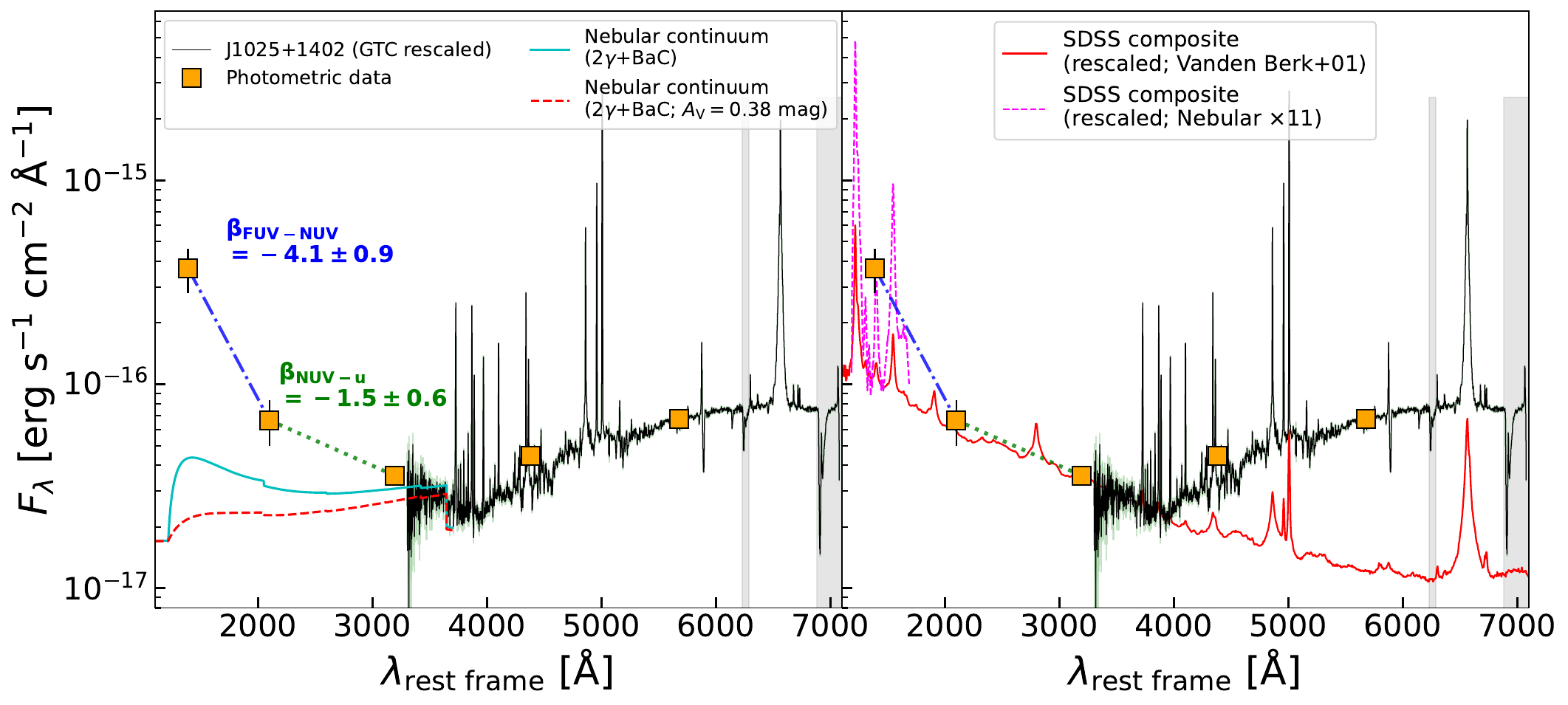}
    \caption{
    Unusually steep rest-frame FUV slope of \target.
    \textit{Left:} comparison between the predicted nebular continuum based on the measurements of optical narrow lines without dust extinction (solid cyan) and with dust extinction (dashed red).
    The zero-point continuum level of the nebular continuum is shifted to match the continuum of the GTC spectrum measured at 3800 \AA.
    Overall the nebular continua are too weak and flat to explain the UV slopes of \target.
    \textit{Right:} comparison between the composite spectrum of SDSS quasars from \citet{vandenberk2001} and observations of \target in FUV, NUV, u bands.
    The NUV-u slope of \target is consistent with that of the SDSS quasar ($\beta _{\rm UV}=-1.56$) within $1\sigma$.
    To explain the FUV observations, the nebular lines (partially or entirely) covered by the band, including Ly$\alpha$, Si\,{\sc iv}, \civ, and \heii\ need to be upscaled by a factor of 11.
    The required strength of high-ionization UV lines is in tension with the weak high-ionization optical lines (e.g., \heii$4686$).
    }
    \label{fig:fuv_crisis}
\end{figure*}

\begin{figure}
    \centering
    \includegraphics[width=\columnwidth]{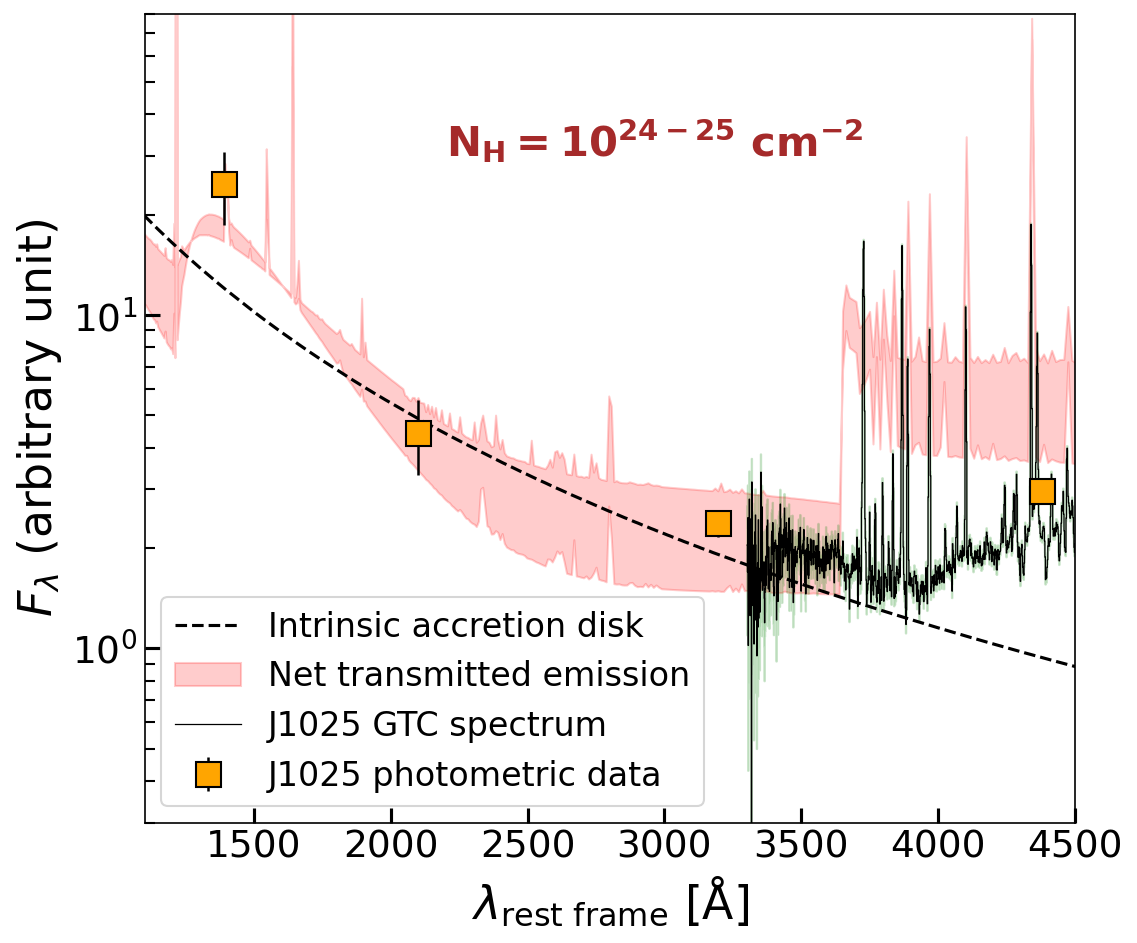}
    \caption{
    Photoionization model of a gas-enshrouded accreting black hole.
    The shaded red region encloses models predicted by \cloudy where the AGN accretion disc (dashed line) is obscured by a spherically symmetric gas shell with a covering faction of unity and a column density in a range of $N_{\rm H}=10^{24-25}~{\rm cm^{-2}}$.
    Both the intrinsic disc emission and the net transmitted emission are normalized by the same scaling factor.
    The net transmitted emission is steepened towards the FUV due to a combined effect of continuum absorption and reemission.
    While the steepened FUV could provide a fit to the photometric data of \target, the optical Balmer break is not seen in \target, suggesting the AGN component cannot dominate the full UV-to-optical regime.
    }
    \label{fig:fuv_reflec}
\end{figure}

The UV continuum of \target is currently only sampled by three photometric points, which are \galex FUV, NUV, and SDSS $u$.
The \galex UV fluxes we queried from \textsc{VizieR} are taken from \galex All Sky Imaging Survey (AIS) within the DR5 \citep{Bianchi_2011}.
A noticeable feature of the UV continuum is a steepening towards the \textit{GALEX} FUV band.
The NUV-u band slope combining the \galex and SDSS photometry is \redtxt{$\beta _{\rm NUV-u}=-1.5\pm 0.6$}, which is comparable to the UV slopes of reddened quasars \citep{vandenberk2001} and those of high-$z$ LRDs \citep{hainline_lrd_2024}.
However, the FUV-NUV slope of \target purely based on \galex is \redtxt{$\beta _{\rm FUV-NUV}=-4.1\pm 0.9$}, which is surprisingly steep for a standard and unattenuated AGN accretion disc \citep[$\beta_{\rm UV}=-2.33$,][]{ss73} or typical stellar populations.
Recent observations of early galaxies by \jwst have revealed that extreme UV slopes reaching $\beta=-2.6$\,-\,$-3.0$ exist at early times \citep{Topping_2022,Cullen_2024,Morales_2024,Baker_blueubz8d5_2025}, which requires very young stellar populations ($t_{\rm age}<10$ Myr) and they must be nearly dust-free \citep{Ferrara_2024}.
Still, these high-$z$ ``Blue Monsters'' are marginally shallower in the UV compared to \target and the much shallower NUV-u band region cannot be explained simultaneously.
Therefore, we discuss some other possible scenarios below to steepen the FUV slope.

In Figure~\ref{fig:fuv_crisis}, we plot the UV part of \target as a best-fit broken-power law.
Compared to the optical part, the FUV is not only extremely blue but also bright.
One possibility to make a shallow NUV-u slope and a steep FUV-NUV slope is to have a nebular dominated UV spectrum, where the FUV region is dominated by the two-photon continuum (plus nebular emission such as Ly$\alpha$) and the region redwards to the NUV is flattened by a Balmer continuum \citep[see e.g.,][]{cameron_9422}.
However, this scenario is easily ruled out, as shown in the left panel of Figure~\ref{fig:fuv_crisis}.
The strength of the Balmer continuum as well as the two-photon continuum should be closely tied to the strength of the Balmer lines \citep{peimbert1967,bottorff2006}.
Based on the measured strength of Balmer lines and assuming $T_{\rm e}=10^4$ K and $n_{\rm e}=10^3~{\rm cm^{-3}}$, we used \pyneb to compute the maximum nebular continuum in the UV with no dust attenuation and plotted it as the solid cyan line.
The nebular continuum is extremely weak and cannot explain the UV photometric points of \target.

As another possibility, the FUV part could be boosted by nebular emission lines, especially if there are lines excited by AGN.
In the right panel of Figure~\ref{fig:fuv_crisis}, we checked such a possibility based on the composite quasar spectrum of \citet{vandenberk2001} constructed using SDSS quasars.
The composite quasar spectrum has $\beta _{\rm UV}=-1.56$, which is marginally steeper than the NUV-u band slope of \target.
We rescaled the composite quasar spectrum with the best-fit normalization factor derived from a minimum $\chi^2$ method by matching the NUV- and u-band fluxes to those of \target.
With this normalization, the FUV flux is significantly lower than the FUV flux of \target.
We then upscaled the nebular emission by subtracting a power-law continuum with $\beta _{\rm UV}=-1.56$ and multiplied the residual flux by a constant factor, and re-added the flux to the power-law continuum.
We then calculated the synthetic \galex\ FUV band flux of the modified quasar spectrum.
We found that the nebular emission of the composite quasar spectrum needs to be upscaled by a factor of $11\pm 4$ to match the observed FUV band flux of \target.
The resulting spectrum within the FUV band is plotted as the dashed magenta line, where strong enhancement in Ly$\alpha$, Si\,{\sc iv}, \civ, and \heii\ can be clearly seen.

While the extremely strong nebular emission of the above lines can indeed boost the FUV flux, the required emission line strengths might become unphysical.
For example, the flux ratio of \heii$\lambda 1640$/\heii$\lambda 4686$ is roughly a constant of 7\,-\,8 over a wide range of physical conditions, and thus upscaling the UV \heii\ by a factor of 11 would lead to unphysical flux relative to the optical \heii$\lambda 4686$, especially given that the observed flux of \heii$\lambda 4686$ in \target is lower than that of typical quasars (see Figure~\ref{fig:bpt}).
Still, one could argue that \heii\ is not the dominant line within the FUV band.
As another example, the Case B value for Ly$\alpha$/\hb is roughly 26 at $T_{\rm e}=10^4$ K and $n_{\rm e}=10^3~{\rm cm^{-3}}$.
Comparing the requiring Ly$\alpha$ flux based on the composite quasar template with that of the \textit{total} \hb flux measured from the GTC spectrum, we have $\rm Ly\alpha/H\beta=83\pm 30$, significantly beyond the Case B value.
A potential mechanism to reach such high Ly$\alpha$/\hb is through collisional excitation as shown by \citet{Ferland_1985}.
In such a case, the FUV nebular emission is likely dominated by a BLR with strong collisionally excited lines, and the high broad-line \ha/\hb ratio is caused by strong collisional excitation rather than dust attenuation.
The implied little dust attenuation of the BLR as well as the accretion disc would support the scenario where the UV light is dominated by an AGN accretion disc, similar to the assumption made by \citet{burke_abs_2021}.

As a third possibility, we checked the recently proposed model of \citet{Inayoshi_maiolino_2025}, the intrinsic UV spectrum of LRDs could be heavily obscured by dense dust-free neutral gas that produces strong hydrogen continuum absorption.
The dense gas ``atmosphere'' might also produce the X-ray weakness of LRDs through Compton scattering \citep{Maiolino2024_Xrays}.
We considered a simple model where a standard \citet{ss73} accretion disc is fully covered by a spherical shell with a covering factor of unity using \cloudy \citep{ferland2017}.
In this simple practice, we fixed the ionization parameter to $\log (U)=-2$ and the density to $n_{\rm H}=10^{10}~{\rm cm^{-3}}$, and considered a range of column density of $N_{\rm H}=10^{24-25}~{\rm cm^{-2}}$.
The final results are plotted in Figure~\ref{fig:fuv_reflec} and compared with the observations.
Due to the continuous hydrogen opacity in the UV and the reemission from the spherical shell \citep{Inayoshi_maiolino_2025}, the FUV slope is steepened with respect to the intrinsic slope, and the NUV-u band slope is instead flattened.
The resulting shape of the UV spectrum can roughly match the observed photometric points, but the model predicts a strong Balmer break that is not observed in the optical, {which is produced by the large Balmer continuum opacity in the gas shell}.
While adding an additional continuum component including a Balmer continuum emission might alleviate the optical tension, this simple exercise suggests that the gas-obscured AGN model cannot dominate the full UV-to-optical regime.

In summary, while the steep FUV slope of \target might be caused by extremely strong nebular emission from a BLR or a heavily gas-obscured accretion disc, none of these scenarios are standard interpretations for local AGN.
Spectroscopic observations of the UV with, for example, \textit{HST}, will be the key to understanding the origin of the UV light.
Next, we discuss the red end of the SED of \target, where an ``MIR-crisis'' occurs in high-$z$ LRDs.

\subsection{Low MIR emission}
\label{subsec:cicada}

\begin{figure*}
    \centering
    \includegraphics[width=\columnwidth]{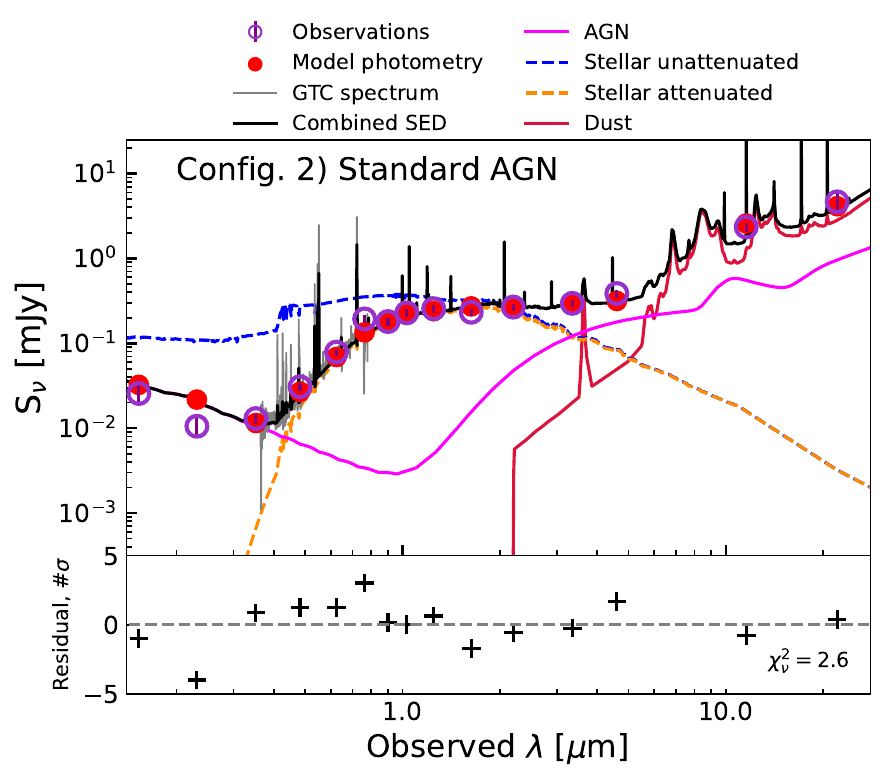}
    \includegraphics[width=\columnwidth]{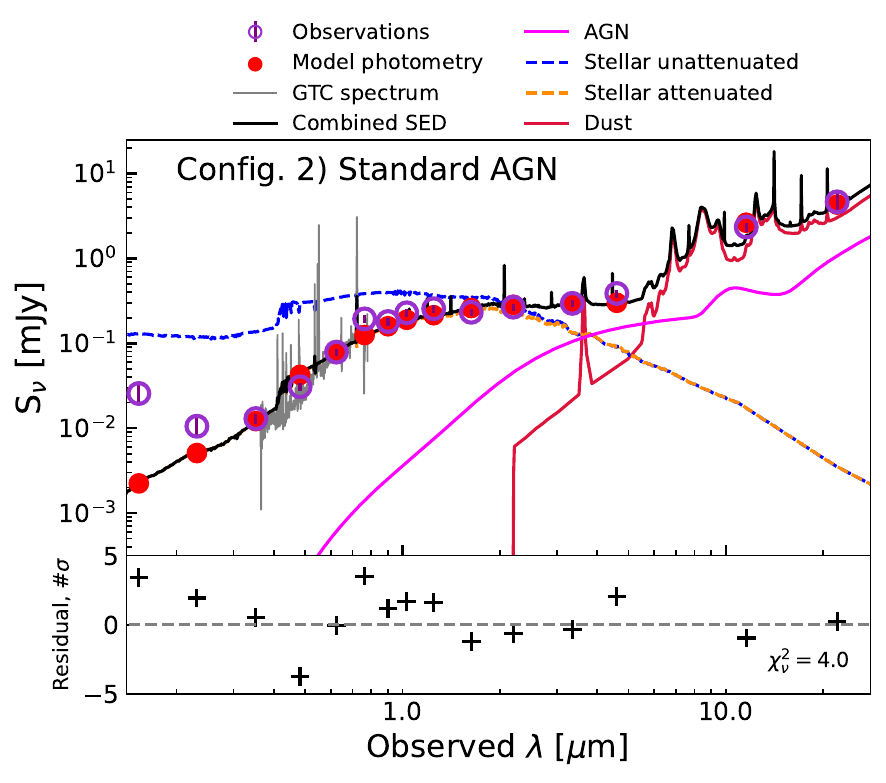}
    \caption{
    Best-fit \textsc{cigale} SED models for \target from FUV to MIR, where an AGN component and a stellar component are included.
    We plot both observed photometric points and the GTC spectrum and compare them with model SEDs.
    \textit{Left:} the UV continuum is fitted by an intrinsically steep AGN SED.
    \textit{Right:} no steepening in the AGN SED is assumed.
    In both cases, the NIR of the best-fit model around $\rm 5~\mu m$ is dominated by hot dust emission from the torus whereas the MIR emission is mainly from dust heated by stellar populations.
    The UV continuum is poorly fitted even with an unphysically steep AGN continuum.
    Also the stellar attenuation is significantly higher than that inferred from the Balmer decrement.
    In configuration 2, the Balmer break is too strong and not observed in the GTC spectrum.
    While the MIR crisis seems not as severe in \target as in other high-$z$ LRDs, the best-fit model is not satisfying due to the above mismatches.
    }
    \label{fig:cigale_fit0}
\end{figure*}

As discussed recently by \citet{williams_lrd_2024,akins_lrd_2024,Setton_lrddust_2025,Chenli_2025}, there is a lack of hot and cold dust emission in the NIR-MIR regime of high-$z$ LRDs, suggesting that the optical continuum of LRDs could be intrinsically red.
As shown in Figure~\ref{fig:full_sed}, the NIR spectrum of \target is similar to those of high-$z$ LRDs until $5~\mu {\rm m}$ in the rest frame.
Redwards of $5~\mu {\rm m}$, \target has the strongest constraint on the fluxes for wavelengths up to about $20~\mu {\rm m}$.
This sets a strong constraint on the allowable dust reemission if one assumes energy balance.

To characterize the NIR-MIR emission, we used the SED fitting code \textsc{cigale} \citep{cigale} that has built-in energy-balance calculations.
We note that a similar fit has been done by \citet{burke_abs_2021} using \textsc{cigale}, who found that the dust is mainly heated by stellar populations and the AGN is subdominant over the optical-MIR regime and only visible in the UV.
We reexamined the fit by assuming different AGN contributions in the UV.
For the fitting, we include photometric data from \galex FUV to \textit{WISE} MIR, and we include an additional uncertainty of 10\% as model systematics.
We modelled the AGN component using the SKIRTOR model \citep{skirtor}, where the dust attenuation curve is assumed to be an SMC curve \citep{gordon2003}.
We considered two configurations for the intrinsic AGN component.
In the first configuration, we allowed the AGN accretion disc to have a steeper slope compared to the default \citet{Schartmann_2005} model, which is achieved by adding a power-law index down to $-2$.
In the second configuration, we only allowed the default disc to be steepened by at most a power-law index of $-0.36$.
The stellar population component uses stellar population synthesis model of \citet{bc03} with a \citet{Salpeter1955} initial mass function (IMF) and a delayed-$\tau$ star formation history.
The stellar dust re-emission is modelled using the \citet{Dale_2014} model with a variable slope of $\alpha=1.3-4$.
The stellar attenuation curve is assumed to follow that of \citet{calzetti2000} and the ratio between the stellar-  and nebular- dust attenuation is set to $\rm E(B-V)_\star /E(B-V)_{\rm gas}=0.44$.

Figure~\ref{fig:cigale_fit0} shows the best-fit models under the two configurations.
The first configuration (left panel of Figure~\ref{fig:cigale_fit0}) produces a fit similar to that done by \citet{burke_abs_2021}, where the whole UV is dominated by the AGN continuum.
The AGN produces dust emission that is stronger than the stellar dust emission in the NIR around $\rm 5~\mu m$ but becomes subdominant in the MIR.
The fitting result for the NIR part is similar to the results presented by \citet{juodzbalis_rosetta_2024} for the Rosetta Stone, although in that case there is no longer wavelength coverage to determine the dominant contributor for the MIR emission.
The fitting results suggest the NIR-to-MIR emission can be explained by a combination of AGN and stellar dust emission, if the optical continuum is star dominated.
The best-fit stellar mass is $M_{\star}=10^{10}~M_\odot$ dominated by a 2-Gyr old evolved stellar population, consistent with the result of \citet{burke_abs_2021}.
However, there are several caveats associated with this interpretation.
First, even with a significantly steepened UV, the FUV-NUV part is not well fitted.
As discussed in Section~\ref{subsec:fuv_crisis}, additional physical scenarios such as extreme nebular emission or gas-obscured accretion disc might need to be considered, which is not included in the SED modelling.
This makes the calculation of the AGN dust re-emission uncertain.
Second, the best-fit dust attenuation is $A_{\rm V}\approx2$ mag, significantly higher than the nebular attenuation derived from the Balmer decrement.
Indeed, it is quite atypical for evolved stellar populations to be more dust attenuated compared to the nebular emission \citep{Calzetti_1994,Charlot_2000,Wild_2011,liniu_2021}.
The second configuration, as shown in the right panel of Figure~\ref{fig:cigale_fit0}, produces a worse fit due to the significantly underpredicted FUV and NUV.
The best-fit stellar mass reaches a very high value of $M_{\star}=10^{11}~M_\odot$ and it is still dominated by a 2-Gyr old evolved stellar population.
Compared to the GTC spectrum, the Balmer break predicted by the model is too strong.
In addition, this model has a problem with the broad-line EW.
The AGN continuum in this model is highly reddened and contributes to less than 1\% of the light near \ha, which poses a great challenge to the energy balance for the broad line.
Even considering that the covering factor of the BLR in \target can reach unity, which would lead to a factor of 2\,-\,10 increase in the broad \ha EWs compared to the normal AGN population with covering factors of 10\,-\,50\% \citep{netzer_1990,ferland2020}, the intrinsic broad \ha\ EW of $>30,000$ \AA\ inferred from this best-fit \textsc{cigale} model is still hard to reconcile with the observed AGN and quasars \citep{lusso_sdssagn_2020,Maiolino2024_Xrays}.

Recently, \citet{Liu_speddlrd_2025} proposed an alternative gas-obscured model for LRDs, where the accretion disc is embedded in turbulent accretion flows.
This model predicts a cool optical continuum resembling a G-to-K star atmosphere with $T_{\rm eff}\sim 5000$ K that dominates the optical continuum.
While similar to the gas-obscured models of \citet{Inayoshi_maiolino_2025,ji_lrdbreak_2025} in terms of producing the optical Balmer break, the model of \citet{Liu_speddlrd_2025} requires little dust attenuation and thus would presumably produce low NIR-MIR dust re-emission from the AGN torus.
\citet{linxiaojing_locallrd_2025} have presented a comparison between the turbulent accretion flow model and the optical spectrum of \target and found a good match in the overall shape.
A full investigation of the alternative AGN continuum model will benefit from a careful comparison between identified absorption lines in the observed spectrum and those predicted by models, which we will investigate in future work.
On the other hand, a stellar-dominated model also resolves the MIR crisis but would face the same issue of reproducing the optical absorptions.

\section{Discussion}
\label{sec:discuss}

Thus far, we have compared the observations of \target and high-$z$ LRDs and shown their similarities.
Also, we have shown the new information extracted from the SED of \target.
In this section, we discuss additional implications for the general LRD population based on \target.

\subsection{Stellar mass of the host galaxy}


A debating question of high-$z$ LRDs is the contribution from their host galaxies to the observed light, since some recent studies have suggested their entire optical spectra can be black-hole dominated \citep{ji_lrdbreak_2025,degraaff_lrd_2025,naidu_lrd_2025}.
While the presence of metal emission lines certainly requires stellar evolution and enrichment, there are also weak-line LRDs with nearly chemically pristine environments \citep{Maiolino_metalpoorlrd_2025}.
\target is clearly more chemically evolved compared to the weak-line LRDs and actually overmassive compared to the MZR given its nominal stellar mass of $M_\star\approx10^{10}~\Msun$.

A useful constraint on the host galaxy contribution is from the dynamical mass, which requires measurements of both the galaxy size and the stellar or gas kinematics.
In Appendix~\ref{appendix:gemini_reduction}, we show that the Gemini slit image of \target indicates its optical light is unresolved and has an effective radius of $R_{\rm e}<620$ pc.
Regarding the kinematics,
{using the high-resolution ($R\approx 3600$) spectrum from MMT/Binospec, \citet{linxiaojing_locallrd_2025} found that the narrow emission lines of \target remain unresolved and their best-fit emission-line models have a small velocity dispersion of $\sigma _{\rm n}=17\pm 1$ \kms.
}
From the high-resolution Gemini/GMOS spectrum ($R\approx 3400$), however, we measured a higher narrow-line velocity dispersion of {$\sigma _{\rm n}=39.1\pm 0.4$ \kms}, where the lines are marginally resolved.
The difference potentially arises from the decomposition of \ha.
The narrow-line \ha\ luminosity we measured from the Gemini spectrum is $L_{\rm H\alpha,~narrow}=(2.37\pm 0.02)\times 10^{41}$ \ergs without dust attenuation correction, which is roughly two times the value measured by \citet{linxiaojing_locallrd_2025}.
During the spectral decomposition, the lower the relative flux of the narrow \ha, the narrower the line profile is fitted since the narrow component is pushed to the core of the observed profile.
Alternatively, if the resolution of the Gemini spectrum is actually $R\approx 2500$, we can recover \citet{linxiaojing_locallrd_2025}'s velocity dispersion.
But in this case, the resolution would be inconsistent with our interpretation based on the slit image as well as our measurement of the sky line width.
Regardless, both velocity dispersions are low and we would have a stronger limit on the dynamical mass if we adopt \citet{linxiaojing_locallrd_2025}'s measurement.

To estimate the dynamical mass in an unresolved system such as \target, we start with the virial theorem following \citet{Binney_tremanine_1987}
\begin{equation}
    M_{\rm dyn}= 3\left(\frac{R_{\rm g}}{R_{\rm e}}\right)\frac{\sigma^2_\star R_{\rm e}}{G}\equiv\eta \frac{\sigma^2_\star R_{\rm e}}{G},
\end{equation}
where $\sigma _\star$ is the LOS stellar velocity dispersion, $R_{\rm e}$ is the effective (half-light) radius, and $R_{\rm g}$ is the gravitational radius.
To take into account that the gas velocity dispersion could be lower than the stellar velocity dispersion, we used the relation of $\sigma_\star \approx \sigma _{\rm n}\times10^{0.18}$ \citep{Bezanson18b}.
The effective radius is set by the Gemini slit image, $R_\mathrm{e}<620$ pc.
The remaining part is the coefficient, $\eta$, which is taken to be $3/0.45\approx6.7$ by \citet{Binney_tremanine_1987} as an average over \citet{King_1966} models.
From observations, \citet{Cappellari06} found $\eta \approx 5$ by fitting local early-type galaxies, similar to the case of a uniform density sphere \citep{MacLaren_1988}, and the same coefficient has been adopted for high-$z$ galaxies (\citealp{Pettini_2001,Shapley_2004,erb_2006}, see also \citealp{degraaff_lsf_2024,ubler2023a} who adopted smaller $\eta$ for resolved sources).
Finally, for dispersion-dominated systems, one can have $\eta\approx 3.4$ \citep{Diaz-santos_2021}.
With the most conservative $\eta=6.7$, we have $M_{\rm dyn} < 3.3\times10^9~\Msun$ using our measured $\sigma _{\rm n}=39.1\pm 0.4$ \kms, or $M_{\rm dyn} < 6.3\times10^8~\Msun$ using $\sigma _{\rm n}=17\pm 1$ \kms\ measured by \citet{linxiaojing_locallrd_2025}.

The most conservative upper limit of the dynamical mass is $2.9\sigma$ lower than the nominal stellar mass estimated from the \textsc{cigale} fitting result by \citet{burke_abs_2021}, which is $M_\star=7.9^{+2.1}_{-1.6}\times10^{9}~M_\odot$, and significantly lower compared to our fitted range of $M_\star=10^{10-11}~M_\odot$.
A stellar mass higher than the dynamical mass is unphysical as the latter also includes the gas mass and the dark matter mass in addition to the stellar mass.
The same mass discrepancy has been pointed out for several high-$z$ LRDs \citep{juodzbalis_rosetta_2024,wangbingjie_2024,ji_lrdbreak_2025,deugenio_qso1_2025,Akins_ci_2025}, which serve as a key evidence that the optical continuum cannot be dominated by stellar populations alone, or at least the typical evolved stellar population fitted by standard SED modelling codes.

{Notably, \citet{juodzbalis_jadesagn_2025} recently found that in a sample of \jwst-selected broad-line AGN at $2<z<7$, there is a relation between the narrow \hb luminosities and narrow-line velocity dispersion consistent with the trend found in local giant \hii\ regions as well as high-$z$ compact SF galaxies \citep[e.g.,][]{Terlevich_1981,Melnick_2017,chavez_2014,chavez_2025}, where both $L_{\rm H\beta}$ and $\sigma _{\rm n}$ are connected to the total stellar mass for virialized ionized gas.
For \target, it has a dust-attenuation corrected \hb luminosity of $\log L_{\rm H\beta}/[{\rm erg~s^{-1}}]=41.01\pm 0.01$ and a velocity dispersion of $\sigma _{\rm n}=39.1\pm 0.4$ \kms, making it consistent with the $L_{\rm H\beta}$\,-\,$\sigma _{\rm n}$ relation derived by \citet{chavez_2025} within $1\sigma$ scatter.
This implies that the NLR of \target could be virialized and follow a scaling relation similar to that of young star clusters.
}
To further understand the actual host galaxy contribution, one needs to measure the spatial extensions of the observed light at different wavelengths, which is not available from current observations but will be key for future follow-up observations.


\subsection{Total energy budget}

Since there is a good coverage of the panchromatic SED of \target from photometric data, one can examine the typical assumptions made for energy conversions across different wavelengths.
A relevant parameter is the intrinsic bolometric luminosity of the AGN.
Based on the broad-line Balmer decrement, the inferred bolometric luminosity reaches a high value of $L_{\rm bol}\sim 10^{45}$ \ergs\ given \redtxt{$A_{\rm V}=5.1\pm 0.1$} mag.
We performed a simple consistency check by interpolating the FUV-to-MIR photometric SED as broken power laws using the \textsc{interp1d} function from the \textsc{scipy} package and obtained an integrated luminosity of $L_{\rm FUV-MIR}\approx 6.2\times 10^{43}$ \ergs.
This integrated luminosity is significantly lower than the nominal bolometric luminosity with dust attenuation corrections.
Given the non-detection in X-ray, it appears implausible that over 98\% of the energy is found at $\lambda <\lambda _{\rm Ly\alpha}=1216$ \AA.
In fact, based on the empirical AGN SEDs constructed by \citet{Jin_2012,Jin_2017} covering sub-Eddington to super-Eddington sources at low redshift, the fraction of total energies at $\lambda <1216$ \AA is typically 40\,-\,50\%.
This estimation supports the bolometric conversion with little or no dust attenuation, with the latter giving \redtxt{$L_{\rm bol}\approx 1.0\times 10^{44}$} \ergs.
If \target is a representative case for LRDs, which is supported by the highly similar panchromatic photometric SEDs of \target and high-$z$ LRDs shown in Figure~\ref{fig:full_sed}, then it suggests the BLRs of LRDs are not heavily dust obscured and the bolometric luminosities are best estimated from the direct measurements of broad-line luminosities.
The little dust obscuration further suggests that if the UV light of LRDs is not from the AGN continuum, strong opacity, possibly from neutral hydrogen, must exist to extinguish the UV.

If the BLR is not dust obscured, the high broad-line Balmer decrement of $\rm H\alpha/H\beta>20$ in \target as well as high-$z$ LRDs \citep{juodzbalis_rosetta_2024} could be a result of collisional excitation and radiative transfer effects.
In the case of the collisional excitation, the broad-line $\rm H\alpha/H\beta$ is enhanced simply because it is easier to collisionally excite hydrogen to $n=3$ than to $n=4$.
Based on this argument, qualitatively, one expects Paschen decrements are more consistent with Case B since they are produced from $n>3$ levels, which is indeed found by \citet{linxiaojing_locallrd_2025} based on the Magellan/FIRE spectrum in the NIR of \target.
In the case of radiative transfer effects, it is possible that \hb\ is more optically thick compared to \ha\ due to obscuration by neutral hydrogen along the LOS.
Since Paschen series likely have lower optical depths, this scenario would again predict less Paschen decrements.
By doing a simple test with \cloudy photoionization models, we found that with a standard accretion disc and a BLR with $N_{\rm H}=10^{23}~{\rm cm^{-2}}$, $n_{\rm H}=10^9~{\rm cm^{-3}}$, and $\log U=-2$, the predicted $\rm H\alpha/H\beta$ is 8.4 and 6.9 for calculations with and without collisional excitation, suggesting both effects should be considered. The predicted $\rm H\alpha/H\beta$ increases to 10 for a higher column density of $N_{\rm H}=10^{24}~{\rm cm^{-2}}$, which matches the observed broad-line Balmer decrement of the Rosetta Stone if it is completely dustless \citep{juodzbalis_rosetta_2024}.

As another possibility, the AGN in \target could be heavily dust obscured but the bolometric conversion factor based on the broad \ha\ from \citet{sternlaor_2012} (which is 130) is significantly overestimated.
This could happen if the BLR has a high covering fraction, but only accounts for a factor of 2\,-\,10 if we consider the typical covering fraction in local AGN where the bolometric conversion is derived is 10\,-\,50\% \citep{ferland2020}.
To further lower the estimated $L_{\rm bol}$ and make it compatible with the integrated photometric luminosity, one needs to increase the fraction of photons near the hydrogen ionizing edge at $\lambda =912$ \AA\ (where the photoionization cross-section is highest for hydrogen) in the SED of the accretion disc seen by the BLR.
Such a condition might be met if the temperature cut-off set by the inner disc radius corresponds to an energy close to 1 Ryd, which is $T_{\rm cut-off}\approx 1.6\times 10^5$ K.
This can be achieved for a black hole mass of $M_{\rm BH}\sim 10^7~\Msun$ based on simple disc SED models from \textsc{xspec} with certain constraints on the black hole spin and the inclination angle.
Since the thermal disc component peaks around the hydrogen ionizing edge, there is little energy beyond the ionizing edge of $\rm He^+$ (i.e., 54.4 eV), which naturally leads to relatively weak high-ionization lines.

In reality, all of the above effects might be at play in modifying the observed Balmer decrements in LRDs.
Deciphering the physical conditions that produce the observed line ratios are key to recovering the correct black hole parameters.
We will present a more comprehensive investigation of the broad-line ratios observed in LRDs and high-$z$ AGN in future work.

\section{Conclusions}
\label{sec:conclude}

In this work, we present an independent investigation on the brightest and lowest redshift \redtxt{LRD} discovered to date, \target, at $z=0.1$, by exploring its \redtxt{SED} from rest-frame X-ray energies to MIR.
We summarize our conclusions in the following.

\begin{itemize}
    \item The UV-optical SED of \target has a characteristic V shape and the optical continuum peaks around \ha similar to high-$z$ LRDs, which implies a cool atmosphere of $T_{\rm eff}\sim 5000$ K in the optical.
    Unlike some high-$z$ LRDs with strong Balmer breaks, \target does not show a clear Balmer break and the turnover point of the V-shaped spectrum is visually found at the location of \caii\ K absorption at 3940 \AA.
    The case of \target \redtxt{suggests} many photometric LRDs at high-$z$ might not have a strong Balmer break despite the fact that the SED turnover point is close to the Balmer limit.
    \item \target shows clear broad Balmer emission lines with a best-fit broad-line width of $\rm 
    FWHM_{H\alpha}=934\pm 10$ \kms, which imply an accreting black hole with a mass of \redtxt{$\mbh = 10^{6.49-7.22}~M_\odot$} and an Eddington ratio of \redtxt{$\lambda _{\rm Edd}=0.26-1.4$}, where the exact values depend on whether the BLR emission is heavily dust attenuated or not.
    The black hole parameters are similar to the broad-line AGN selected by \jwst at high redshift.
    In addition, the metal-poor [$\rm 12+\log(O/H)=7.73^{+0.21}_{-0.14}$] and compact ($R_{\rm e}<620$ pc) nature of \target makes it a close counterpart of high-$z$ broad-line AGN selected by \jwst.
    \item Archival Gemini/GMOS spectrum shows strong Balmer absorption and NaD absorption in the spectrum of \target. By fitting these features, we found \ha and part of the NaD absorption likely originate in slowly outflowing gas rather than the stellar atmospheres.
    However, \hb exhibits a redshifted absorption in the SDSS spectrum, suggesting another inflow component with a lower optical depth where \hb originates.
    \item We obtained GTC/OSIRIS spectroscopic observation for \target and discovered rich absorption lines and bands in the rest-frame optical. We found typical stellar absorption including \caii\ K, G-band, Mgb, and NaD.
    We also made tentative identifications of a series absorption lines potentially associated with low-ionization ionic or atomic transitions such as \feii, Ti\,{\sc ii}, Ba\,{\sc ii}, Na\,{\sc i}, Fe\,{\sc i}, and Ca\,{\sc i}.
    Many of the absorption strengths are underpredicted by empirical stellar templates, suggesting potential contribution from the \redtxt{ISM} or outflows.
    We found an unknown and strong absorption feature at 4570 \AA, which is clearly seen in high-$z$ LRDs as well.
    Among the absorption lines, the G-band absorption is the only one produced by a molecule (CH), which poses a challenge for the current model that describes the optical continuum as gas-enshrouded black hole accretion discs, as stellar atmosphere-like gas with extremely high density of $n_{\rm H}\gtrsim 10^{14}~{\rm cm^{-3}}$ and warm temperature of $\sim 5000$ K is required.
    Still, typical stellar templates do not reproduce the observed set of absorptions as well.
    \item \target \redtxt{exhibits} an emission-line spectrum indicative of a soft ionizing SED.
    While there is detection of \heii$\lambda 4686$, the flux ratio of \heii$\lambda 4686$/\hb is consistent with star-forming galaxies in the local Universe.
    There are also a series of weak lines associated with the forbidden and permitted transitions of \feii. The density diagnostic based on \feii\ suggests a high density of $n_{\rm H}=10^{5-7}~{\rm cm^{-3}}$. {The different kinematics of [\feii]$\lambda 7155$ and [\caii]$\lambda 7291$ compared to other optical narrow lines revealed by the Gemini/GMOS spectrum further suggest they originate from a spatially distinct region.}
    We identified several [\feii] lines in an LRD at $z=2.26$ that match the lines seen in \target, suggesting a high gas density of $n_{\rm H}\gtrsim 10^{6}~{\rm cm^{-3}}$ in this source as well.
    \item By analyzing the change in the equivalent widths of \ha\ from spectroscopic observations over 19 years in the rest frame of \target, we found a small variation of 9\% but it is very significant ($>5\sigma$). This is consistent with the previous finding by \citet{burke_abs_2021} and suggest \target hosts an accreting black hole with low variability.
    \item \target exhibits an extremely steep FUV slope of \redtxt{$\beta =-4.1\pm 0.9$} from archival \galex data, atypical for AGN accretion disc or stellar populations.
    The shape can be roughly reproduced if the FUV has extreme nebular emission, potentially requiring strong collisional excitation.
    Alternatively, a gas obscured standard AGN accretion disc can produce a steepened FUV spectrum and a flattened NUV spectrum, but the optical part of disc would have a strong Balmer break that is not seen in the observed spectrum.
    \item With archival \chandra and new \nustar observations, we found a high 90\% confidence limit for the bolometric-to-X-ray luminosity ratio of $L_{\rm bol}/L_{\rm 2-10~keV}>336$ assuming no gas obscuration, meaning \target is extremely X-ray weak even compared to high-$z$ broad-line AGN selected by \jwst.
    If the X-ray weakness of \target is produced by Compton scattering of gas obscuring the line-of-sight, the non detection in $\rm 10-30~keV$ from the \nustar observation indicates a large gas column density of $\log[N_{\rm H}/({\rm cm^{-2}})]=24.1-24.6$.
    In this scenario, the same obscuring gas could be responsible for producing the absorption lines seen in the optical as well as the optical continuum shape.
    \item The \redtxt{MIR} data of \target set by far the most stringent constraint on the dust re-emission in LRDs.
    Using the SED fitting code \textsc{cigale} that assumes energy balance, we found that the MIR can be well fitted with models where the optical is dominated by dust-obscured and evolved 2-Gyr old stellar population and the \redtxt{NIR} is partly contributed by emission from the dusty torus of AGN.
    However, the models either require an unphysically steep UV slope for the AGN accretion disc, or fail completely in the UV and overpredict the strength of the Balmer break.
    The models also require the stellar populations to be more dust reddened by about 1.4 mag in the V band compared to the nebular emission, which is atypical for evolved populations.
    \item While the nominal stellar mass obtained through SED fitting is $M_\star=10^{10-11}~M_\odot$, the actual stellar mass in \target is likely lower. This is because based on the size limit and the width of narrow emission lines, the most conservative limit we derived for the dynamical mass is {$M_{\rm dyn}<3.3\times10^{9}~M_\odot$}, which is significantly lower than the nominal stellar masses. This implies the optical continuum is unlikely to be dominated by stellar populations alone, or at least a typical, evolved stellar population.
    \item The integrated photometric luminosity from FUV to MIR of \target is two orders of magnitude lower than the dust attenuation corrected bolometric luminosity inferred from broad Balmer lines.
    This implies a combined effect of an intrinsically high Balmer decrement due to an optically thick BLR and collisional excitation, and a non-standard bolometric conversion modified by a high covering fraction and a soft ionizing SED.
\end{itemize}

Our new GTC and \nustar data confirm \target as a local LRD, and all evidence suggests that it has a complex gaseous environment. The strong ionic, atomic, and molecular absorptions revealed by the new data are difficult to explain with typical stellar and AGN models.
As a concluding remark, the case of \target opens an interesting new window where we can study previously identified sources in the local Universe within the new context of \jwst discoveries.
Despite the fact that its physical nature still remains unclear, the low redshift and high brightness of \target would allow for efficient follow ups that constrain the full SED of this LRD in greater detail and it can serve as a template for the search of other LRD candidates in the local Universe.

\section*{Acknowledgements}
We thank our referee, Mar Mezcua, whose thoughtful suggestions improved the clarity of this work.
This work is based on observations made with the GTC telescope, in the Spanish Observatorio del Roque de los Muchachos of the Instituto de Astrofísica de Canarias, under Director’s Discretionary Time. 
We also thank Fiona Harrison for approving the \nustar\ Director’s Discretionary Time.
We acknowledge invaluable help from Antonio Luis Cabrera Lavers in designing the GTC/OSIRIS observations, and we thank the GTC staff for their immediate response once the Director’s Discretionary Time was approved.
We thank Stefano Carniani, Benjamin D. Johnson, and Max Pettini for helpful discussions.
XJ, FDE, IJ, RM, and JS acknowledge ERC Advanced Grant 695671 ``QUENCH'' and support by the Science and Technology Facilities Council (STFC) and by the UKRI Frontier Research grant RISEandFALL.
RM acknowledges funding from a research professorship from the Royal Society.
CRA acknowledges support from the Agencia Estatal de Investigaci\'on of the Ministerio de Ciencia, Innovaci\'on y Universidades (MCIU/AEI) under the grant ``Tracking active galactic nuclei feedback from parsec to kiloparsec scales'', with reference PID2022$-$141105NB$-$I00 and the European Regional Development Fund (ERDF).

\redtxt{This work used observations made with the NASA/ESA/CSA James Webb Space Telescope. The data are available at the Mikulski Archive for Space Telescopes (MAST) at the Space Telescope Science Institute, which is operated by the Association of Universities for Research in Astronomy, Inc., under NASA contract NAS 5-03127 for JWST.}

\section*{Data Availability}

All \jwst\ observations used in this paper are available through the MAST portal.
The Gemini observations are available at the 
\href{https://archive.gemini.edu/searchform/publication=2021MNRAS.504..543B}{Gemini Observatory Archive}.
\redtxt{We have made our reduced Gemini/GMOS spectrum publicly available through \href{https://doi.org/10.5281/zenodo.17749122}{this link}.}
The GTC observations will be shared on the GTC archive after the 12-month proprietary period.
\redtxt{We have uploaded the reduced GTC/OSIRIS spectrum and made it publicly available through \href{https://doi.org/10.5281/zenodo.17235199}{this link}.}
All analysis results of this paper will be shared on reasonable request to the corresponding author.




%
   \bibliographystyle{mnras} 
   \bibliography{ref} 
%

\appendix

\section{Gemini GMOS data reduction}
\label{appendix:gemini_reduction}

Rest-frame optical spectroscopy from Gemini was obtained as part of
the programme GN-2020A-FT-204 (PI C.~Burke), using the Gemini Multi-Object
Spectrograph \citep[GMOS;][]{hook+2004} in long-slit mode. The observations used the
R831\_G5302 grating, with the central wavelength set to 7230~\AA, covering the nominal
wavelength range 6050--8410~\AA with 0.38-\AA pixels. For a uniformly
illuminated slit of width $0.\!\!''5$, this grating gives a spectral resolution
at 7570~\AA of $R_{0.\!\!''5} = 4396$ (see below for the effective resolution).
The observations of \target were taken on May 30\textsuperscript{th} 2020;
they consist of two integrations of 450~s each, with 1$\times$2 detector binning.

\begin{figure}
\includegraphics[width=\columnwidth]{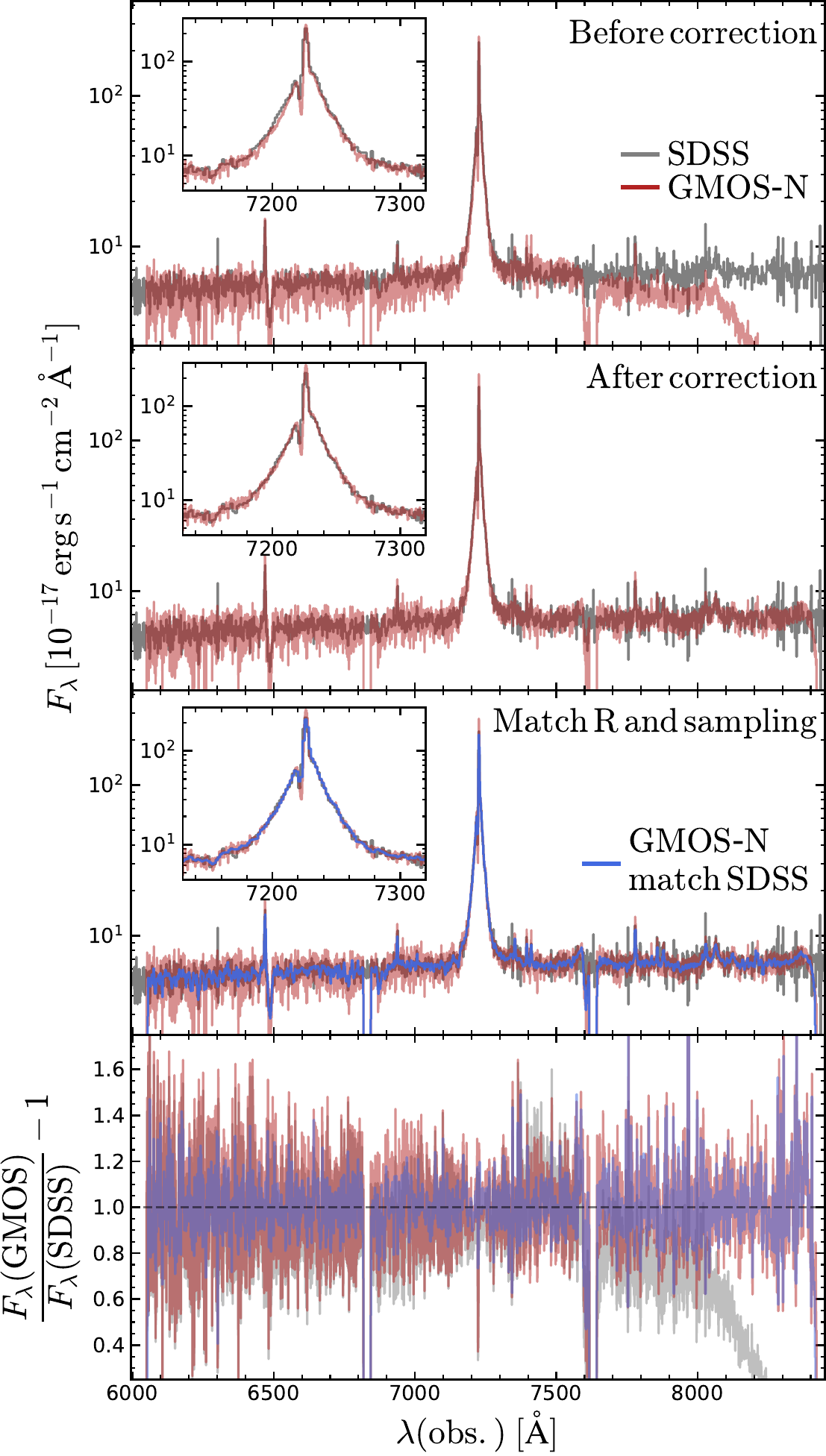}
\caption{
Correction for the Gemini/GMOS spectrum based on the SDSS spectrum. The top two panels highlight the GMOS spectra before and after the correction.
The bottom two panels illustrate the goodness of the correction by convolving the GMOS spectrum to the SDSS spectral resolution, resampling them to the common wavelength grid, and calculating the relative flux difference.
}\label{f.fluxcal}
\end{figure}

For the data reduction, we used the \textsc{dragons} data reduction pipeline
\citep{labrie+2023}. We
performed the standard calibration tasks of subtracting the bias current using
a master bias, and flat fielding the observations using a lamp-flat observations
taken immediately before and after the two science integrations. The wavelength
solution was calibrated using the standard arc-lamp exposures. After completing the data
reduction, we correct the wavelengths from air to vacuum values, and find good agreement
with the SDSS wavelength solution.
We subtracted the sky with the standard routine, which fits the continuum and telluric
lines along the spatial direction of the slit, with a gap at the location of \target.
A modification of the standard algorithm was used to save the sky spectrum as a separate
extension.
For the flux calibration, we initially used the standard star O-type star Feige~66 \citep{feige1958},
from the long-term baseline calibration programme.
Since this star was observed four days before \target, the resulting calibration is rather
uncertain. We apply a further calibration correction by matching the shape of the continuum
between GMOS and SDSS. The resulting calibration is shown in Figure~\ref{f.fluxcal}, where we
compare SDSS (grey line) to the initial flux calibration (red, top panel), and to the
continuum-matched calibration (middle panel). The flux ratio between the GMOS and SDSS data is
shown in the bottom panel. We also show the GMOS data, after matching the spectral resolution
and sampling of SDSS (blue, third panel).
Each calibration step was visually inspected, to assess the quality of, for example, the flat-field
model, the wavelength solution, the extraction and sky apertures, and the spectral trace.

To gauge the seeing, we extract several slit images by taking the median of the slit image
across several windows in wavelength, of width 500 pixels each. We model these images as
a Moffat function plus a line representing the background, and find a FWHM of $0.\!\!''64$, with a dispersion of $0.\!\!''01$, and a maximum variation of 10~percent between
the blue and red ends of the spectrum. For the Moffat exponent, we find a somewhat low value
$\beta = 2.18\pm0.05$. The fit residuals show no evidence for a host galaxy, neither in the
continuum nor around emission lines, hence we conclude that \target is unresolved and that
the Moffat profile traces the seeing of the observations, with $\rm FWHM=0.\!\!''64$.
This implies that \target has half-light radius $R_\mathrm{e} < 0.\!\!''32$, a more 
stringent limit than that provided by SDSS and Legacy Survey imaging.

Given the slit width of $1''$, the spectral resolution for a uniformly
illuminated slit should be $R = 0.5 \cdot R_{0.\!\!''5}$, based on the
grating equation. However, the PSF FWHM is narrower than
the slit width, so the effective spectral resolution is higher than the 
nominal value, and is of order {$R \sim 0.\!\!''5 / \text{FWHM} \cdot R_{0.\!\!''5} \approx 3430$.}
We verify this value by fitting the brightest telluric lines. Since the
PSF is narrower than the slit, the sky lines have a profile shape that reflects both
the slit image (which we assume to be a top-hat function) and the line spread function 
(which we assume to be Gaussian). For this reason, we model each line with a difference of
error functions, using a simple $\chi^2$-minimization technique. The $\sigma$ parameter
of the error functions is $0.86\pm0.09$~\AA, with no clear trend with wavelength
(although we remark that we measure only two emission lines at wavelengths $\lambda < 
7,200~\AA$). With this $\sigma$ value, we obtain an effective resolution at 7570~\AA of
$R = 7570 / (\sqrt{\ln 256} \, \sigma) \approx 3740$, which is within
10~percent of the expected value.

\section{GTC OSIRIS data reduction}
\label{appendix:gtc_reduction}

The GTC observations were carried out during the night of May $\rm 21^{st}$, 2025, divided into two OBs of $3 \times 1200$ s each, using a $10''$ offset to improve sky correction, and we used $2\times 2$ detector binning.
The data reduction of the GTC/OSIRIS spectrum was done using the \textsc{PypeIt} pipeline, which permits a semi--automated processing of spectroscopic data. \textsc{PypeIt} is developed as a Python package based on well-tested algorithms \citep{pypeit:joss_arXiv,pypeit:joss_pub,pypeit:zenodo} and configured to be used in a large list of spectrographs, including OSIRIS. 
The data reduction starts with the standard basic image processing, ie., bias subtraction, trimming, and flat-fielding. Following these steps, cosmic rays are identified and masked. The wavelength calibration is done automatically using an algorithm based on pattern matching the detected lines with the expected ones read from a linelist database. In this case, we use ThAr and Ne lamps. The wavelength calibration is provided in vacuum frame. At this point, object detection is executed in both science and standard star spectral frames. The detected objects are extracted following optimal extraction and at the same time the sky subtraction is performed. The last steps are flux calibration and spectra combination. 
The sensitivity function was determined using a spectrum of the spectrophotometric standard Ross 640 and modelled as a B-spline function. Note that the standard star spectrum was taken with the $2.\!\!''5$ slit, whereas the science spectra were taken using the $0.\!\!''6$ slit. The last step was to coadd the extracted science spectra from the six science exposures, each one of 1200s.
We used the coadded spectra as the fiducial GTC spectrum throughout the manuscript.
In Section~\ref{subsec:var}, we also examined coadded spectra from the first three and the last three science exposures separately, corresponding to two different seeing conditions.

\section{Slit-loss correction for the GTC spectrum}
\label{appendix:slit_loss}

\begin{figure}
    \centering
    \includegraphics[width=\columnwidth]{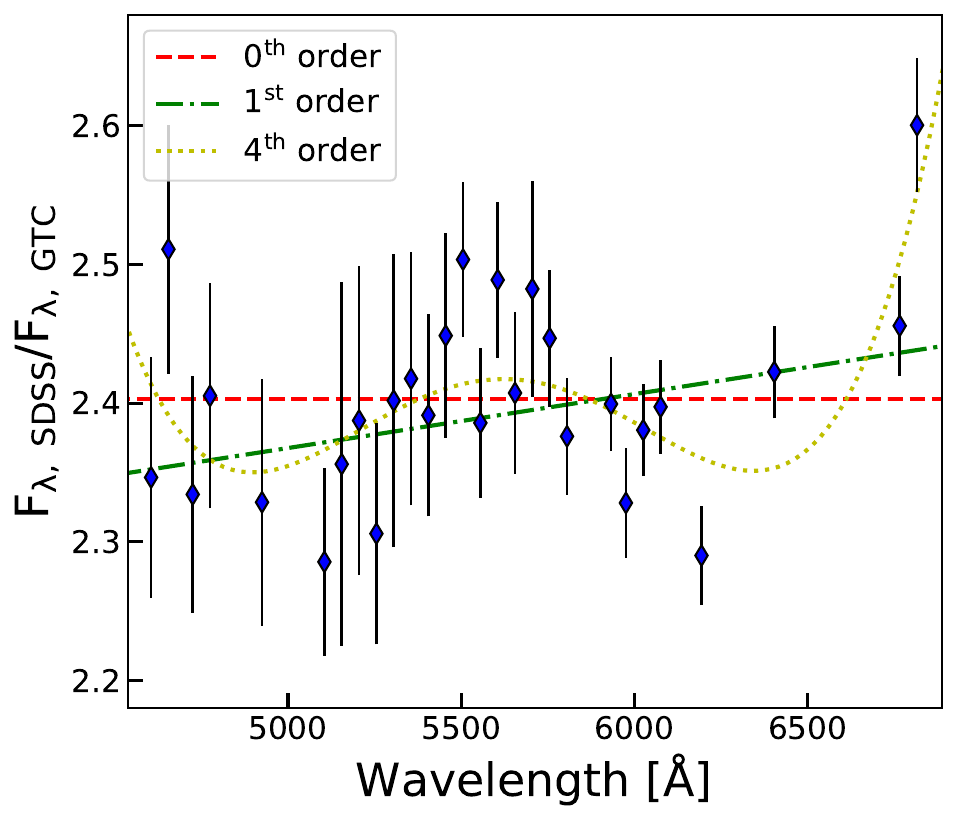}
    \caption{Ratios between the flux densities of the SDSS spectrum and the GTC spectrum in 28 different spectral windows.
    The average values as well as the $1\sigma$ uncertainties in individual spectral windows are shown as blue diamonds with errorbars.
    The best-fit zeroth-, first-, and fourth-order polynomials for the flux density ratio as a function of the wavelength are plotted as dashed red, dash-dotted green, and dotted yellow lines, respectively.
    }
    \label{fig:slit_loss}
\end{figure}

The GTC/OSIRIS spectrum we obtained has clear slit losses that needs to be corrected.
While as we discussed in Appendix~\ref{appendix:gemini_reduction}, the effective radius of the source should be smaller than $0.\!\!''32$, similar to the half width of the slit used for the GTC observations, the seeing of the GTC observations is roughly $0.\!\!''8$ for the first OB and $1.\!\!''2$ for the second OB, which should lead to significant slit losses.

To correct for the slit losses, we used the SDSS spectrum as a reference.
We defined 28 non-overlapping spectral windows with a full width of 50 \AA each, covering $4600~\AA < \lambda < 6800~\AA$ in the rest frame of \target and avoiding strong emission lines including \hb, \oiii, \hei, \ha, and \sii.
For each spectral window, we calculated the average ratio between the flux densities of the SDSS spectrum and the GTC spectrum.
The results are plotted in Figure~\ref{fig:slit_loss}.
The inverse-variance weighted correction factor across the full wavelength range is $2.40\pm 0.01$ as indicated by the dashed red line.
Given the variation of the correction factor across the wavelength range, we also fitted a first-order polynomial and a fourth-order polynomial to describe the correction factor as a function of the wavelength.
Overall, the deviation of these functions from the zeroth-order correction is within 2\% for $\lambda < 6700~\AA$, which has negligible impact on the derivation of the dust attenuation from the GTC spectrum.
For the variability analysis, since we used EWs of emission lines and only focused on small wavelength ranges, the variation in the slit-loss correction is also negligible (which has been taken into account nonetheless).

\section{Fitting an exponential profile to broad \texorpdfstring{\ha}{Ha}}
\label{appendix:gemini_exp}

Recently, \citet{Rusakov_escattering_2025} showed 
that most high-redshift LRDs have broad lines that can be explained with exponential wings. They
interpret this as evidence for electron scattering
of the broad lines, which would imply a much narrower
intrinsic FWHM for the BLR, and up to two orders of
magnitude smaller SMBH masses. The effect is more 
pronounced if one uses a single Gaussian as
benchmark, as done by \citet{Rusakov_escattering_2025}. In contrast,
our double-Gaussian approach yields already a
relatively narrow FWHM, so the expected reduction in 
mass when using an exponential profile should be less
than two orders of magnitude.

To test this hypothesis, we repeat the Gemini/GMOS line
modeling by replacing the double-Gaussian parameterization with a single Gaussian, which we
then convolve with an exponential function. 
We assume that the intrinsic BLR profile is dominated by virial motion and is described by a Gaussian.
The ratio between the intrinsic emission, the
transmitted emission, and the scattered emission are set by
the scattering optical depth $\tau_{\rm sc}$, such that the transmitted
and scattered fractions are $\exp(-\tau_{\rm sc})$ and
$1-\exp(-\tau_{\rm sc})$, respectively. 
The exponential
kernel has a width of $W \equiv (a \tau_{\rm sc} + b) \sqrt{T_{\rm e}/10,000~\mathrm{K}}$, where the coefficients $a$ and $b$ are
from \citet{Rusakov_escattering_2025}.

\begin{figure}
{\phantomsubcaption\label{f.gmos.expfit.a}
 \phantomsubcaption\label{f.gmos.expfit.b}
 \phantomsubcaption\label{f.gmos.expfit.c}
 \phantomsubcaption\label{f.gmos.expfit.d}
 \phantomsubcaption\label{f.gmos.expfit.e}
 \phantomsubcaption\label{f.gmos.expfit.f}}
\includegraphics[width=\columnwidth]{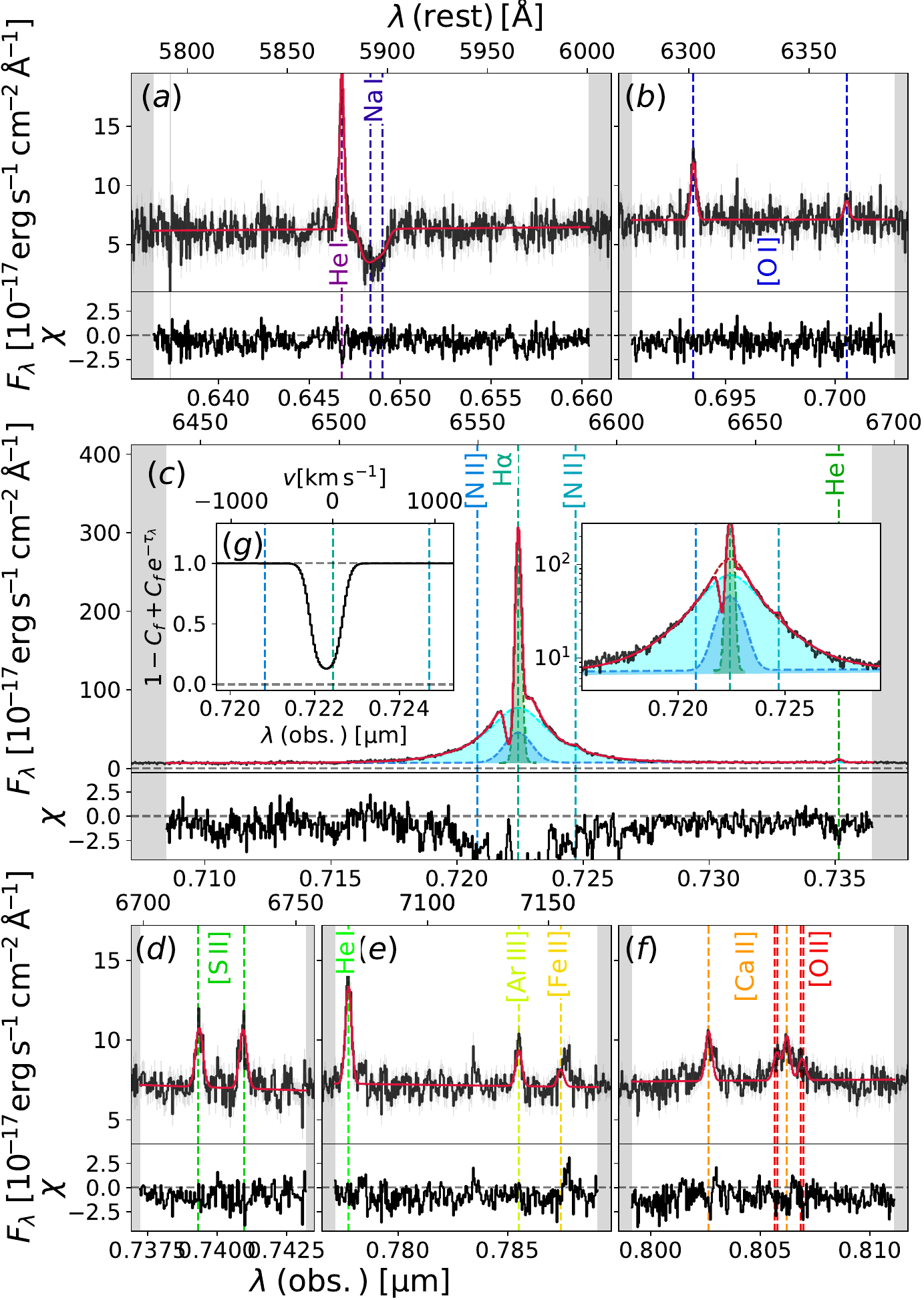}
\caption{
Alternative fit of the Gemini/GMOS spectrum assuming the broad component in \ha\ is described by exponential wings made by electron scattering.
The rest of the parametrization is the same as the double-Gaussian fit shown in Figure~\ref{f.gmos.gaussfit}.
The exponential model provides a statistically better fit compared to the double-Gaussian model, with $\Delta \text{BIC}=199$.
However, the physical interpretation of the exponential profile remains unclear for \target.
}
\label{f.gmos.expfit}
\end{figure}

The resulting model is shown in Figure~\ref{f.gmos.expfit}, with the same meaning as
Figure~\ref{f.gmos.gaussfit}. 
The best-fit model parameters as well as the derived black hole parameters are listed in Table~\ref{tab.gmosexp}.
We infer $\tau_{\rm sc} = 1.79\pm0.05$, and $T_\mathrm{e}=2,500$~K. The latter value is much lower than typical ISM conditions, where $T_\mathrm{e}\sim 10,000$~K. The model suggests that the observed width of the 
exponential profile is too narrow for the inferred scattered fraction at typical ISM $T_\mathrm{e}$, thus requiring lower temperature. On one hand, this resonates with the cool
envelope hypothesis \citep{linxiaojing_locallrd_2025}, while on the other, the discrepancy
could be due to the simplistic assumption of a simple scattering screen
\cite[which are known to yield to inconsistent line profiles in LRDs;][]{Brazzini_2025}.
More importantly, we find a lower $\rm FWHM_{b}(H\alpha) = 520\pm10~\kms$, resulting in a smaller
black hole mass of \redtxt{$\log(M_\bullet/\Msun) = 5.97\pm0.02$},
or $6.0\pm0.3$, after accounting for the
uncertainties in the calibration (and assuming no dust attenuation). While lower than
the fiducial model, the difference is not
dramatic.

\begin{table}
\centering
\caption{
Best-fit parameters for the \ha\ emission and absorption as well as the NaD absorption assuming the broad \ha\ has an exponential profile. We also list the derived BH mass and Eddington ratio.}\label{tab.gmosexp}
\begin{tabular}{ccc}
\hline
\hline
$v_\mathrm{b}$  & $1_{-1}^{+2}$ & [\kms]       \\
$F_\mathrm{b}({\rm H\alpha})$ & \redtxt{$3100_{-20}^{+20}$} & [\fluxcgs[-17][]] \\
$\rm FWHM_{\mathrm{b}}(\ha)$ & $520_{-10}^{+10}$ & [\kms]       \\
$\tau_\mathrm{e}$        & $1.79_{-0.05}^{+0.05}$ & [---]        \\
$T_\mathrm{e}$  & $0.25_{-0.01}^{+0.02}$ & [$10^4\,\mathrm{K}$] \\
$f_\mathrm{scatt}$ & $0.833_{-0.009}^{+0.007}$ & [---]        \\
$W$             & $0.00137_{-0.00001}^{+0.00001}$ & [$\mu \mathrm{m}$] \\
$v_\mathrm{abs}$ & $-57_{-1}^{+1}$ & [\kms]       \\
$\sigma_\mathrm{abs}$ & $86_{-2}^{+2}$ & [\kms]       \\
$C_f$           & $0.92_{-0.02}^{+0.03}$ & [---]        \\
$\tau_0(\mathrm{\ha})$ & $3.0_{-0.2}^{+0.2}$ & [---]        \\
$v_\mathrm{Na\,I}$ & $-40_{-20}^{+20}$ & [\kms]       \\
$\sigma_\mathrm{Na\,I}$ & $100_{-20}^{+20}$ & [\kms]       \\
$C_{f,\mathrm{Na\,I}}$ & $0.49_{-0.05}^{+0.07}$ & [---]        \\
$\tau_0(\mathrm{Na\,I})$ & $2.2_{-0.9}^{+1.2}$ & [---]        \\
$\rm EW_\mathrm{abs}(\ha)$ & $6.3_{-0.1}^{+0.1}$ & [\AA]        \\
$\rm EW_\mathrm{b}(\ha)$ & $-342_{-3}^{+3}$ & [\AA]        \\
$\rm EW_\mathrm{abs}(\mathrm{Na\,I})$ & $6.2_{-0.6}^{+0.6}$ & [\AA]        \\
$\rm FWHM_{\mathrm{b}}(\ha)$ & $520_{-10}^{+10}$ & [\kms]       \\
$L_{\mathrm{n}}({\rm H\alpha})$ & \redtxt{$0.265_{-0.002}^{+0.002}$} & [$10^{42} \rm{erg\,s^{-1}}$] \\
$L_{\mathrm{b}}({\rm H\alpha})$ & \redtxt{$0.845_{-0.004}^{+0.006}$} & [$10^{42} \rm{erg\,s^{-1}}$] \\
$\log \mbh$   & \redtxt{$5.97_{-0.02}^{+0.02}$} & [\Msun]      \\
$\ledd$       & \redtxt{$0.94_{-0.06}^{+0.07}$} & [---]        \\

\hline
\end{tabular}


\end{table}

\bsp	
\label{lastpage}
\end{document}